\documentclass[onecolumn]{els-mrw} 
\usepackage{amsmath,amssymb,amsfonts,amsthm,makeidx,graphicx}
\usepackage{hyperref}
\usepackage{txfonts}
\usepackage{helvet}
\usepackage{physics}
\usepackage{bbm}
\newcommand{\beq}{\begin{equation}}
\newcommand{\eeq}{\end{equation}}
\newcommand{\kt}{\rangle}
\newcommand{\br}{\langle}
\newcommand{\ld}{\lambda}
\newcommand{\mcH}{\mathcal{H}}
\newcommand{\mcU}{\mathcal{U}}
\newcommand{\mcA}{\mathcal{A}}
\newcommand{\mbI}{\mathbb{I}}
\newcommand{\mbS}{\mathbb{S}}
\begin{document}

\chapter{\ Quantum Chaos and Quantum Information:\ 
\\ 
\ Interactions and Implications} % and Resources}
% the spaces above are placed so a space after sign :
%in the running title appears in later pages... 
%\chapter{Quantum Chaos and Quantum Information:\\ 
% An entropic approach}

\label{chap1}
\author[1]{Arul Lakshminarayan}%
\author[2,3]{Karol {\.Z}yczkowski}%
\address[1]{\orgname{Indian Institute of Technology Madras}, \orgdiv{Department of Physics}, \orgaddress{ Chennai 600036, India}}
\address[2]{\orgname{Jagiellonian University}, \orgdiv{Institute of Theoretical Physics}, \orgaddress{ ul. {\L}ojasiewicza 11, 30--348 Krak\'ow}, Poland}
\address[3]{\orgname{Polish Academy of Sciences}, \orgdiv{Center for Theoretical Physics}, \orgaddress{Al. Lotnik\'{o}w 32/46, 02-668 Warszawa, Poland}}

%\articletag{Chapter Article tagline: update of previous edition, reprint..}
\articletag{Chapter Article tagline: April 9, 2026}

\maketitle

\begin{glossary}[Keywords]
Quantum chaos, quantum information, entropy dynamics, random matrices, concentration of measure

\end{glossary}

\begin{abstract}[Abstract]
The notion of Shannon entropy is crucial for the theory of classical information. In quantum information theory, an analogous key role is played by the von Neumann entropy: quantum information processing is closely related to entropy dynamics. This reveals a direct link with the theory of quantum chaotic systems, which can be characterized by a positive entropy production. Furthermore, noise, which inevitably affects any quantum system, can be modeled by a random quantum operation or by coupling to an environment in a generic chaotic state. In this contribution, we emphasize the universality of quantum chaotic dynamics and discuss its implications for quantum information processing.

\end{abstract}

\section{Introduction}

\hskip 5.0cm {\sl
Quantum chaotic systems are all alike; every 
regular system is regular in its own way.\footnote{Inspired by Leo Tolstoy, {\sl Anna Karenina},  Russkiy Vestnik, Moscow, 1878.}
}

\medskip

The notion of entropy plays a pivotal role in the theory of classical and quantum information processing \cite{NielsenChuang}. As entropy is used to measure the amount of information in a given system, it is important to describe entropy changes during the time evolution of a quantum system interacting with an environment \cite{Wa18}. Without detailed knowledge of this process, it is convenient to assume that the interaction is generic and can be modeled by random quantum structures.
On the other hand, classical deterministically chaotic systems can also be characterized by a positive {\sl dynamical entropy} of Kolmogorov and Sinai \cite{Sin09}, 
measuring generation of information along the orbit \cite{ozorio1989,ott-2002}.
Studies of quantum analogues, pioneered about half a century ago \cite{Berry_Balazs_QMaps,CasChiIzrFor1979},
revealed universal spectral properties of such systems \cite{BGS84}. Independently of the specific features of the underlying classical dynamics, the corresponding quantum systems display generic properties that can be described by suitable ensembles of random matrices \cite{Wi51,Mehta,Haake}.
From this perspective all quantum chaotic systems are alike, as the only factor which matters is the symmetry class, a given system belongs to \cite{Wi32}.
 There exist three universality classes discovered by Dyson \cite{Dy62},
 later extended to the 'ten-fold way' by Altland and Zirnbauer \cite{AZ97}.

The unitary time evolution of typical quantum systems in the classically chaotic regime is described by matrices with properties similar to random unitary matrices distributed according to the Haar measure on the unitary group. As a simple example, one can consider the quantum kicked pendulum or rotor with broken time-reversal symmetry \cite{Izrailev-1990}.
Genuine randomness of the uniform Haar measure of states and unitaries need exponential resources in the number of qubits $n$ making up the system with $N=2^n$ states. 
Remarkably, deterministic quantum chaotic systems, such as the kicked rotor, can be  efficiently simulated \cite{GeoShe2001}
with a polynomial number of two-qubit gates, $O(n^3)$. Another paradigmatic model of quantum chaos, the baker's map, is also efficiently simulable using the quantum Fourier transform \cite{Schack1998}. 
These pave the way for a quantum chaotic route to approximate Haar randomness that can be  characterized by notions of quantum $t-$designs \cite{Renes_ScottCaves_Designs,Roberts2017Chaos} and psuedorandomness \cite{EWSL03}.

 For typical values of its parameters, a suitable deterministic quantum chaotic system yields a Floquet operator $U$ of size $N$, describing its discrete-time evolution, which provides fair approximations \cite{ZS01,HLW06,BCSZ09} of various structures, including state and operator entanglement, crucial in quantum information theory:
\smallskip

\noindent
I) Haar-{\bf random unitary}  $\sim$ evolution operator,
    $U \in U(N)$;

\noindent
II) Haar {\bf random pure quantum state},  $\sim$ $|\psi\rangle = U|0\rangle \in {\cal H}_N$, 
where $|0\rangle$ denotes an arbitrary fixed state.

\smallskip
\noindent
Consider now a unitary operator $V$ of order $N^2$ acting on a bipartite space, ${\cal H}_A \otimes {\cal H}_B$,
which leads to 

\noindent
III) a random {\bf Ginibre matrix} $G$ $\sim $ defined via the bipartite state,
$|\phi_{AB}\rangle = V|00\rangle = \sum_{i,j=1}^N G_{ij}|i,j\rangle$;

\smallskip
\noindent
Making use of the notion of partial trace we arrive at

\noindent
IV) random {\bf mixed quantum state} (Wishart matrix) $\sim$
$\rho ={\rm Tr_B} |\phi_{AB}\rangle \langle \phi_{AB}| = GG^{\dagger}$, with partial trace over
$N$-dimensional 
subspace ${\cal H}_B$,

 \noindent
V) random {\bf quantum operation} \cite{BCSZ09} $\sim$ 
$\rho'=\Psi(\rho)= {\rm Tr_B}  
(V(\rho \otimes |0\rangle \langle 0|)V^{\dagger})$,
  with   $V\in U(N^3)$ and  
  $|0\rangle \in {\cal H}_B$ of dimension $N^2$.

\smallskip
\noindent
Hence, quantum chaotic systems are useful for mimicking generic quantum noise \cite{Sh01,EWSL03}. In an analogous manner, one can construct further structures, including quantum supermaps, random measurements, and random Lindblad generators \cite{DLTCZ19}, whose statistical
properties can also be modeled using appropriate ensembles of random matrices \cite{Haake}.

Various techniques relying on random matrices are often used in quantum information theory \cite{CollisNechita2015}. Exemplary applications include randomized benchmarking \cite{EAZ05,RandomBench2007,HROWE22},
random measurements \cite{vEB12,EKHBKDCZV20,EFHKKPVZ23,Cieslinski2024},
shadow tomography \cite{Huang2020},
data hiding, quantum encoding or  encryption \cite{Hayden2004randomizing} and decoupling theorem \cite{Hayden_Decoupling_2008}.
The theory of random matrices allows one to obtain exact analytic results concerning mean values averaged over the unitary group. By making use of the concentration-of-measure phenomenon, one can derive useful bounds that limit the probability of deviations from the mean value certain functions of  a random state \cite{HLW06}. These techniques allowed Hastings to prove the violation \cite{Ha09} of the conjectured additivity of the minimum output entropy of a quantum channel. A counterexample to this conjecture could, in principle, be constructed using a quantum chaotic system, although the dimension $N$ of a random
operation from item V) above would have to be very large.

The goal of this contribution is to demonstrate the strong links and cross-fertilization between two important fields of science: the theory of quantum chaos and quantum information. 
Considering the pedagogical value of simple models that are easy to implement, we have used the quantum kicked rotor or standard map on the torus and the baker's map to illustrate basic concepts, and hope that this makes the subject accessible even to relative novices. Although we aimed to make this review up to date, due to lack of space, we had to limit the topics and  references to a small number, which does not reflect the vast recent literature on the subject.

This work is organized as follows. In the next section, we analyze various links between randomness and chaos in classical and quantum settings, using simple models to illustrate the applicability of various random matrix measures.
%{\color{blue} Perhaps this long section could be split into smaller sections?}
In Sec. 3 we discuss Haar random state and unitaries, the notion of $t-$designs and employ quantum chaotic models to motivate how they give rise to approximate designs. We also discuss various implications for quantum information and computing, from protocols and simulation to quantum algorithms. Section 4 is dedicated to the implications of random matrix theory to entanglement, both in bipartite and multipartite settings. In Sec. 5 we discuss entanglement generated by quantum chaotic dynamics to compare with that of random states and unitaries. This is again a vast area of many-body quantum theory, and we only give, an admittedly biased, glimpse into the topic. Section 6. discusses quantum channels and chaos, including
random channels and 
concentration of measure. 
We end with an outlook and open questions in Sec. 7.

%The generation of quantum entanglement by quantum chaotic dynamics is presented in Section 3. 
%Quantum channels and designs are discussed in Section 4.

%{\color{blue} to be extended and improved..,}

%{Sr94} M. Srednicki, Phys. Rev. E 50 (1994)
%Thermalization and Eigenstate Thermalization Hypothesis

%{Sh01} D. L. Shepelyansky, Quantum Chaos & Quantum Computers
% Physica Scripta, T90, 112 (2001),

%  a) The onset of chaos causes strong entropy growth, 
% which leads  to degradation of peformance of a quantum processor
%  b) Quantum Information Scrambling

% Quantum chaos stands behind information scrambling, 
% as local information is spread across all degrees of 
% freedom in a many-body system

% Current quantum processors are used to simulate chaotic systems
% that are too complex for classical computers, 

% Quantum Error Correction can be used to fight against
% increase of entropy induced by chaos

\section{Randomness from Chaos}
\label{chap1:sec1}

In this section we introduce the basic notions needed to connect quantum chaos with quantum information. Emphasizing how deterministic classical dynamics can generate effective randomness, we briefly review key ideas from classical chaos, with a focus on Hamiltonian systems and their discrete-time (stroboscopic) realizations. Two paradigmatic examples are the standard map (or kicked rotor) on the torus, which exhibits a generic transition from integrability to chaos, and the baker’s map, which provides an exactly solvable model of complete chaos.
Upon quantization, the signatures of classical chaos manifest themselves in universal spectral and dynamical properties that are well described by random matrix theory. We illustrate this connection using the quantum standard map, employing standard diagnostics such as the nearest-neighbor level spacing distribution and the spectral form factor.
\subsection{Classical chaos}
Classical chaos in deterministic systems can serve as a statistical proxy for randomness \cite{LicLie2013,ott-2002,ORNSTEIN1989}. An ergodic hierarchy
\[ \mbox{Ergodic} \;\Leftarrow\; \mbox{Mixing} \;\Leftarrow\; \mbox{K-systems} \;\Leftarrow\; \mbox{Bernoulli}\]
organizes dynamical systems according to increasing degrees of stochasticity \cite{LicLie2013,ott-2002,ArnAve1989}.
Ergodic systems admit an invariant measure with the property that time averages along almost all trajectories coincide with phase-space (ensemble) averages. Mixing systems exhibit a loss of memory in the sense that correlations between sufficiently regular observables decay to zero at long times.
K-systems are characterized by positive Kolmogorov–Sinai dynamical entropy \cite{Sin09,ozorio1989}, reflecting intrinsic dynamical randomness; in smooth hyperbolic systems \cite{ArnAve1989} this is often accompanied by positive Lyapunov exponents \cite{LicLie2013} and exponential instability of trajectories. At the apex of the hierarchy, Bernoulli systems are measure-theoretically isomorphic to sequences of independent random variables, making them the closest deterministic analogues of ideal random processes \cite{ORNSTEIN1989}.
The arrows indicate implication: Bernoulli systems are K-systems, which are necessarily mixing, and mixing systems are ergodic.

Real-world 
dynamical systems, however, rarely fall cleanly into purely regular or purely chaotic categories. Instead, they typically exhibit a complex coexistence of regular (stable) and chaotic (unstable) dynamics within the same phase space \cite{TelGru2006}. For Hamiltonian systems, the transition to chaos is commonly analyzed through variation of control parameters, starting from an integrable limit, in which the number of independent constants of motion at least equals the number of degrees of freedom. As these integrals of motion are progressively destroyed, invariant tori break up, chaotic regions form, and the system develops a mixed phase space consisting of regular islands embedded within a chaotic sea.

\begin{figure}[h!]
    \centering
\includegraphics[width=\linewidth]{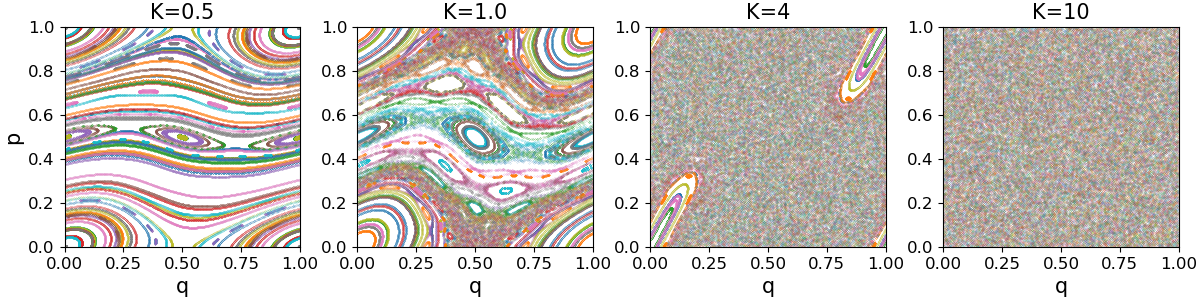}
    \caption{Phase space of the standard map, or kicked rotor, on the torus. 
    As the kicking strength is changed from the integrable case $K=0$ it undergoes a transition to almost complete chaos for $K\gg 5$. 
    Second panel, with mixed phase space,
    represents the case just above the critical value, $K>K_c \approx 0.9716$, at which  the last rotational KAM torus breaks, leading to a global diffusion in momentum. 
    Right most panel shows
      the case $K=10$, for which no discernible stable islands are present.}
    \label{fig:class_stand_map}
\end{figure}

The simplest class of Hamiltonian systems capable of exhibiting chaos consists of systems with a single degree of freedom subject to periodic forcing. When the forcing takes the form of a Dirac delta train, the dynamics can be reduced analytically to a discrete-time map, or canonical transformation, whose iterations fully describe the evolution \cite{Berry_Balazs_QMaps,CasChiIzrFor1979}.
A paradigmatic example is the kicked rotor (or kicked pendulum) \cite{Chirikov-1979,LicLie2013,ott-2002}, described by the Hamiltonian
\begin{equation}
H\,=\, \frac{p^2}{2} \, -\, \frac{K}{4 \pi^2}\cos(2 \pi q)\sum_{n=-\infty}^{\infty} \delta\left(\frac{t}{T}-n\right).
\label{eq:kickham}
\end{equation}
Integrating the equations of motion from immediately before the 
$n$th kick, $n^-$, to immediately before the next, $(n+1)^-$, yields the stroboscopic map known as the standard map. If $M_S(q_n,\, p_n)=(q_{n+1},\, p_{n+1})$,
\begin{equation}
\begin{split}
q_{n+1} &= q_n + p_{n+1} \; (\mathrm{mod}\,1),\\
p_{n+1} &= p_n - \frac{K}{2\pi}\sin(2\pi q_{n}).
\end{split}
\label{eq:standmap}
\end{equation}
Here we have chosen the kick period as the unit of time $T=1$.The periodic boundary condition in position (angle) naturally follows from the periodicity of the potential. Due to the boost symmetry $p \mapsto p+1$
this periodicity can be extended to the conjugate momentum variable as well. Therefore, it is convenient to formulate the dynamics on the unit torus $(q,p) \in [0,1]\times [0,1]$. Some representative dynamical regimes are illustrated in Fig.~\ref{fig:class_stand_map}.
The map is area-preserving and  represents the typical Poincar\'e section of a two-degree-of freedom bounded Hamiltonian system. The fixed point at the origin $(0,0)$ loses stability at $K=4$, paving the way for global chaos. The Lyapunov exponent for $K\gg 5 > K_c$ is approximately given by 
\begin{equation}
    \lambda=\ln(\frac{K}{2}) +\frac{1}{K^2-4}+O(K^{-6}),
    \label{eq:Chirikov}
\end{equation}
with the leading term, $\ln(K/2)$, derived by Chirikov \cite{Chirikov-1979} and the corrections by Tomsovic \cite{Tomsovic2007}.

\begin{figure}
    \centering
    \includegraphics[width=\linewidth]{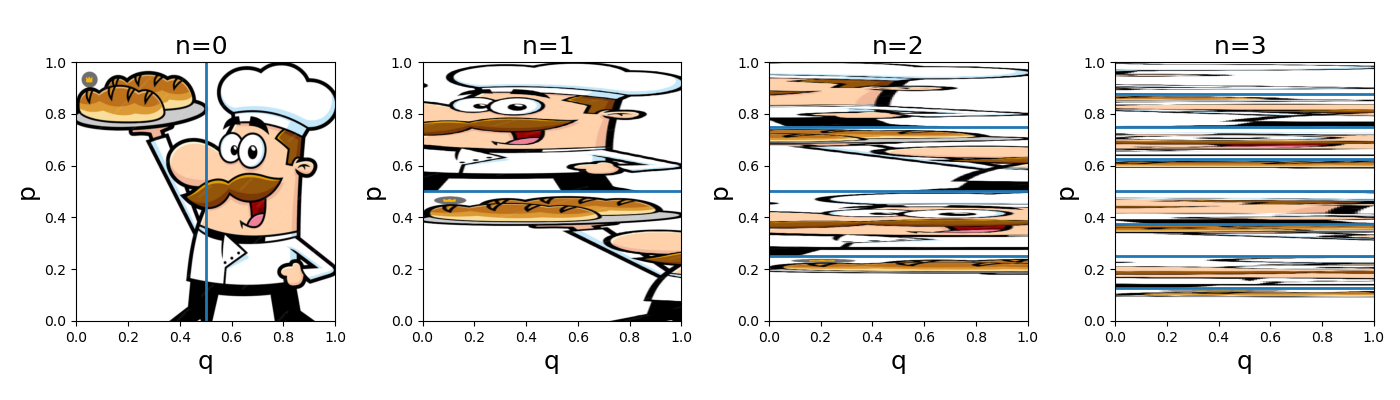}
    \caption{Action of the baker's map on the image of a professional. The left vertical half at time $n=0$ gets stretched horizontally by a factor of 2 and compressed by the same factor vertically and is put in the bottom horizontal half at time $n=1$. Similarly the right vertical half gets mapped to the top horizontal half. Also shown are the iterations at times $n=2$ and $n=3$.}
    \label{fig:cl_baker_map}
\end{figure}

An archetypal Bernoulli system is the baker’s map on the unit square,
\begin{equation}
M_B(q_n, \,p_n)=(q_{n+1},\,p_{n+1}) = 
\begin{cases}
\left(2q_n,\, \dfrac{p_n}{2}\right), & 0 \le q_n < \tfrac12,\\[6pt]
\left(2q_n-1,\, \dfrac{p_n+1}{2}\right), & \tfrac12 \le q_n \le 1,
\end{cases}
\label{eq:cl_bakers_map}
\end{equation}
a solvable, invertible, area-preserving map that provides the simplest model of fully developed Hamiltonian chaos \cite{LicLie2013}. The dynamics is uniformly hyperbolic, with constant expansion and contraction rates, yielding a Lyapunov exponent,
$\lambda=\ln 2$, in this case
equal to the Kolmogorov--Sinai dynamical 
entropy \cite{ott-2002}. 
The chaotic mechanism is most transparent in the binary representation:
writing $q_0=0.a_0a_1a_2\cdots$ and $p_0=0.a_{-1}a_{-2}\cdots$, one finds
$q_1=0.a_1a_2\cdots$ and $p_1=0.a_0a_{-1}a_{-2}\cdots$. Thus the baker’s map
acts as a left shift on the bi-infinite symbolic sequence,
\begin{equation}
(q_0,p_0)=\cdots a_{-2}a_{-1}\bullet a_0a_1a_2\cdots
\;\longrightarrow\;
(q_1,p_1)=\cdots a_{-1}a_{0}\bullet a_1a_2\cdots ,
\end{equation}
with $q$ encoded to the right of the $\bullet$ and $p$ to the left. Repeated
stretching, compression, and stacking produce strong mixing, as illustrated
in Fig.~\ref{fig:cl_baker_map}. 
% nice remark I was not aware of. 
% Let us tentatively include it in!  (KZ)
The map was invented by 
E~Hopf in 1937 and apparently christened the "baker's map", based on its relation to the making of pastry dough, by von Neumann; see \cite{BakersMap} for original references and some historical details.
%%% OK, I had commented it for saving references, but now it is $\infty +1$!

Despite its abstract nature, the baker’s map
is invaluable as a fully analyzable model of chaos and concretely realizes the symbolic  left-shift underlying homoclinic chaos \cite{TelGru2006,ott-2002}. The quantum versions of the standard map and the baker's map have been studied extensively.

\subsection{Quantum chaos}
 
Quantum chaos originated in the study of quantum dynamics corresponding to classically chaotic systems, see \cite{GVZ1989} for a collection of lectures on key developments till 1989. It has since expanded considerably to include many-body systems via the Eigenstate Thermalization Hypothesis \cite{Sr94,Alessio2016},  many-body localization \cite{Prosen_Challenge2020}, cold-atom physics \cite{Raizen:2011}, and holography \cite{Holography_Chaos}, just to highlight a few topics.
%Several books on the subject have also appeared \cite{Haake, ozorio1989,Wimberger2014,stockmann}.

Interactions with quantum information and computing is the topic of this article, which will necessarily contain a restricted sampling of a rapidly growing body of literature. We focus on illustrating concepts using the quantum standard map and the quantum baker’s map, whose classical counterparts were discussed in the previous section. It is emphasized that this is convenient, as they are simple models: the standard map showing a smooth transition from integrable to chaotic dynamics, and the baker map being a solvable model of hard chaos and its quantum version being built of Fourier transforms. The results they illustrate are, naturally, of much broader applicability.

\subsubsection{Quantum standard and baker's map}

The quantum standard map is defined as the unitary Floquet operator that propagates the system over one kick period, and is the quantum analogue of the classical map $M_S$. It may be written as
\begin{equation}
U_S = U_F\, U_K, \qquad
U_F = \exp\!\left(-\frac{i}{2\hbar}\hat{p}^2\right), \quad
U_K = \exp\!\left(\frac{iK}{4\pi^2\hbar}\cos(2\pi\hat{q})\right),
\label{eq:qsm1}
\end{equation}
where $U_F$ describes free evolution and $U_K$ the kick. The time between kicks has been set to unity. The map admits three natural topologies: the plane, the cylinder (periodic in $q$), and the torus. While the cylindrical case underlies much of the literature on dynamical localization \cite{CasChiIzrFor1979,Izrailev-1990}, we adopt the torus quantization, which yields a finite-dimensional Hilbert space \cite{Izrailev-1990} and is particularly suited to quantum-information applications.

We briefly summarize the quantization on the unit torus \cite{Saraceno1990}. The position states $\ket{n}$, $n=0,\dots,N-1$, form an orthonormal basis with boundary conditions specified by a phase $\beta$,
\begin{equation}
\ket{n+N} = e^{-2\pi i \beta}\ket{n}, \qquad 0\le \beta <1.
\end{equation}
Momentum states $\ket{\tilde m}$ ($m=0, \cdots, N-1)$ are obtained via a discrete Fourier transform,
\begin{equation}
\ket{\tilde m} = F_N^{\dagger} \ket{m}, \qquad
\braket{n}{\tilde m} = \frac{1}{\sqrt{N}}
\exp\!\left[\frac{2\pi i}{N}(n+\alpha)(m+\beta)\right]\equiv (F_N^{\dagger})_{nm},\qquad {\rm and}\qquad \ket{\widetilde{m+N}} = e^{2\pi i \alpha}\ket{\tilde m}.
\label{eq:DFT}
\end{equation}
The phases $\alpha$ and $\beta$ determine the quantum boundary conditions on the torus and the Fourier transform $F_N$ depends on these (although not explicitly indicated). If the phases are $0$ (or $1/2$), there are periodic- (or antiperiodic-) boundary conditions.

Let $T_q$ denote the position translation operator, $T_q\ket{n}=\ket{n+1}$, so that $T_q^N=e^{-2\pi i\beta}\mathbbm{1}_N$, and let $T_p$ be the momentum boost operator with $T_p^N=e^{2\pi i\alpha}\mathbbm{1}_N$. The momentum states $\ket{\tilde m}$ are eigenstates of $T_q$ with eigenvalues $e^{-2\pi i(m+\beta)/N}$, while the position states $\ket{n}$ are eigenstates of $T_p$ with eigenvalues $e^{2\pi i(n+\alpha)/N}$.
The "phase-space lattice" is thus given by
$(q,p)=\bigl((n+\alpha)/N,(m+\beta)/N\bigr)$. The effective Planck constant is
\begin{equation}
\hbar = \frac{1}{2\pi N},
\end{equation}
reflecting the unit phase-space area. The classical limit corresponds to $N\to\infty$, where the grid becomes dense and $\hbar\to 0$. 
For a detailed treatment of quantizing a toral phase space see work by Schwinger \cite{Schwinger}.

The quantum standard map, Eq.~(\ref{eq:qsm1}), adapted on the torus is then given in the position representation as
\begin{equation}
    U_S=F_N^{\dagger} D_F F_N D_K.
    \label{eq:quantum_standard_map}
\end{equation}
Here $D_F$ and $D_K$ are diagonal unitary matrices:
\begin{equation}
    \bra{n'}D_K\ket{n}=\exp\left\{\frac{i NK}{2 \pi} \cos\left[ \frac{2\pi}{N}\left(n+\alpha\right) \right]\right\} \delta_{nn'},\qquad \bra{m'}D_F\ket{m}=\exp\left[-i\frac{i \pi}{N}(m+\beta)^2\right]\delta_{m'm}.
\label{eq:quantum_standard_map_Ds}
\end{equation}

This structure is generic to kicked maps with a potential $V(q)$ and enables efficient numerical implementation by  matrix multiplication
based on fast Fourier transform.
The Floquet operator $U_S$ depends on four parameters:
\begin{enumerate}
    \item \emph{Kick strength $K$.} 
    For $K=0$ the classical and quantum dynamics are integrable. Increasing $K$ leads to widespread classical chaos, and for sufficiently large $K$ the spectrum of $U_S$ displays fluctuations consistent with random-matrix universality \cite{Mehta,Haake}, in accordance with the Bohigas--Giannoni--Schmit (BGS) conjecture \cite{BGS84}.
    
    \item \emph{Phase $\beta$.} 
    An Aharonov--Bohm--type phase controlling time-reversal (TR) symmetry; TR invariance is present for $\beta=0$.
    
    \item \emph{Phase $\alpha$.} 
    For $\beta=0$, parity symmetry is preserved when $\alpha=0$ or $\alpha=1/2$.
    
    \item \emph{Hilbert-space dimension $N$.} 
    The effective inverse Planck constant, with $N\to\infty$ defining the classical limit.
\end{enumerate}
%{\color {blue}
The quantum kicked rotator, especially on the cylinder with momentum diffusion, and several generalizations of this model have been used as a main working horse
of the entire field of quantum chaos
for more than  4 decades, see \cite{Santhanam2022} for a recent review. The current state of the art is
presented in an accompanying contribution
\cite{BC26}.
%}

Unlike the standard map, the baker’s map does not arise from a generating Hamiltonian. Its quantization therefore proceeds directly from the classical canonical transformation in Eq.~\eqref{eq:cl_bakers_map}. On the unit torus, the quantum baker’s map \cite{Balazs1989,Saraceno1990} may be written as
\begin{equation}
    U_B= F_N^{\dagger}
    \begin{pmatrix}
        F_{N/2} & 0 \\
        0 & F_{N/2}
    \end{pmatrix}
    = F_N^{\dagger}\left(\mathbbm{1}_2 \otimes F_{N/2}\right),
    \label{eq:quantum_baker}
\end{equation}
where $F_N$ denotes the discrete Fourier transform, see Eq.~(\ref{eq:DFT}). For antiperiodic boundary conditions in both canonical variables ($\alpha=\beta=1/2$), the quantum  map retains both the classical parity and time-reversal (TR) symmetries, providing a particularly clean setting for quantum--classical correspondence \cite{Saraceno1990}. The Hilbert space dimension $N$ is taken to be even, allowing for an exact partition of the Hilbert space into two equal subspaces, a quantum version of the vertical cut at $q=1/2$ in the classical baker's map, see Fig.~\ref{fig:cl_baker_map}.

The tensor-product structure makes explicit a bipartite interpretation: a qubit coupled to a qudit of dimension $N/2$. 
% qudit of a qunit?
% do we assume d=N/2 ?
% perhaps:.. coupled to an ancilliary system of dimension $N/2$. 
In this representation, the dynamics consists of a Fourier transform acting solely on the qudit, followed by the global inverse transform $F_N^{\dagger}$, which entangles the two subsystems. This viewpoint naturally generalizes to hierarchies of quantum baker maps constructed via partial Fourier transforms on multiple qubits \cite{SchackCaves2000}.

More broadly, both the generic standard map and the fully chaotic baker’s map admit compact finite-dimensional unitary representations that permit detailed analytical and numerical study. For instance, rigorous results concerning quantum ergodicity, implying that typical eigenstates equidistribute in the classical limit $N \rightarrow \infty$,
have been obtained for the quantum baker's map \cite{DegliEsposti2006,Shou2025}.  Closely related paradigmatic models include the quantum baker on the sphere \cite{Prot_baker_sphere_1999}, the quantum kicked top \cite{Haake1987_KickedTop,Haake} — whose classical phase space is $S^2$ and which admits a realization in the totally symmetric subspace of many qubits—and the cat map, a linear toral automorphism \cite{ArnAve1989} whose quantum version \cite{Hannay_Berry_CatMap}, while fully chaotic classically, exhibits non-generic spectral statistics. Another extensively studied area-preserving map is the Harper map \cite{Lima_kickedHarper}, whose cosine dispersion in both $q$ and $p$ renders the torus topology particularly natural and provides an alternative route to quantum chaos.

\subsubsection{Random-matrix fluctuations from quantum chaos}

The spectrum of Hamiltonians for time independent systems and unitary maps for Floquet systems are of interest. If the eigenvalue problems are
\begin{equation}
    H\ket{E_k}=E_k\ket{E_k}, \qquad U\ket{\phi_k}=e^{i \phi_k}\ket{\phi_k},
\end{equation}
quantum chaos is found to be reflected in the properties of $\{E_1, E_2, \cdots E_N\}$ or $\{\phi_1, \phi_2, \cdots \phi_N\}$ and the corresponding eigenstates. $N$ could be the dimension of the Hilbert space or a finite truncation of it. The eigenangles $\phi_k$ are sometimes referred to as quasienergies.

As alluded to above, the Bohigas--Giannoni--Schmit (BGS) conjecture \cite{BGS84} asserts that quantum systems whose classical counterparts are chaotic exhibit universal spectral fluctuations described by random-matrix ensembles determined solely by symmetry. We briefly recall the relevant elements of random-matrix theory (RMT) \cite{Mehta,Bohigas-LesHou,Haake,livan2018} and spectral statistics widely used, and illustrate them with the quantum standard map in Eq.~\eqref{eq:quantum_standard_map}.
The fundamental symmetry classes are the Gaussian orthogonal, unitary, and symplectic ensembles (GOE/GUE/GSE) of Hermitian matrices, together with their unitary counterparts, the Dyson circular ensembles (COE/CUE/CSE). The defining features of these ensembles, constituting Dyson’s threefold way, are summarized in Table~\ref{tab:dyson_threefold_way} 
inspired by Dyson \cite{Dy62}. 

The GOE/COE applies to time-reversal symmetric bosonic (or spinless) systems with $\mathcal{T}^2=+1$, where $\mathcal{T}$ is an anti-unitary time-reversal operator, and is invariant under conjugation by the orthogonal group $\mbox{O}(N)$. The GOE consists of real symmetric $N \times N$ matrices. The GUE/CUE describes systems without time-reversal symmetry and is invariant under conjugation by the unitary group $\mbox{U}(N)$, with the GUE comprising complex Hermitian matrices. The GSE/CSE is relevant for time-reversal symmetric fermionic systems with $\mathcal{T}^2=-1$, incorporates Kramers degeneracy, and is invariant under conjugation by the unitary symplectic group $\mbox{USp}(2N)$; the GSE may be viewed as consisting of Hermitian matrices with quaternion-real entries. All Gaussian ensembles are defined by independently distributed matrix elements with Gaussian weights, leading, up to a choice of overall scale, to the probability density
\begin{equation}
P_{\beta}(H)\;\propto\;\exp\!\left[-\frac{\beta}{2}\,\mathrm{tr}\,H^{2}\right],
\end{equation}
where the Dyson index $\beta=1,2,4$ corresponds to the GOE, GUE, and GSE respectively. For large $N$, all Gaussian ensembles lead to  the Wigner semicircle law,
$\br \rho(E)\kt\sim \sqrt{2N-E^2}/\pi$,
which characterizes  the average density of states \cite{Wi51,Mehta,Haake,livan2018}.

The circular ensembles of unitary matrices naturally arise in Floquet systems, such as the kicked rotor and the baker’s map, as well as in the original motivation of Dyson-- unitary scattering ($S$-matrix) theory. Unlike Gaussian ensembles, the matrix elements of unitaries are not independent, and their construction requires group-theoretic machinery, which is by now standard and implemented in widely used software packages.
The central ensemble is the circular unitary ensemble (CUE), consisting of unitary matrices drawn uniformly from the compact group $\mbox{U}(N)$, i.e.\ with respect to the unique Haar measure on $\mbox{U}(N)$. Starting from an unstructured ensemble of complex matrices from the Ginibre ensemble (real and complex parts of all entries are independently and identically distributed according to the standard normal distribution), the CUE can be obtained via their QR decomposition \cite{Mezzadri2007}. The circular orthogonal ensemble (COE) consists of symmetric unitary matrices and may be constructed from a CUE matrix $W$ via
\begin{equation}
U = W W^{T},
\end{equation}
where $W^{T}$ denotes the transpose of $W$.

\begin{table}[t]
\centering
\renewcommand{\arraystretch}{1.25}
\setlength{\tabcolsep}{6pt}
\begin{tabular}{p{2.9cm} p{3.2cm} p{4.7cm} p{4.5cm}}
\hline
\textbf{Ensemble} & \textbf{Invariance (conjugation)} & \textbf{Applicability (symmetry class)} & \textbf{NNS $p(s)$ (Wigner surmise)}\\
\hline
\textbf{GOE\;/\;COE} 
\hskip 1.4cm
$\beta=1$
&
\begin{minipage}[t]{\linewidth}\vspace{2pt}
GOE: $H\mapsto O H O^{T}$,\\ $O\in \mbox{O}(N)$\\
COE: $U\mapsto V U V^{T}$,\\ $V\in \mbox{U}(N)$; equivalently \\$U=W W^{T}$ \\with $W$ Haar in $\mbox{U}(N)$
\vspace{2pt}\end{minipage}
&
\begin{minipage}[t]{\linewidth}\vspace{2pt}
Time-reversal symmetric (TRS) systems with $\mathcal{T}^2=+1$ (``spinless'' or integer spin, no Kramers degeneracy).\\
Typical: real-symmetric Hamiltonians / TR-invariant Floquet operators with appropriate basis choice.
\vspace{2pt}\end{minipage}
&
\begin{minipage}[t]{\linewidth}\vspace{2pt}
$p_{\beta=1}(s)=\dfrac{\pi}{2}\,s\,e^{-\pi s^{2}/4}$\\
level repulsion:\ $p(s)\sim s$ as $s\to 0$
\vspace{2pt}\end{minipage}
\\
\hline
\textbf{GUE\;/\;CUE} 
\hskip 1.4cm
$\beta=2$
&
\begin{minipage}[t]{\linewidth}\vspace{2pt}
GUE: $H\mapsto U H U^{\dagger}$,\\ $U\in \mbox{U}(N)$\\
CUE: $U\mapsto V U V^{\dagger}$,\\ $V\in \mbox{U}(N)$\\$U$ Haar in $\mbox{U}(N)$
\vspace{2pt}\end{minipage}
&
\begin{minipage}[t]{\linewidth}\vspace{2pt}
No time-reversal symmetry (TRS broken), e.g.\ magnetic field / vector potential / TR-breaking phases in toral maps.\\
Generic complex Hermitian Hamiltonians or generic unitary Floquet operators.
\vspace{2pt}\end{minipage}
&
\begin{minipage}[t]{\linewidth}\vspace{2pt}
$p_{\beta=2}(s)=\dfrac{32}{\pi^{2}}\,s^{2}\,e^{-4 s^{2}/\pi}$\\
level repulsion:\  $p(s)\sim s^{2}$
\vspace{2pt}\end{minipage}
\\
\hline
\textbf{GSE\;/\;CSE} 
\hskip 1.4cm
$\beta=4$
&
\begin{minipage}[t]{\linewidth}\vspace{2pt}
GSE: $H\mapsto S H S^{\dagger}$,\\ $S\in \mbox{USp}(2N)$\\
CSE: $U\mapsto V U V^{\dagger}$,\\ $V\in \mbox{USp}(2N)$
\vspace{2pt}\end{minipage}
&
\begin{minipage}[t]{\linewidth}\vspace{2pt}
Time-reversal symmetric with $\mathcal{T}^2=-1$ (half-integer spin with strong spin-orbit; Kramers degeneracy).\\
Spectral statistics usually discussed after removing the twofold Kramers degeneracy.
\vspace{2pt}\end{minipage}
&
\begin{minipage}[t]{\linewidth}\vspace{2pt}
$p_{\beta=4}(s)=\dfrac{2^{18}}{3^{6}\pi^{3}}\,s^{4}\,e^{-64 s^{2}/(9\pi)}$\\
level repulsion:\  $p(s)\sim s^{4}$
\vspace{2pt}\end{minipage}
\\
\hline
\end{tabular}
\caption{Dyson's threefold way for Hamiltonians (Gaussian ensembles) and Floquet/unitary spectra (circular ensembles). Here $s$ denotes unfolded nearest-neighbour spacings (mean spacing $=1$). The listed $p(s)$ are the Wigner surmises, which are
exact for $N=2$ but  also provide fair approximations
to the asymptotic density for large $N$.}
\label{tab:dyson_threefold_way}
\end{table}

\paragraph{Nearest neighbor spacings and ratio of spacings.}

Although the smooth density of states, $\overline{\rho(E)}$, for a generic Hamiltonian, chaotic or not, is system specific and does not follow the Wigner semicircle of the Gaussian ensembles, the fluctuations about the smooth density are a universal signal of quantum chaos and agree well with RMT. In the case of Floquet systems, though, even the density of states (eigenangles) is generically uniform,
$\overline{\rho(\phi)}=2\pi/N$, matching that of  circular ensembles. However, this can be the case whether the dynamics is chaotic or not, and the signatures are manifest only in the fluctuations about this smooth structureless density. To compare fluctuations from various background densities, we first unfold the eigenvalues so that the mean spacing is unity. As we focus on the unitary case, assuming a uniform distribution of eigenangles, the unfolded values and the corresponding nearest neighbor spacings (NNS) are 
\begin{equation}
x_k=\frac{N}{2\pi}\phi_k, \qquad s_k=x_{k+1}-x_k,
\label{eq:unfolded_angles}
\end{equation}
where we have assumed that $x_k$ are ordered so that $x_{k+1}\geq x_k.$ A widely used measure of fluctuations is the NNS distribution $p(s)$ \cite{Bohigas-LesHou}.  Wigner found $p(s)$ for $2\times 2$ GOE and GUE random matrices, and these were shown to be accurate approximations to large $N$ results \cite{Mehta}. They also apply equally to the circular unitary ensembles, all fluctuations being identical in the bulk of the spectrum. The form of the Wigner surmises for the NNS $p(s)$ are shown in Table~\ref{tab:dyson_threefold_way}. 

\begin{figure}
    \centering
    \includegraphics[width=\linewidth]{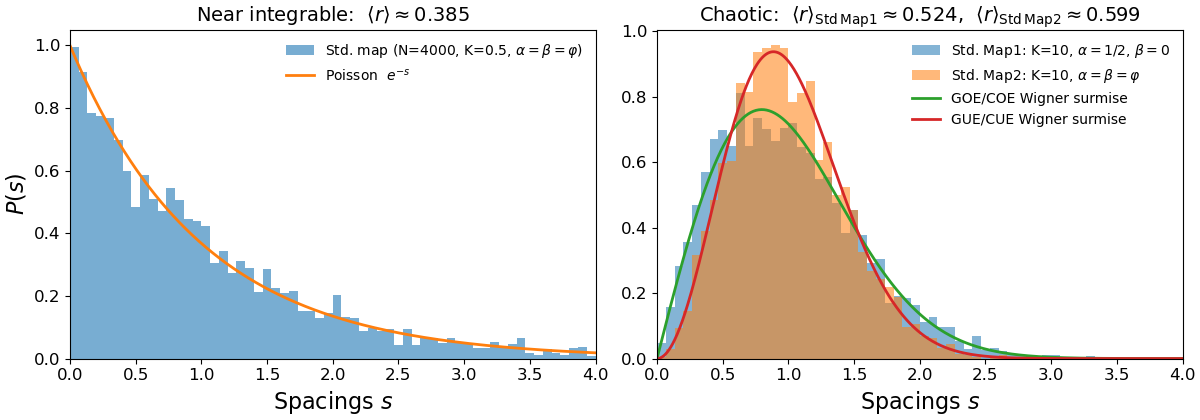}
    \caption{The nearest-neighbor spacing distribution for the quantum standard map, for $N=4000$. On the left is a case of near integrable dynamics, when $K=0.5$ and the NNS distribution is close to that of the Poisson. On the right is shown the case of fully developed chaos $K=10$, and two values of the boundary phases such that when $(\alpha=1/2,\,\beta=0$) there is both parity and time symmetry. The even and odd parity states are separately analyzed and combined in the histogram. The GOE surmise fits the statistics well. When the phases break time-reversal (parity is broken also, $\varphi=(\sqrt{5}-1)/2$), the GUE surmise fits well. Also shown at the top of the Figure are the corresponding values of the average of ratios and these may be compared to the expected values from RMT in Eq.~\eqref{eq:avg_r}. For the corresponding classical phase-spaces, see Fig.~\ref{fig:class_stand_map}. }
    \label{fig:SM_NSS}
\end{figure}

In contrast to the RMT fluctuations, applicable for quantum chaotic systems, the spectra of integrable systems are known to be generically statistically characterized, to good accuracy, by the Poisson point process. In this description, the energy levels are placed independently and uniformly, with a unit rate, leading to the NNS distribution
\begin{equation}
    p_0(s)=\, \exp(-s).
\end{equation}
Qualitatively, integrable systems have a large probability for small spacings, while quantum chaotic systems, if they obey RMT, have a vanishing probability as $s\rightarrow 0$. Thus, level-repulsion is a hallmark of quantum chaos. It is convenient to refer to the Poisson case as having a Dyson index $\beta=0$ (not to be confused with the phase $\beta$ used to break time-reversal in maps, the context will make the distinction clear).

%{\color{blue}
Investigation of the level spacing distribution $P(s)$ 
 is applicable only for unfolded spectra only,
but the ratio $r$ of consecutive spacings
introduced in \cite{OH07},
\begin{equation}
r_k=\mbox{min}\left(\frac{s_{k+1}}{s_k}, \frac{s_k}{s_{k+1}}\right)
\end{equation}
 does not require unfolding.
Analytic expressions 
for the density $P(r)$,
analogous to Wigner surmise,
were obtained for
$N=3$ for all three ensembles \cite{AtaBogGir}, 
but it is possible to differentiate 
between ensembles by comparing the average
values.
The exact results, derived for $N=3$
 along with the Poisson value expected for integrable systems ($\beta=0$) \cite{AtaBogGir} 
 are widely used,
\begin{equation}
    \langle r \rangle_{\rm Poisson}=2\ln2-1\approx 0.386, \;  \langle r\rangle_{\rm GOE}= 4-2\sqrt{3} 
    \approx 0.536(31), \; \langle r\rangle_{\rm GUE}= \frac {2\sqrt{3}}{\pi} -\frac{1}{2}\approx 0.603(00), \, \langle r\rangle_{\rm GSE}=\frac {32 \sqrt{3}}{15 \pi}-\frac{1}{2} \approx 0.676(74), 
    \label{eq:avg_r}
\end{equation}
as they describe well the asymptotic values 
obtained numerically for large $N$ corresponding to two last digits of each number (in brackets).
%}

Figure~\ref{fig:SM_NSS} shows the NNS distributions for the quantum standard map on the torus, for two values of the parameter $K$. At $K=0.5$, dynamics is still predominantly regular (see Fig.~\ref{fig:class_stand_map})
and the spacing distribution is a good fit to the Poisson curve. In this domain, it is irrelevant whether time-reversal is broken or not. However, the same figure also shows in the right panel the case for $K=10$, when there are no discernible regular orbits. The cases when TR is present and broken are clearly differentiated and there is good agreement with the RMT curves. Thus the quantum standard map with a definite matrix structure has the same properties of the CUE or COE.
The quantum standard map with a definite matrix structure has properties similar to those of CUE or COE. The ratio of spacings in the standard map spectrum, including in the mixed -partially regular and partially chaotic - regime has been studied in \cite{Yan2025}.

\paragraph{Two-point correlations and the spectral form factor.}

While NNS distribution and ratio are short-ranged, long-range statistics such as number variance $\Sigma^2(L)$ (the variance in the number of levels in an unfolded energy window of length $L$) have been used to probe fluctuations of the order of many mean spacings. These are defined in terms of two-point density-density correlations
\begin{equation}
    R_2(x,y)=\langle \rho(x) \rho(y)\rangle-\delta(x-y)\langle \rho(x) \rangle.
\end{equation}
Assuming stationarity in the bulk, $R_2(x,y)$ is only a function of $x-y=s$. We state the $N \rightarrow \infty$ result for the important case of the GUE, 
which holds also
for CUE,
\begin{equation}
    R_2^{\mbox{\tiny{GUE}}}(s)=\frac{\sin^2\pi s}{\pi^2 s^2}.
\end{equation}
It may be mentioned in passing that the nontrivial zero's of the Riemann zeta function, on unfolding, are found to obey this same law to an excellent approximation; see, for example \cite{Odlyzko1987distribution}, bolstering the expectation that there is a corresponding TR breaking quantum chaotic system \cite{Berry1986ZetaChaos}. In contrast for integrable systems, the Poisson process implies that there is no correlation between levels at $x$ and $y\neq x$, thus $R_2(x,y)=\langle \rho(x)\rangle \langle \rho(y)\rangle =1$, assuming unfolding to unit level density, hence 
\begin{equation}
    R_2^{\mbox{\tiny{Poisson}}}(s)=1.
\end{equation}

\begin{figure}
    \centering
    \includegraphics[width=\linewidth]{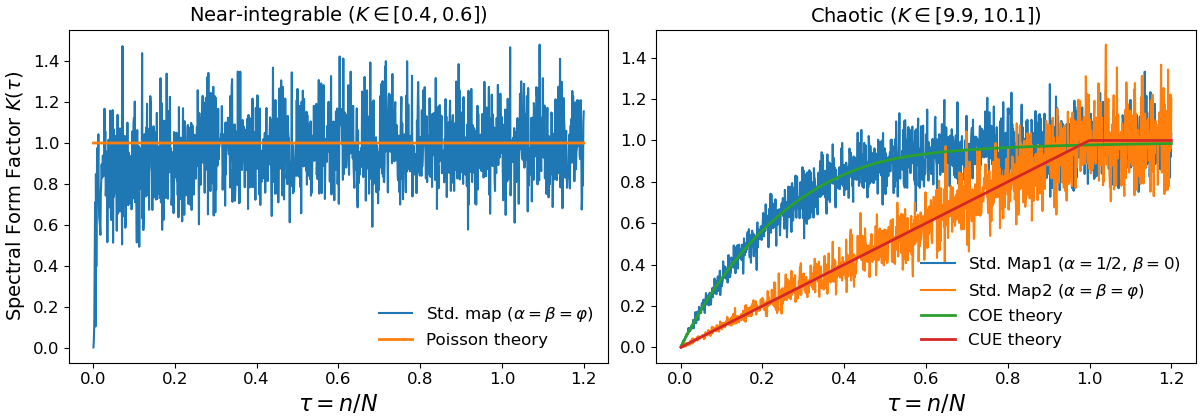}
    \caption{The spectral form factor defined in Eq.~\eqref{eq:SFF_defn} as a function of scaled time $\tau=n/N$ for: a) nearly integrable; b) chaotic case. The SFF is averaged over 50 samples of $K$ in the stated intervals. $N=1000$, and for the COE case with parity, the reduced spectrum with $N=500$ levels are considered. The center of the $K$ intervals coincide with those in Fig.~\ref{fig:SM_NSS} and details of the symmetries are also provided.}
    \label{fig:SM_SFF}
\end{figure}

While the number variance and related measures \cite{Mehta,Bohigas-LesHou} are in the energy domain, the spectral form factor (SFF) is in the time domain and related to the Fourier transform of the two-point function $R_2(s)$ \cite{Bohigas-LesHou,Haake}.
Once again focusing on Floquet systems or unitary operators, the SFF is defined as
\begin{equation}
K(n) =\left\langle
\frac{1}{N}\left|
\tr U^n
\right|^2
\right\rangle =1+\left\langle \frac{1}{N}
\sum_{k\neq l =1}^N \exp\left(\frac{2\pi i n}{N} (x_k-x_l)\right)
\right\rangle=1+\int_{-\infty}^{\infty}R_2(s) \exp\left(\frac{2\pi i n s}{N}\right) \, ds,
\label{eq:SFF_defn}
\end{equation}
where we $x_k$ are the unfolded eigenangles from Eq.~\eqref{eq:unfolded_angles}.
Using $R_2(s)$ for various ensembles, and defining the scaled time $\tau=n/N$, the RMT expectations of the spectral form factor for $\beta=0,1,2$ are as follows \cite{Bohigas-LesHou}:
\begin{equation}
\begin{split}
    K_{\mbox{\tiny{Poisson}}}(\tau)=1,\;\; \tau>0, 
    \ \ \ \
    K_{\text{\tiny{GOE}}}(\tau)=\begin{cases}
2 \tau-\tau \ln (1+2 \tau)& 0 < \tau \leq 1 \\
2-\tau \ln\left(\dfrac{2\tau+1}{2 \tau-1}\right)& \tau>1
,
\end{cases}
\ \ \ \ 
K_{\text{\tiny{GUE}}}(\tau)=\begin{cases} \tau & 0 <\tau \leq 1\\ 1 & \tau>1 . \end{cases}
\end{split}  
\label{eq:SFF_RMT}
\end{equation}
Fig.~\ref{fig:SM_SFF} shows the spectral form factor averaged over a window of the parameter $K$, as it is not self-averaging. It is seen that there is good correspondence with predictions ~\eqref{eq:SFF_RMT}
of RMT. 
Thus, for the quantum standard map, in parameter regimes where there is classical chaos, not just short-range statistics, but also long-range statistics are seen to agree reasonably well with RMT. In particular, this implies that for the CUE case, the standard map provides a psuedo-random realization of the Haar measure.

\section{Haar random states and unitaries}\label{chap1:subsec1}

 From the perspective of quantum information \cite{NielsenChuang,Bengtsson2007}, eigenstates and time-evolved states in chaotic systems are accurately modeled by Haar-random states, while the propagator itself, as we saw above, has properties of random unitaries. Hence, we review the relevant concepts of Haar-distributed unitaries and states, and the notion of unitary designs \cite{Renes_ScottCaves_Designs,Dankert_2designs2009}, which provides a graded framework to understand the emergence of randomness in quantum chaotic dynamics.
%Notions of state and unitary t-designs. Implication for the theory of Quantum Information \cite{NielsenChuang}:
 %chaotic systems approximate pseudorandom unitaries, with applications to benchmarking, cryptography, and simulation hardness. Applications to Quantum Fisher information and metrology.

The eigenstates of quantum chaotic systems display a characteristic coexistence of randomness and structure. While a small subset of states may exhibit pronounced scarring associated with unstable classical periodic orbits \cite{Heller1984}, the overwhelming majority are dominated by features indistinguishable from randomness. Focusing on this generic behavior, typical eigenstates of fully chaotic quantum systems are well approximated by Haar-random states, or equivalently by eigenvectors of random matrix ensembles. Likewise, time-evolved states under chaotic dynamics rapidly scramble and approach similarly structureless statistics. To quantify this effective randomness, several measures have been employed in traditional studies of quantum chaos, including the participation ratio and the Shannon entropy. It is well appreciated, however, that such measures are inherently basis dependent.

%More recently, particularly in the context of many-body systems, the closeness of quantum states to randomness has been explored using the notion of quantum $t$-designs, which provide a basis-independent and operational characterization of pseudorandomness.

Let $\ket{\psi}$ be an eigenvector of a random matrix of dimension $N$, with components
$x_j = \braket{j}{\psi}$.  
For matrices drawn from the Gaussian or Circular Orthogonal Ensembles (GOE/COE), the components $x_j$ are real, and their joint probability density function (j.p.d.f.) is uniform on the normalization sphere \cite{BroFloFre1981,Wootters1990_randomstates,Haake},
\begin{equation}
P(x_1,\cdots,x_N)
=
\frac{\Gamma(N/2)}{\pi^{N/2}}
\,
\delta\! \, \left(1-\sum_{j=1}^N x_j^2\right).
\label{eq:rmt_GOE_eigenvector_dist}
\end{equation}
As a consequence, the scaled intensities $N x_j^2$ fluctuate about unity. In the limit $N \to \infty$, the joint distribution of any fixed number $k \ll N$ of components converges to
\begin{equation}
P_k(x_1, \cdots, x_k)
\xrightarrow{N \to \infty}
\left(\frac{N}{2\pi}\right)^{k/2}
\exp\, \!\left(-\frac{N}{2}\sum_{j=1}^k x_j^2\right),
\label{eq:rmt_evec1comp_Normal}
\end{equation}
This implies
that the components become asymptotically independent, identically distributed Gaussian random variables with zero mean and variance $1/N$.
For the Gaussian or Circular Unitary Ensembles (GUE/CUE), the eigenvectors are generically complex. Writing their components as $z_j$, and the intensities as $t_j=|z_j|^2$ their j.p.d.f. read
\begin{equation}
P(z_1,\ldots,z_N)
=
\frac{(N-1)!}{\pi^N}
\,
\delta\, \!\left(1-\sum_{j=1}^N |z_j|^2\right), \qquad P(t_1,\ldots,t_N)
=
(N-1)!
\,
\delta\, \!\left(1-\sum_{j=1}^N t_j\right).
%\label{eq:Haar_random_state}
\label{eq:rmt_GUE_intensity_jpdf}
\end{equation}
%The corresponding intensities $t_j = |z_j|^2$ are uniformly distributed over the simplex
%$\sum_{j=1}^N t_j = 1$, $t_j \ge 0$, with
In the large-$N$ limit, any fixed subset of $k \ll N$ intensities become practically independent,
\begin{equation}
P_k(t_1, \cdots, t_k)
\xrightarrow{N \to \infty}
N^k
\exp \, \! \left(-N\sum_{j=1}^k t_j\right),
\label{eq:rmt_evec1comp_dist_C}
\end{equation}
so that each intensity is exponentially distributed with mean $1/N$, in agreement with the   
%This exponential law is commonly referred to as the
Porter--Thomas distribution.

 \begin{figure}
     \centering
     \includegraphics[width=\linewidth]{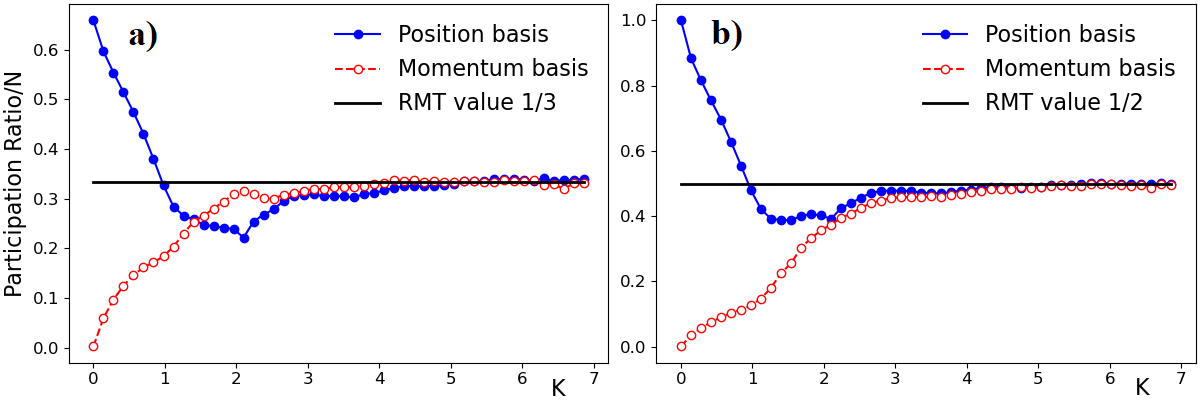}
     \caption{Spectrum averaged participation ratio (scaled by $1/N$) for eigenfunctions of the quantum standard map 
     for: a) system with generalized time reversal symmetry; b) system without such 
     a symmetry,
     The data are collected for $N=750$
     in both the position and momentum basis 
     as a function of the kicking strength $K$.
For small values of $K$ the states are localized in momentum and extended in position space, but as $K$ is increased they come together 
to the random state values of $1/3$ and $1/2$, for
 systems with time reversal symmetry (panel a) and with symmetry broken (panel b), respectively.
 }
     \label{fig:PR_standard_map}
 \end{figure}

\subsection{Measures of state randomness: examples from quantum maps}

Eigenstate localization plays a central role in transport phenomena: localized states suppress conductivity, while extended states may support it \cite{Haake}. Localization is typically examined in a physically motivated basis, such as position or momentum. In contrast, for strongly chaotic states almost any generic basis reveals simwithoutilar statistical features. Among the many possible measures of localization and randomness, we focus on the participation ratio and the Shannon entropy, including its coherent–state (Wehrl) version.

\paragraph{Participation ratio.}
Let $\ket{\psi}=\sum_{j=1}^N x_j\ket{j}$ be a normalized real state in some orthonormal basis. The inverse participation ratio (IPR) and participation ratio (PR) are
\beq
{\rm IPR}=\sum_{j=1}^N x_j^4, 
\qquad 
{\rm PR}=\frac{1}{\rm IPR}.
\eeq
Large IPR (small PR) indicates localization, while small IPR (large PR) signals delocalization. 
A basis state has ${\rm IPR}={\rm PR}=1$, representing maximal localization. 
If all $N$ intensities are equal, ${\rm IPR}=1/N$ and ${\rm PR}=N$, corresponding to maximal delocalization (up to phases). Typical states lie between these extremes.

For real random (GOE) states,
\beq
\br {\rm IPR}\kt_R=\frac{3}{N+2}
= \frac{3}{N}+O\, \!\left(\frac{1}{N^2}\right),
\qquad 
\br {\rm PR}\kt_R\approx \frac{N}{3},
\eeq
while for complex random (GUE/Haar) states with amplitudes $z_j$,
\beq
\br {\rm IPR}\kt_C=\left\br \sum_{j=1}^N |z_i|^4 \right\kt=\frac{2}{N+1}
= \frac{2}{N}+O\!\left(\frac{1}{N^2}\right),
\qquad 
\br {\rm PR}\kt_C\approx \frac{N}{2}.
\eeq
Thus, in time-reversal symmetric systems roughly one-third of the components contribute significantly, while in symmetry-broken cases about one-half do.

Figure~\ref{fig:PR_standard_map} shows the spectrum-averaged ${\rm PR}/N$ for eigenstates of the quantum standard map in Eq.~\eqref{eq:quantum_standard_map}, plotted in both momentum and position bases, and for symmetry-preserving and symmetry-breaking cases. For $K\gtrsim 5$, both bases converge toward the RMT predictions ($1/3$ or $1/2$). However, for small $K$, the behavior reflects classical structures visible in Fig.~\ref{fig:class_stand_map}: rotational KAM tori below the critical $K_c$ localize momentum while the angle variable remains extended.  As these tori break, more global momentum transport sets in. This leads to the momentum PR increasing and the position PR decreasing, with non-monotonic features associated with resonance structures. For sufficiently large $K$, both fluctuate around their RMT values. 

\paragraph{Shannon and Wehrl entropies.}
For intensities $t_j= x_j^2$ or $t_j=|z_j|^2$, the Shannon entropy (in nats) is
\beq
H(\psi)=-\sum_{j=1}^N t_j \ln t_j.
\eeq
For random states, using the j.p.d.fs above, one finds for large $N$ \cite{Wootters1990_randomstates}
\beq
\br H \kt_R=\ln N -(2-\gamma-\ln 2)
\approx \ln N -0.72963,
\eeq
\beq
\br H \kt_C=\ln N -(1-\gamma)
\approx \ln N -0.42278,
\eeq
where $\gamma \approx 0.5772$ denotes the Euler constant.
The dominant $\ln N$ scaling arises from the typical $t_j\sim 1/N$ behavior, while the subleading constant distinguishes real and complex ensembles. Complex states have larger average entropy, consistent with their greater delocalization. Exact finite-$N$ expressions can be written in terms of harmonic numbers or digamma functions. Early studies employed the quantum kicked top to study the entropy of coherent states in the  eigenstate basis \cite{Kus1988,Zyczkowski_eigenstates}.

The Shannon entropy in the overcomplete coherent-state basis defines the Wehrl entropy, providing a phase-space measure of spreading. For coherent states $\ket{q_i p_j}$,
\beq
W_{ij}=\frac{1}{N}\left|\braket{q_i p_j}{\psi}\right|^2,
\qquad 
H_{W}[\psi]=-\sum_{i,j=0}^{N-1} W_{ij}\ln W_{ij}.
\eeq
If $\ket{\psi(n)}=U^n\ket{\psi(0)}$ evolves from an initially localized coherent state, $H_{qp}[\psi(n)]$ quantifies phase-space scrambling. For chaotic dynamics, the initial entropy growth rate is governed by the local Lyapunov exponent $\lambda$ and persists up to the Ehrenfest time \cite{Berry_Balazs_QMaps,Shepelyansky:2020}
\beq
t_E\sim \frac{\log N}{\lambda}.
\label{eq:Ehrenfest}
\eeq
Beyond $t_E$, interference effects dominate and the entropy saturates close to the RMT value $\approx 2\ln N-0.422$, reflecting the effective phase-space dimension $N^2$.

\begin{figure}
     \centering
     
     \includegraphics[width=\linewidth]{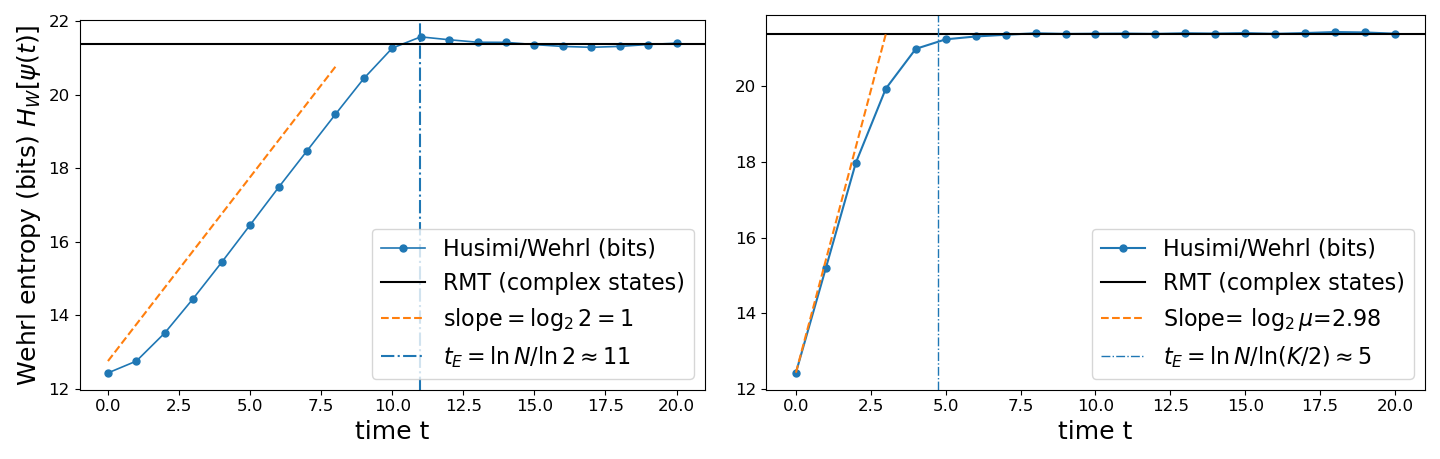}

     \caption{The Wehrl entropy (evaluated in bits, log base-2) of a time evolving state in the quantum baker's map (left) and the quantum standard  map (right) for $K=10$, both for $N=2038$. The initial states are coherent states localized at the period-2 point $(1/3,2/3)$, for the baker and at the fixed point $(0,0)$ for the standard map. The local unstable direction eigenvalues $2$ and $\mu =4+\sqrt{15}$ for the two classical maps, determine the initial rate of entropy production.
      Vertical dashed lines are at the Ehrenfest time $t_E=\ln N/\ln \lambda$  and  the Lyapunov exponents are $\lambda=\ln 2$, and $\ln(K/2)$ for the baker and standard map, respectively.
       }
     \label{fig:Wehrl_entropy}
 \end{figure}

This behavior is illustrated in Fig.~\ref{fig:Wehrl_entropy} for both the quantum baker and standard maps. 
For the baker map, a coherent state centered at the period-2 point $(1/3,2/3)$ exhibits linear entropy growth at one bit per iteration, in precise correspondence with classical stretching, up to $t_E\approx \log_2 N$ (about $11$ for $N=2038$, chosen so that $N/2$ is prime and avoids special arithmetic effects). 
For the quantum standard map, richer features arise from nonuniform hyperbolicity. With the initial state centered at $(0,0)$ and $K=10$, the initial slope equals the local Lyapunov exponent $\log_2(4+\sqrt{15})\approx 2.98$, but decreases as the state spreads into regions of weaker instability. Saturation occurs near the Ehrenfest time predicted using the global Lyapunov exponent from Eq.~\eqref{eq:Chirikov}. 

%\textcolor{blue}{
While Fig.~\eqref{fig:Wehrl_entropy} is
obtained 
for a unitary dynamics, non-unitary dynamics resulting from measurements or the effect of an environment has been studied by defining appropriate entropies in analogy with the classical Kolmogorov-Sinai entropy. See \cite{RobertAlicki2004} for a formalism and treatment of the
entropy production in the model
of quantum baker map.%}

\subsection{State t-designs and dynamical approach to 2-design behavior}

Let $\mathcal{E}=\{p_a, \,\ket{\psi_a}\}_{a=1}^M$ be an ensemble of $M$ normalized pure states in a $N$-dimensional Hilbert space $\mathcal{H}_N$. The ensemble is said to form a (complex projective) \emph{$t$-design} \cite{Renes_ScottCaves_Designs,Ambainis2007quantumtdesigns} if its $t$-th tensor moment matches that of the Haar measure \cite{Collins2006,Mele2024introductionHaar}, namely,
\begin{equation}
\rho_t(\mathcal{E})=\sum_{a=1}^M p_a\,
\left(\ket{\psi_a}\bra{\psi_a}\right)^{\otimes t}
=
\int d\psi_{\rm Haar}\,
\left(\ket{\psi}\bra{\psi}\right)^{\otimes t}=\frac{\Pi_{\rm sym}}{d_t}. \qquad d_t=\binom{N+t-1}{t}.
\label{eq:t_design_def}
\end{equation}
Here, $\Pi_{\rm sym}$ is the projector onto the totally symmetric subspace of $\mathcal{H}^{\otimes t}$ and $d_t=\Tr \Pi_{\rm sym}$. We consider the case of $p_a=1/M$ for simplicity. With $t=1$, this reduces to isotropy of the ensemble,
\begin{equation}
\frac{1}{M}\sum_{a=1}^M \ket{\psi_a}\bra{\psi_a}=\frac{\mathbb{I}}{N},
\end{equation}
while $t=2$ fixes all  amplitude correlations with 4 terms, such as $z_iz_j z_k^*z_\ell^*$, and determines, for example, inverse participation ratios, when all indices are equal. The  average of the second tensor moment over the 
Haar ensemble (denoted $H$)
reads \cite{Collins2006,Mele2024introductionHaar}
\begin{equation}
\rho_{2}(H)
=
\int d\psi_{\rm Haar}\,
\left(\ket{\psi}\bra{\psi}\right)^{\otimes 2}
=
\frac{\mathbb{I}+\mathbb{S}}{N(N+1)},
\label{eq:Haar_second_moment}
\end{equation}
where $\mathbb{S}$ is the swap operator on $\mathcal{H}\otimes\mathcal{H}$. It is defined  as $
\mbS|\phi_A\kt |\phi_B \kt = |\phi_B\kt |\phi_A\kt, \, \text{or}\, \mbS(u_A\otimes
u_B)\mbS=u_B\otimes u_A,
$
for arbitrary states $|\phi_{A,B}\kt $ and operators $u_{A,B}$.

A convenient scalar diagnostic for $t$-design behavior is the $t$-frame potential \cite{Renes_ScottCaves_Designs,Gross2007_Designs},
\begin{equation}
F_t(\mathcal{E})
=\|\rho_t(\mathcal{E})\|_2^2=
\frac{1}{M^2}\sum_{a,b}
|\braket{\psi_a}{\psi_b}|^{2t}.
\label{eq:F2_def}
\end{equation}
It attains the lowest value when $\rho_t(\mathcal{E})$ is the isotropic mixed state in the support of the totally symmetric subspace of $\otimes^t \mathcal{H}$. Hence, it satisfies the inequality
\begin{equation}
F_t(\mathcal{E}) \ge \frac{1}{d_t}=F_t(H),
\label{eq:F2_bound}
\end{equation}
with equality if and only if the ensemble is a complex projective $t$-design, justifying referring to $F_t(\mathcal{E})$ as a potential. Thus, attaining the Haar value of $F_t$ is necessary and sufficient for $\mathcal{E}$ to be an exact $t$-design. If $M<d_t$, $F_t(\mathcal{E})\geq 1/M$, which follows simply by retaining only the diagonal terms, $a=b$,
in Eq.~\eqref{eq:F2_def}. Hence, an exact design must have at least $d_t$ elements. The upper bound $F_t(\mathcal{E})\leq 1$ follows from the normalization of the states in the design. In contrast to exact designs, $\epsilon$~-~approximate t-designs satisfy Eq.~\eqref{eq:t_design_def} approximately --  ratio of the ensemble average to the Haar average is within $1 \pm \epsilon$, see \cite{Ambainis2007quantumtdesigns} for a more precise definition.  

To illustrate the dynamical emergence of approximate 2-design behavior in quantum chaos, in a very preliminary exploration, we consider ensembles generated by iterating the quantum standard map,
\begin{equation}
\ket{\phi(n)}
=
U(K_n)\ket{\phi(n-1)},
\end{equation}
where $K_n\in [K-\delta K/2, K+\delta K/2]$ is randomly chosen from a small interval $\delta K$ around $K$. From a fiducial state $\ket{\phi(0)}$ we construct an ensemble
\begin{equation}
\mathcal{E}
=
\left\{\ket{\phi(n_1)},\ket{\phi(n_2)},\dots,\ket{\phi(n_M)}\right\},
\end{equation}
with optional burn-in (discarding an initial set of states, minimizing initial state memory) and subsampling (with a constant stride of $n_{k+1}-n_k$) to reduce transient and short-time correlations, respectively. We study
a quantity
$\Delta_2$
measuring
the deviation from a perfect $2$-design,
\[
\Delta_2
=
d_2\,F_2^{\rm off}-1, \;\; {\rm where}\;\; F_2^{\rm off}
=
\frac{1}{M(M-1)}
\sum_{a\neq b=1}^M
|\braket{\psi_a}{\psi_b}|^4.
\]
$F_2^{\rm off}$ removes the trivial diagonal terms $a=b$ from $F_2$ and for $M \gg 1$ also approaches 
$1/d_2=2/N(N+1)$ for a good design when $\Delta_2 \sim 0$. 

\begin{figure}[ht]
    \centering
    \includegraphics[width=.49\linewidth]{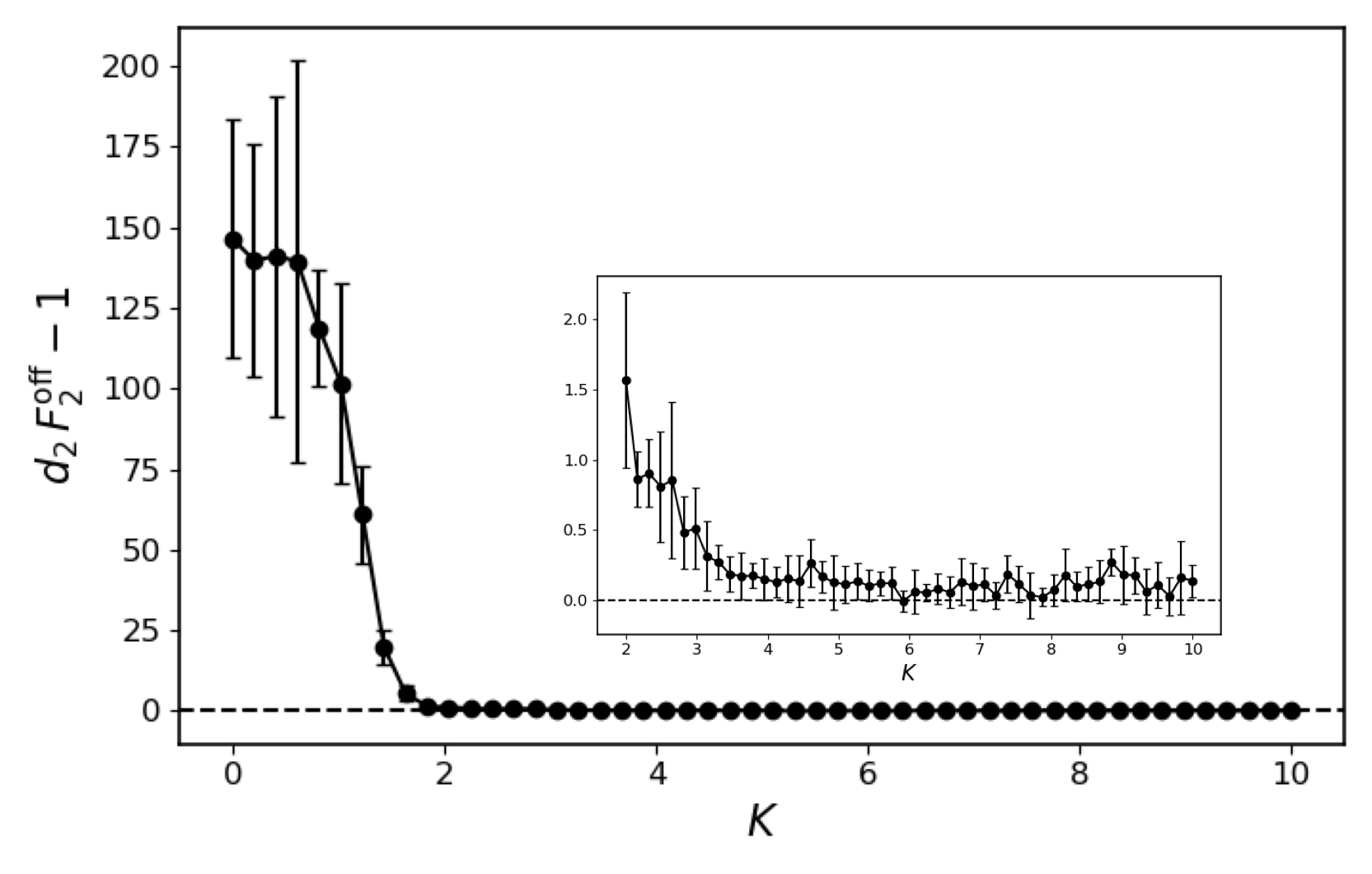}
\includegraphics[width=.49\linewidth]{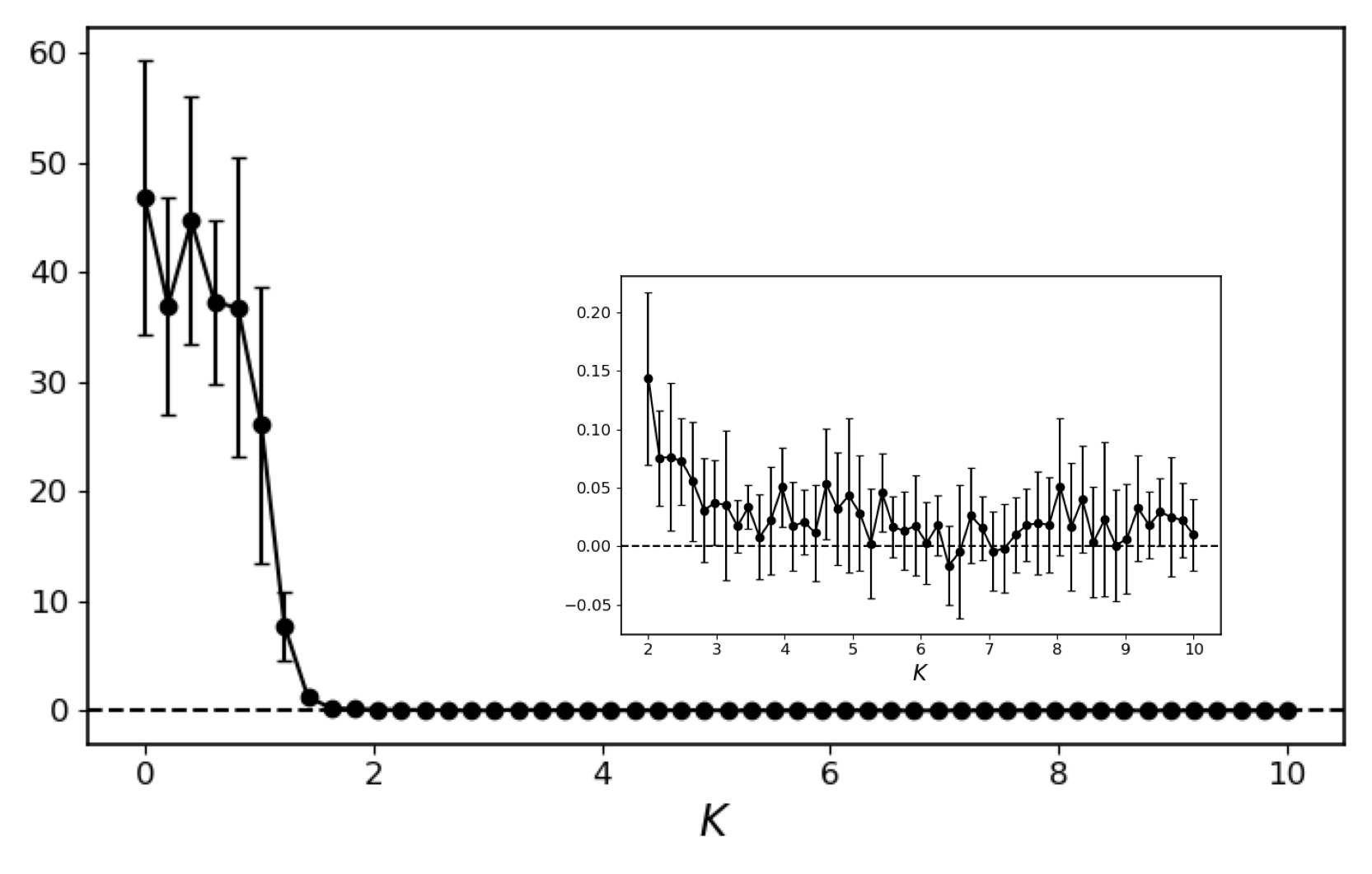}
    \caption{Deviation $\Delta_2=d_2 F_2^{\rm off}-1$
    from a perfect $2$-design
    for an ensemble of time-evolving states in the quantum standard map with kicking strength $K$. %which vanishes for exact 2-designs. 
    The initial state is a coherent state at $\ket{1/3,2/3}$, and $\delta K=0.05$. $N=128$, and the number of states in the design is also $M=N$. Insets show zoomed in values for $K>2$. The design consists of time evolving states, after dropping the first 10 time steps. The left corresponds to all consecutive time steps, with a stride$=1$, while on the right a stride $=10$ keeps every tenth state.  The error bars indicate variance from 10 different histories of the kicking strength in $K \pm \delta K/2$, see text for further discussions. }
    \label{fig:2Design}
\end{figure}

Figure~\ref{fig:2Design} shows $\Delta_2$ as a function of the central value of $K$, with a window of $\delta K=0.05$, and when the fiducial state is a coherent state centered at a generic point. The ensemble consists only of $M=N$
states, and hence cannot be an exact design.
Several key observations emerge:
\begin{enumerate}
\item \textbf{Lack of approach to 2-design in the mixed regime.}
For coherent-state fiducials, $\Delta_2 \gg 0$ for $K<2$,   reflecting slow decay of Hilbert-space overlaps and long-lived dynamical correlations associated with mixed classical phase space, especially when there are rotational KAM tori and remnant cantori. 

\item \textbf{Rapid convergence in the globally chaotic regime.}
The insets zoom into $K \in [2,10]$ showing that for sufficiently large $K$ (typically $K\gtrsim 4$--$5$ for the parameters studied), $\Delta_2$ approaches zero indicating that time-evolved states behave as approximate 2-designs.

\item \textbf{Effect of subsampling (stride).}
While the left panel is for consecutive times, stride $=1$,
using stride $s>1$ suppresses short-time correlations and reveals longer-time mixing properties. This is shown in the right panel for a stride $=10$, every $10^{\rm th}$ time step is kept. The persistence of nonzero $\Delta_2$ confirms that deviations in the mixed regime are not purely short-time artifacts. For large $K$, we see that increasing the stride takes the ensemble closer to the exact 2-design.

\item 
\textbf{Fiducial state dependence in near-integrable regimes.} Not shown in the figures is that 
for small $K$, particularly near the mixed phase-space regime around the breakup of invariant tori, the deviation is strongly dependent on the initial state. Momentum eigenstates remain highly non–2-design at small $K$, while position states may exhibit partial isotropization due to dephasing.
\end{enumerate}

The 2-frame potential thus provides a rigorous, representation-independent measure of state randomness. It goes beyond the IPR and entropy statistics that is usually found, as it probes other correlations involving 4 state amplitudes. Higher-order diagnostics (e.g., the 3-frame potential) are expected to be increasingly stringent tests of chaos. In this sense, the approach of a dynamical system toward design behavior
quantifies the emergence of information-theoretic chaos.

\subsection{Haar random unitaries and t-designs}
The Haar measure $d\mu_H(U)$ on $\mathcal{U}(N)$ is the unique probability measure invariant under left and right multiplication:
\beq
d\mu_H(U)= d\mu_H(W_L U)= d\mu_H(U W_R),
\qquad W_L,W_R\in \mathcal{U}(N).
\label{eq:Haar_CUE}
\eeq
As an ensemble this is the Circular Unitary Ensemble (CUE), relevant for quantum chaotic systems without time-reversal symmetry. Its bulk spectral correlations coincide with those of the GUE.

Moments of matrix elements $u_{k\ell}$ with respect to $d\mu_H$ are evaluated using the Weingarten calculus. The lowest moments illustrate the invariance:
\beq
\langle u_{k\ell} \rangle = 0, 
\qquad 
\langle u_{k\ell} u_{k'\ell'}^* \rangle = \frac{1}{N}\,\delta_{k k'}\delta_{\ell \ell'}.
\eeq
The vanishing of the first moment follows from left invariance under diagonal phase rotations 
$W_L=\mathrm{diag}(e^{i\theta_1},\dots,e^{i\theta_N})$, which implies 
$\langle u_{k\ell}\rangle = e^{i\theta_k}\langle u_{k\ell}\rangle$ for arbitrary $\theta_k$. 
The second moment (in matrix elements) follows from unitarity $UU^\dagger=I$ and invariance, which enforce isotropy across matrix elements. This Haar average admits a compact operator expression:
\beq
M_1=\int_{\mathcal U(N)} U \otimes U^* \, d\mu_H(U)
=
\ket{\phi^+}\bra{\phi^+},\;\; {\rm where}\;\; \ket{\phi^+} = \frac{1}{\sqrt N}\sum_{j=1}^N \ket{j}\otimes\ket{j}
\label{eq:M1}
\eeq
is the maximally entangled state.
This follows from left–right invariance and Schur's lemma, since
$\ket{\phi^+}$ is the unique invariant vector of $U \otimes U^*$. Equivalently using the vectorization $\ket{ABC}=(A\otimes B^T)\ket{C}$, and $\ket{\mathbb I}=\sqrt{N} \,\ket{\phi^+}$ results in the useful twirl
\beq
\int_{\mathcal U(N)} U X U^{\dagger} d\mu_H(U) =\frac{\Tr X}{N} \mathbb{I},\qquad X \in \mathcal{B}(\mcH)
\label{eq:twirl1}
\eeq
which maybe interpreted as the Haar averaging resulting in a maximally depolarizing channel setting all states to the isotropic, maximally mixed state. 

\paragraph{Fourth moment and permutation structure.}
All odd-order moments of matrix elements vanish, and the even ones that do not vanish have an equal number of $u$ and $u^*$ terms.
The first nontrivial correlations arise at fourth order. Haar invariance and
unitarity imply that $\langle 
u_{i_1 j_1} u_{i_2 j_2}
u_{i'_1 j'_1}^* u_{i'_2 j'_2}^*
\rangle$
must be expressible solely in terms of Kronecker deltas contracting row and
column indices. By Schur–Weyl duality, the only tensors allowed are those
associated with permutations in the symmetric group of two elements $S_2$ \cite{Collins2006}, hence
\beq
\langle 
u_{i_1 j_1} u_{i_2 j_2}
u_{i'_1 j'_1}^* u_{i'_2 j'_2}^*
\rangle
=
\sum_{\sigma,\tau \in S_2}
\delta_{i_1 i'_{\sigma(1)}}\delta_{i_2 i'_{\sigma(2)}}
\delta_{j_1 j'_{\tau(1)}}\delta_{j_2 j'_{\tau(2)}}
\,\mathrm{Wg}(\sigma^{-1}\tau;N),
\label{eq:Weingarten_S2}
\eeq
where $\mathrm{Wg}(\pi;N)$ are the Weingarten functions.
For $S_2=\{e,(12)\}$ these coefficients are
$\mathrm{Wg}(e;N)=\frac{1}{N^2-1}$, 
$\mathrm{Wg}((12);N)= -\frac{1}{N(N^2-1)}.$

Explicitly,
\beq
\langle 
u_{i_1 j_1} u_{i_2 j_2}
u_{i'_1 j'_1}^* u_{i'_2 j'_2}^*
\rangle
=
\frac{1}{N^2-1}
\left(\delta_{i_1i_1'}\delta_{i_2 i_2'} \delta_{j_1 j_1'}\delta_{j_2 j_2'}+\delta_{i_1i_2'}\delta_{i_2 i_1'} \delta_{j_1 j_2'}\delta_{j_2 j_1'}
\right)
-
\frac{1}{N(N^2-1)}
\left(
\delta_{i_1i_1'}\delta_{i_2 i_2'} \delta_{j_1 j_2'}\delta_{j_2 j_1'}+\delta_{i_1i_2'}\delta_{i_2 i_1'} \delta_{j_1 j_1'}\delta_{j_2 j_2'}
\right).
\label{eq:Wein2_deltaform}
\eeq
Consider the second moment operator
\beq
M_2^{\rm Haar} \;:=\;\int_{\mathcal U(N)} U\otimes U\otimes U^*\otimes U^*\, d\mu_H(U),
\qquad
M_2:\;(\mathbb{C}^N)^{\otimes 4}\to(\mathbb{C}^N)^{\otimes 4},
\eeq
as acting on 4 parties: $1,2$ for $U\otimes U$ and $3,4$ for $U^*\otimes U^*$, such that the matrix elements of $M_2$ in the computational basis forms the L.H.S. of Eq.~\eqref{eq:Wein2_deltaform}.
Defining two pairing states in terms of standard maximally entangled states:
$|\phi_e\rangle := |\phi^+\rangle_{13}\otimes|\phi^+\rangle_{24}$, and
$
|\phi_{(12)}\rangle := |\phi^+\rangle_{14}\otimes|\phi^+\rangle_{23},
$
the Haar $t=2$ moment can be written compactly as the $2\times 2$ Weingarten matrix
\beq
M_2^{\rm Haar}
=
\frac{N^2}{N^2-1}\Big(
|\phi_e\rangle\langle\phi_e|
+
|\phi_{(12)}\rangle\langle\phi_{(12)}|
\Big)
-\frac{N}{N^2-1}\Big(
|\phi_e\rangle\langle\phi_{(12)}|
+
|\phi_{(12)}\rangle\langle\phi_e|
\Big).
\label{eq:M2_pairing_form}
\eeq
As a consequence, the equivalent of the twirl \cite{Mele2024introductionHaar} in Eq.~\eqref{eq:twirl1} for $t=2$ is
\beq
\int_{\mathcal U(N)} U^{\otimes 2} X U^{\dagger \otimes 2} d\mu_H(U) =\frac{1}{N(N^2-1)}\left[\left(N \Tr X-\Tr(X\mbS)\right)\; \mbI +\left( N \Tr(X \mbS)-\Tr  X\right)\;\mbS \right].
\label{eq:twirl2}
\eeq

Generalizing the moment operators $M_1$ and $M_2$, the Haar measure is characterized by its tensor-power moments
\beq
M_t^{\rm Haar}
:=
\int_{\mathcal U(N)}
U^{\otimes t}\otimes U^{*\otimes t}
\, d\mu_H(U)=
\sum_{\sigma,\tau\in S_t}
\mathrm{Wg}(\sigma^{-1}\tau;N)
\,P_\sigma \otimes P_\tau,
\label{eq:Haar_moment_operator}
\eeq
where $P_\sigma$ are permutation operators. A general ensemble of  $\mathcal{E}$ of unitary matrices with measure $d \mu_{\mathcal{E}}$ is called a \emph{unitary $t$-design} if its $t$-th moment matches that of the Haar measure \cite{Gross2007_Designs,Dankert_2designs2009,Roy2009_Designs}:
\beq
M_t^{\mathcal{E}}
=\int_{\mathcal U(N)}
U^{\otimes t}\otimes U^{*\otimes t}
\, d\mu_{\mathcal{E}}(U)=
M_t^{\rm Haar}.
\label{eq:unitary_t_design_def}
\eeq
Thus a $t$-design reproduces all polynomial averages of degree at most $t$ in the matrix elements of $U$ and utmost $t$ in the elements of $U^*$.
A $1$-design reproduces complete depolarization.
A $2$-design reproduces fourth-order correlations in particular,
Haar values of quantities such as out-of-time-ordered correlators (OTOC) \cite{Roberts2017Chaos}. Just as in state-designs a convenient scalar diagnostic of unitary $t$-design behavior is the $t$-th unitary frame potential \cite{Roy2009_Designs,Roberts2017Chaos},
\beq
\mathcal F_t(\mathcal E)
:=
\int_{\mathcal U(N)}
\, d\mu_{\mathcal{E}}(U)d\mu_{\mathcal{E}}(V)
\left|
\operatorname{Tr}\!\left(U^\dagger V\right)
\right|^{2t}.
\label{eq:unitary_frame_potential}
\eeq
An ensemble is a unitary $t$-design
if and only if
\(
\mathcal F_t(\mathcal E)
=
\mathcal F_t^{\rm Haar}=t!\; ( {\rm for}\;N\geq t).
\)
Moreover,
\(
\mathcal F_t(\mathcal E)
\ge
\mathcal F_t^{\rm Haar},
\)
with equality precisely for a $t$-design. However, rather than deal directly with this potential, we will see the use of 2-designs in entanglement related measures.

\subsection{Implications for quantum information and computation}
\label{sec:Implications}

%There are many disparate roles that Haar random states and unitaries play in quantum information theory, from estimating  entanglement in typical states to benchmarking so-called quantum supremacy claims.

\paragraph{Quantum protocols}

In high-dimensional Hilbert spaces, Haar-random states are overwhelmingly typical:
reduced density matrices are nearly maximally mixed,
entanglement is near maximal \cite{Page1993,HLW06}, as we will describe in more detail below. The connections and applications to quantum information  are often via approximate state and unitary designs, high entangling power, and scrambling of localized information.
To describe a few concrete applications,  random unitaries play a central role in randomizing channels for privacy and data hiding. If $\mathcal{S}$ is a finite set of $|\mathcal{S}|$ unitaries $U_s$, the 
{\sl mixed unitary}
channel $\rho \mapsto \mathcal{R}(\rho)$ in a $d$ dimension space, where
\beq
\label{R_rho}
\mathcal{R}(\rho)=\frac{1}{|\mathcal{S}|}\sum_{s\in \mathcal{S}}
U_s \, \rho \, U_s^{\dagger},
\eeq
is a secure quantum one-time pad if the channel is completely depolarizing: $\mathcal{R}(\rho)=I/d$ for all $\rho$, and it is known that we need at least $|\mathcal{S}|=d^2$ to accomplish this \cite{Boykin_2003}. This implies that any 1-design provides such a channel, due to the twirl condition in Eq.~\eqref{eq:twirl1}. However, pretty good approximate channels can be found if some error is tolerated using only $d \log d$ random unitaries \cite{Hayden2004randomizing}.
In protocols such as state merging, entanglement-assisted communication,
and generalized superdense coding,
one applies a random unitary to a subsystem and discards part of it,
thereby decoupling \cite{Hayden_Decoupling_2008} it from an external reference system.
Haar typicality guarantees that subsystems become nearly maximally mixed
and that correlations are suppressed \cite{HLW06}.
Importantly, the proofs rely on second-moment properties of the Haar measure,
and approximate unitary $2$-designs  \cite{Dankert_2designs2009}.

\paragraph{Quantum simulations, creating random states and unitaries}

Exact Haar-random unitaries on $n$ qubits are not efficiently implementable:
a generic element of $\mathcal U(2^n)$ has $N^2=O(4^n)$ continuous parameters,
and compiling a typical unitary requires a number of elementary gates that
scales exponentially with $n$ (in the worst case). However, approximate unitary $t$-designs reproduce Haar moments up to order $t$
and can be generated by polynomial-depth local circuits.
Such ensembles provide a notion of computationally accessible randomness,
bridging quantum chaos and quantum computation. In particular dynamical systems such as the quantum standard map and the baker's map are candidates of psuedorandomness and are known to have efficient implementations in the quantum circuit model of computation. The quantum standard map in Eq.~\eqref{eq:quantum_standard_map} is a product of Fourier transforms and structured diagonal unitary gates in finite dimensional Hilbert spaces. There is a well-known efficient quantum Fourier transform (QFT) algorithm that uses $O(n^2)$ two-qubit gates, and hence if the diagonal unitary is implemented efficiently, the quantum standard map can be for all value of the parameter $K$, including those in which it is an approximate unitary design. A pioneering algorithm for doing this has been developed in \cite{Georgeot_Shepel2001,LeviEtal2003} to realize
each map step in $O(n^3)$ elementary gates on a $n$-qubit register. 

The quantum baker's map being only constructed from Fourier transforms, see Eq.~\eqref{eq:quantum_baker}, is algorithmically quite straightforward and implemented with $O(n^2)$ gates efficiently. Since the original proposal in \cite{Schack1998}, an early 3 qubit NMR experimental implementation has been reported in \cite{Weinstein2002}. As quantum baker's maps are quite structured and non-universal when $N=2^n$ is a power of 2, having multifractal states and IPR scaling $1/N^{\gamma}$ with $\gamma<1$ \cite{Meenaksh2005}, efficient implementations of related more Haar like baker's maps are of future interest.
Although further experimental realization remains limited by current quantum hardware, this body of work confirms that pseudo-Haar chaotic dynamics can be generated by efficiently implementable circuits, making them attractive for scrambling, decoupling,
and design diagnostics in quantum information and computation
\cite{Wang2004}.

\paragraph{Quantum chaos and quantum algorithms}

Although core quantum algorithms such as the Grover search, Shor factoring,  and the QFT \cite{NielsenChuang} are highly structured, their cores contain unitary operators with potential quantum chaos. 
An early study explored the Grover and QFT algorithms in this light \cite{braun2002quantum}, and found some evidence of hypersensitivity to perturbations, an information theoretic characterization of quantum chaos \cite{Schack_Caves1996}.
A closer relationship is found in the Shor algorithm, as its core contains the modular multiplication operator that underlies modular exponentiation.
When the multiplier is $2$ \cite{NielsenChuang} this part can be be expressed as a coherent superposition of quantum baker maps, such as in Eq.~\eqref{eq:quantum_baker}, providing an explicit connection between quantum algorithms and quantum-chaotic dynamics \cite{Arul2007}. Recent work identifies chaotic roots in modular multiplication dynamics for arbitrary multipliers, connecting to generalized baker-map decompositions with larger Lyapunov exponents. \cite{Patoary2024}. Related analyzes find compelling evidence of CUE statistics in operators used in Shor's algorithm after appropriate symmetry reduction. \cite{Maity2006}. Interestingly these connections of the quantum baker with the modular exponentiation has resulted in a proposal to implement it efficiently with an algorithm consisting only of QFTs in $O(n^2)$ number of gates, improving on existing methods \cite{Patoary2025}.

\section{Random matrix theory and entanglement}
\label{chap2:sec1}

Before embarking on dynamical systems and entanglement generation from quantum chaos, we review the entanglement in bipartite pure random states, sampled from the Haar measure of the combined space, which proves to be an important and useful benchmark. 
We also mention results related to tripatite and multipartite partitions of random states and the implications of the monogamy of entanglement.

\subsection{Trace-Constrained Wishart Ensembles and Bipartite Entanglement}

The statistical properties of reduced density matrices of Haar-random pure states
are described by trace-constrained Wishart (or Wishart-Laguerre) ensembles \cite{ZS01,livan2018}.
These ensembles play a central role in understanding typical entanglement. Let $|\phi\rangle$ be a bipartite pure state in
$\mcH_{AB}=\mathcal H_A \otimes \mathcal H_B$, with
$N_1=\dim \mathcal H_A \le N_2=\dim \mathcal H_B$.
In a product orthonormal basis,
\begin{equation}
\label{eq:bipartstate}
|\phi\rangle
= \sum_{i=1}^{N_1}\sum_{j=1}^{N_2} a_{ij}\, |i\rangle\otimes|j\rangle,
\end{equation}
where the coefficients $a_{ij}$ form an $N_1\times N_2$ matrix $\mathcal{A}$.

The reduced density matrices are
\begin{equation}
\rho_A = \operatorname{Tr}_B(|\phi\rangle\langle\phi|)=\mcA \mcA^\dagger,
\qquad
\rho_B = \operatorname{Tr}_A(|\phi\rangle\langle\phi|) = (\mcA^\dagger \mcA)^T =\mcA^T  {\bar \mcA}.
\label{eq:rho_AandB}
\end{equation}
Both are manifestly positive semidefinite with identical nonzero eigenvalues
$\{\lambda_j\}_{j=1}^{N_1}$ (the larger subsystem contains $N_2-N_1$ additional zeros). The Schmidt decomposition \cite{NielsenChuang},  which follows from the singular value decomposition of $\mcA$ can be written as
\begin{equation}
\label{eq:DefnSchmidt}
|\phi\rangle
= \sum_{j=1}^{N_1} \sqrt{\lambda_j}\,
|\phi^A_{j}\rangle \otimes |\phi^B_{j}\rangle,
\qquad \ld_1 \geq \ld_2 \cdots \ld_{N_1}, \qquad
\sum_{j=1}^{N_1}\lambda_j=1.
\end{equation}
The Schmidt vectors $\{ \ket{\phi^A_j}\}$ and $\{ \ket{\phi^B_j}\}$ can be chosen as orthonormal bases in their respective spaces.
The state $\ket{\phi}$ is separable if and only if $\lambda_1=1$, when the Schmidt decomposition is of rank one. 
In contrast, $\ket{\phi}$ is maximally entangled if and only if $\rho_A=\mbI/N_1$ is maximally mixed hence, $\lambda_j=1/N_1$ for all $j$.
The entanglement entropy is the von Neumann entropy of the reduced state \cite{NielsenChuang},
\begin{equation}
S_{\rm vN}(\phi) = -\operatorname{Tr}(\rho_A \ln \rho_A)=-\operatorname{Tr}(\rho_B \ln \rho_B)
  = -\sum_{j=1}^{N_1} \lambda_j \ln \lambda_j, \qquad 0 \leq S_{\rm vN} \leq \ln N_1.
\end{equation}
 Another useful measure is the linear entropy:
\beq
S_L(\phi)=1-\Tr \rho_A^2=1-\Tr \rho_B^2=1-\sum_{j=1}^{N_1} \ld_j^2, \qquad 0 \leq S_L(\phi) \leq 1-\frac{1}{N_1}.
\label{eq:Linear_Entropy_Defn}
\eeq
 Indeed a spectrum of entropies are found useful:
\beq
S_q^R=\frac{\ln \Tr \rho_A^q}{1-q}, \qquad S_q^T=\frac{1-\Tr \rho_A^q}{q-1}, \qquad 0 \leq q < \infty,\;\;  q \neq 1.
\label{eq:Renyi-Tsallis_defn}
\eeq
The R\'enyi entropies $S^R_q$ are additive while the Tsallis entropies $S^T_q$ are not, and both are equal to $S_{\rm vN}$ when $q \rightarrow 1$. Note that $S^T_2\equiv S_L$, the linear entropy and the R\'enyi-2 entropy ($q=2$), $S_2^R$ or simply $S_2$, are also widely used. All these measures  attain their minimum and maximum values if and only if the state is a product state or a maximally entangled state, respectively.

If $|\phi\rangle$ is selected from the Haar-random ensemble in dimension $N=N_1 N_2$,
the coefficient matrix $\mcA$ is uniformly distributed on the unit sphere.
Equivalently, $\mcA$ and the reduced density matrix (dropping the subsystem label $A$) is 
\begin{equation}
\mcA = \frac{M}{\sqrt{\operatorname{Tr}(M M^\dagger)}}, \qquad \rho = \mcA \mcA^{\dagger}=\frac{M M^\dagger}{\operatorname{Tr}(M M^\dagger)}.
\end{equation}
where $M$ is an $N_1\times N_2$ matrix drawn from the Ginibre ensemble. The reduced density matrix therefore belongs to the trace-constrained Wishart ensemble of positive semidefinite matrices. 
Assuming the Hilbert-Schmidt norm in the space of matrices, this ensemble induces the joint probability density function (j.p.d.f.) of the Schmidt eigenvalues
$\{\lambda_i\}$ is \cite{Bengtsson2007}
\begin{equation}
\label{jpdf1}
P(\lambda_1,\dots,\lambda_{N_1})
= B_{N_1,N_2}\,
\delta\,\!\left(\sum_{i=1}^{N_1}\lambda_i -1\right)
\prod_{i=1}^{N_1}
\lambda_i^{\frac{\beta}{2}(N_2-N_1+1)-1}
\prod_{j<k}|\lambda_j-\lambda_k|^\beta,
\end{equation}
where $\beta=1,2$ corresponds to real and complex ensembles, and
$B_{N_1,N_2}$ is a known normalization constant.

\paragraph{The Marchenko–Pastur Law}

In the asymptotic regime
$N_2 \geq N_1 \gg 1$ with fixed ratio $Q=N_2/N_1 \ge 1$,
the density of rescaled eigenvalues
\[
x_i = N_1 \lambda_i
\]
follows the Marchenko–Pastur (MP) distribution \cite{livan2018}:
\begin{equation}
\rho_{\mathrm{MP}}^{Q}(x)
=
\frac{Q}{2\pi}
\frac{\sqrt{(x_{+}-x)(x-x_{-})}}{x},
\qquad
x_- \le x \le x_+,\;\; {\rm with}\;\;
x_{\pm}
=
\left(1+\frac{1}{Q}\right)
\pm
\frac{2}{\sqrt{Q}}.
\label{eq:marcenko-pastur-law}
\end{equation}
In particular, for the symmetric case of $N_1=N_2$, $Q=1$ and
\begin{equation}
\rho_{\mathrm{MP}}(x)
=
\frac{1}{2\pi}
\sqrt{\frac{4-x}{x}},
\qquad
0 \le x \le 4,
\label{eq:marcenko-pastur-law_Q1}
\end{equation}
which diverges at the origin.
The MP law is universal for correlation matrices with finite second moments
and is closely related to the Wigner semicircle law.

\paragraph{Average Entanglement: Page and Lubkin Formulas}

For complex ($\beta=2$) Haar-random states,
Page \cite{Page1993} conjectured a remarkable exact formula for the average entanglement entropy:
\begin{equation}
\langle S_{{\rm vN}}  \rangle
=\int
\left(
-\sum_j \lambda_j \ln \lambda_j
\right)
P(\lambda_1,\dots,\lambda_{n_1})
\, d\lambda_1\cdots d\lambda_{n_1}=
\sum_{k=N_1+1}^{N_1 N_2} \frac{1}{k}
-
\frac{N_1-1}{2 N_2}
\approx
\ln N_1 - \frac{N_1}{2 N_2},
\end{equation}
which was proved shortly thereafter \cite{Sen1996}.
The approximation, valid for $N_2 \ge N_1 \gg 1$, follows exactly from the Marchenko-Pastur law in Eq.~\eqref{eq:marcenko-pastur-law}, demonstrating that Haar-random states are nearly maximally entangled.

The linear entropy defined in Eq.~\eqref{eq:Linear_Entropy_Defn} has a Haar average that was found by Lubkin in 1978 \cite{Lubkin1978}. For the bipartite state
$|\psi\rangle$
in Eq.~\eqref{eq:bipartstate} the reduced density matrix $\rho_A$ has the purity
\beq
\Tr \rho_A^2=\sum_{i,k=1}^{N_1} \sum_{j,\ell=1}^{N_2} a_{ij}a^*_{kj}a^*_{i\ell}a_{k\ell}=\sum_{ijk\ell}\bra{ijkl}\; (\ketbra{\phi}{\phi})^{\otimes 2}\; \ket{kji \ell}.
\eeq
Hence, it is Haar average can be found on using Eq.~\eqref{eq:Haar_second_moment}
with $N=N_1N_2$, resulting in the following exact average purity and linear entropy:
\beq
\br \Tr \rho_A^2 \kt = \frac{N_1+N_2}{1+N_1 N_2}, \qquad \br S_L\kt =\frac{(N_1-1)(N_2-1)}{1+N_1N_2} \approx 1-\frac{1}{N_1}-\frac{1}{N_2}
\eeq
The approximation is in the large system size limits and once again if $N_2\gg N_1$, the linear entropy approaches that of the maximally entangled state. For $N_1=N_2=N$, an important special case,
\beq
\br S_{{\rm vN}} \kt \approx \ln N -\frac{1}{2}, \;\; \br S_L \kt \approx 1-\frac{2}{N},
\label{eq:PageLubkin}
\eeq
both consequences of the Marchenko-Pastur law in Eq.~(\ref{eq:marcenko-pastur-law_Q1}). 
The entanglement spectrum of large Haar-random bipartite states
is governed by universal random matrix statistics.
The moments of the purity and its distribution were studied in \cite{Giraud2007}.
The distribution of von Neumann entropy
 \cite{HLW06}
and its concentration
around the mean value $\langle S \rangle$ is discussed
in Section 6 in context of  L\'evy's lemma.
 %can be used to prove concentration of measure around the average value of the entropies and underlies typicality arguments in quantum information \cite{HLW06}.

\subsection{Entanglement in multipartite random states}
Consider a Haar random pure state $|\psi\rangle \in 
\mathcal H_A\otimes \mathcal H_B \otimes \mathcal H_C$
with subsystem dimensions
$
N_1=\dim \mathcal H_A,\;
N_2=\dim \mathcal H_B,\;
N_3=\dim \mathcal H_C.
$
The reduced density matrix of two subsystems
\begin{equation}
\rho_{AB}=\mathrm{Tr}_C |\psi\rangle\langle\psi|
\end{equation}
forms a random mixed state known as a \emph{random induced state} \cite{ZS01}.
Let $N=N_1N_2$.  The ensemble can be represented as
\begin{equation}
\label{Wishart}
\rho_{AB}=\frac{W}{\mathrm{Tr}\,W},
\qquad
W=XX^\dagger ,
\end{equation}
where $X$ is an $N\times N_3$ complex Ginibre matrix.
Thus $W$ is Wishart distributed and the eigenvalues of $\rho_{AB}$
follow the Marchenko--Pastur distribution in the large-dimension limit. The entanglement content in $\rho_{AB}$ is interpreted as that of a composite system interacting with a bath $C$.

While the general problem of separability in mixed states is a NP hard problem, the Peres–Horodecki positive partial transpose (PPT) criterion \cite{Bengtsson2007,HHHH09} provides a simple route to detect entanglement for the two-qubit system. It says that the partial transpose of a separable bipartite state $\rho_{AB}$ must necessarily be positive semidefinite. The partial transpose is defined as $\bra{i \alpha}\rho^{\Gamma_B}\ket{j \beta}=\bra{i \beta}\rho_{AB}\ket{j\alpha}$. In general, the PPT criterion is only a necessary condition except for a two qubits or a qubit and qutrit state. However, if the partial transpose has negative eigenvalues (negative under partial transpose or NPT), the state is necessarily entangled. If it is PPT, it could still be entangled, but it does not possess distillable entanglement \cite{HHHH09}.

If the eigenvalues of $\rho_{AB}^{\Gamma_B}$ are $\{\mu_i\}$, the trace norm is $\|\rho_{AB}^{\Gamma_B}\|_1 =\sum_{i} |\mu_i|$, and the entanglement between $A$ and $B$ can be quantified using the negativity $\mathcal{N}$ or the logarithmic negativity $E_N$ \cite{Karol1998,Vidal2002,Bengtsson2007} defined as
\begin{equation}
\mathcal{N}(\rho_{AB})=\frac{\|\rho_{AB}^{\Gamma_B}\|_1-1}{2}, \qquad  E_{\rm LN}(\rho_{AB})=\log \|\rho_{AB}^{\Gamma_B}\|_1.
\end{equation}
\begin{figure}
    \centering
    \includegraphics[width=0.8\linewidth]{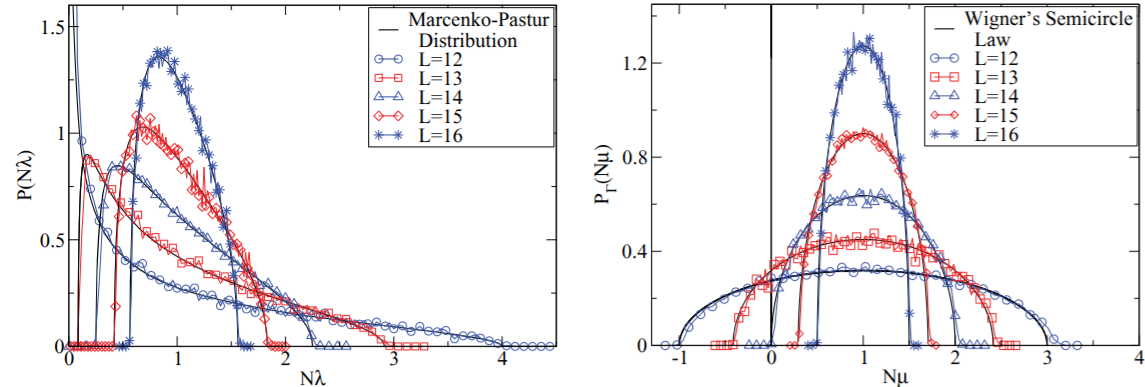}
    \caption{Density of states of $\rho_{AB}$ (left) and its partial transpose  
     $\rho_{AB}^{\Gamma_B}$ (right) for $L_1=L_2=3$, and where $L$ is the total number of qubits. A vertical line at the origin has been shown in the right figure to
draw attention to the negative part of the spectrum. In each case 250 complex random states of dimension $2^L$ are
used. (Borrowed from \cite{Bhosale2012})}
    \label{fig:MP_and_Semicircle}
\end{figure}
 Haar random pure states of $L$ qubits are considered in Fig.~\eqref{fig:MP_and_Semicircle}, with $L_1=L_2=3$, $L_3=L-6$ being the number of qubits in each of the 3 partitions. Shown are the density of states of $\rho_{AB}$ and that of its partial transpose for various values of $L$. Hence, $N_1=N_2=2^3$ and $N_3=2^{L-6}$. The eigenvalues of $\rho_{AB}$ have the expected Marchenko-Pastur distribution from Eq.~\eqref{eq:marcenko-pastur-law} with $Q=2^{L-6}/2^6$. 
 
 In contrast, the eigenvalues of the partial transposed states,
 $\rho_{AB}^{\Gamma_B}$,
 follow a semicircle law with a shifted center.
 To reach this conclusion one can study the  
  moments of this distribution. The first two moments do not change on partial transpose, $\Tr \rho_{AB}^{\Gamma_B}=\Tr \rho_{AB}$, and $\Tr \left(\rho_{AB}^{\Gamma_B}\right)^2=\Tr \rho_{AB}^2$, while the third moment does. It has been shown \cite{Bhosale2012} that for the induced ensemble
 \beq
\left \br \Tr \left(\rho_{AB}^{\Gamma_B}\right)^3 \right \kt =
\frac{N_1^2+N_2^2+N_3^2+3N_1N_2N_3}{(N_1N_2N_3+1)(N_1N_2N_3+2)} \sim \frac{1}{N_1^2N_2^2}+\frac{1}{N_2^2N_3^2}+\frac{1}{N_3^2N_1^2}+\frac{3}{N_1N_2N_3},
\label{eq:TrPT3}
 \eeq
 a manifestly symmetric form related to the third-moment being a local-unitary-invariant. 
 % Recent work has developed
 A diagrammatic random-matrix approach 
 allows one  to  compute \cite{Shapourian2021}
 the higher order moments for large $N_3$, spectral density of the partial transpose, two-point correlators, and the logarithmic negativity. 

\paragraph{Shifted semicircle law for the partial transpose}

The semicircle is determined by its center and its width, and observing that the first two moments on partial transpose are unchanged, a simple RMT model of the partial transpose $\rho_{AB}^{\Gamma_B}\sim I_N/N +A$ was considered in \cite{Bhosale2012}. Thus, the partial transpose is a GUE fluctuation matrix $A$ away from the maximally mixed state $I_N/N$, the properties of $A$ fixed by the first two moments.

This immediately leads to the eigenvalue density of the partial transpose. Writing the rescaled eigenvalues as $x=N_1N_2\,\mu$, one has
\begin{equation}
P_\Gamma(x)\approx
\frac{1}{2\pi \widetilde R^2}
\sqrt{4\widetilde R^2-(x-1)^2},
\qquad x\in [\,1-2\widetilde R,\;1+2\widetilde R\,],\qquad {\rm with} \;\; \widetilde R=2\sqrt{\frac{N_1N_2}{N_3}}.
\end{equation}
Negative eigenvalues typically occur when the lower edge crosses zero, namely when
\begin{equation}
\widetilde R>1
\qquad\Longleftrightarrow\qquad
N_3<4N_1N_2.
\label{eq:Negativity_threshold}
\end{equation}
Thus the semicircle model predicts a PPT/NPT transition at $N_3=4N_1N_2$. These are well-realized in the data shown in Fig.~\ref{fig:MP_and_Semicircle}. Rigorous derivations are based on employing free-probability theory and binary correlations \cite{Aubrun2012, Aubrun2014}. This has been characterized as the typical loss of distillable entanglement when the bath contains at least two additional qubits than the system.

The ensemble-averaged logarithmic negativity may be derived from the shifted semicircle model and for large dimensions $N_i$. Deep in the NPT regime, $\widetilde R\gg 1$, one obtains the simple asymptotic form
\begin{equation}
\langle E_{\mathrm{LN}}\rangle
\approx
\log\!\left(\frac{8}{3\pi}\sqrt{\frac{N_1N_2}{N_3}}\right).
\label{eq:ELN-deep}
\end{equation}
For $\widetilde R<1$, the semicircle support is strictly positive and the model gives $\langle E_{\mathrm{LN}}\rangle =0$, although the true average is small but nonzero because of edge fluctuations.
In fact at the critical dimension
$N_3=4N_1N_2$,
the semicircle support just touches zero. In this regime the negativity is controlled by
the extreme tail of the spectrum, 
asymptotically described by the Tracy--Widom distribution of the smallest eigenvalue. This correctly captures the small but finite
fraction of NPT states and the corresponding residual average log-negativity is $
\langle E_{\mathrm{LN}}\rangle \sim N^{-5/3}$ \cite{Bhosale2012}. It may be emphasized that even the states becoming PPT does not imply that they are separable, there is a broad regime in which random induced states are typically PPT yet still entangled, hence bound entangled \cite{Aubrun2014}. The study of entanglement in many-body systems where Haar random states may be approximately realized as thermalization is of wide interest. The transitions
between entanglement with positive and negative 
partial transpose were analyzed in \cite{Lu2020} for 
triparite systems.

\paragraph{Concurrence in multiqubit random states}

\begin{figure}
    \centering
\includegraphics[width=0.8\linewidth]{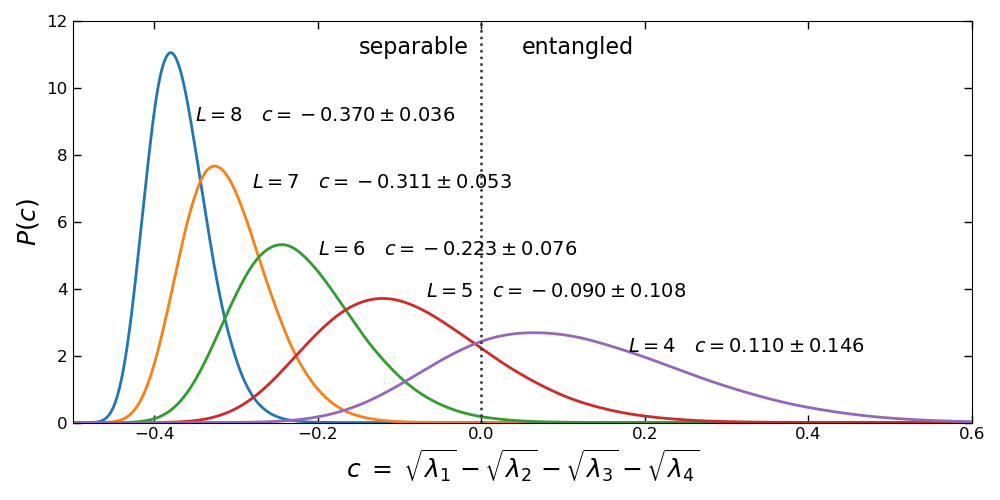}

    \caption{Distribution of two qubit pre-concurrences, $c$, in random $L-$qubit pure states. The entanglement  monotone concurrence between the qubits is $c$ if $c>0$, else it is $0$.  Indicated are the mean and standard deviation of the pre-concurrences. The approximate probability to find pairwise entanglement
between a particular pair of qubits in states with $L=4$ to $L=8$ total qubits is 0.758, 0.0198, 0.0062, 0.00001, and 0 respectively. The distributions
were numerically calculated using 1 million random states 
of size $N=2^L$.
(Essentially a remake of Fig.~3 from Ref.~\cite{Scott2003}).}
 \label{fig:RandomQubitState_Concurrence}
\end{figure}

For insight into multipartite random states and entanglement sharing among the particles consider the entanglement between pairs of qubits in Haar random states of $L$ qubits. The entanglement in two qubit states $\rho_{AB}$ is well understood
\cite{Bengtsson2007}, 
PPT states being separable and hence $\mathcal{N}(\rho_{AB})>0$ is a necessary and sufficient condition for entanglement. The widely used measure of concurrence is a measure of the entanglement of formation and is calculable from the density matrix $\rho_{AB}$ \cite{Wootters1998PRL} 
as 
\beq
C(\rho_{AB})=\text{Max}\left\{0,c\right\}, \;\; c= \sqrt{\lambda_1}-\sqrt{\lambda_2}-\sqrt{\lambda_3}-\sqrt{\lambda_4}.
\eeq
Here $\lambda_i$ are the eigenvalues of $R=\rho_{AB} \widetilde{\rho_{AB}}$ where $\widetilde{\rho_{AB}}=(\sigma_y \otimes \sigma_y)\, \rho_{AB}^*\,(\sigma_y \otimes \sigma_y)$. They are guaranteed to be non-negative and are arranged such that $\lambda_1 \geq \lambda_2 \geq \lambda_3 \geq \lambda_4$. The $*$ is complex conjugation when the state is in the computational ($\sigma_z$) basis, and $c$ is referred to as the "pre-concurrence". Thus $0 \leq C(\rho_{AB}) \leq 1$, with $0$ being attained if and only if the state is separable and $1$ if and only if it is maximally entangled. Thus, a negative value of $c$ indicates separability. Figure~\ref{fig:RandomQubitState_Concurrence} shows the distribution of the pre-concurrence among pairs of qubits from a million samples of Haar random states of $4$ to $8$ qubits. It is seen that there is a small probability that a pair of qubits is entangled in a 6 qubit random state and vanishingly small probability (for the sample size of a million) for a total of 7 or 8 qubits. Taking the threshold in Eq.~\eqref{eq:Negativity_threshold} with $N_1=N_2=2$ gives $N_3<16$ or a total of utmost 5 qubits for nonzero concurrence.  Although this is a large $N_i$ estimate, it already agrees approximately. 

\begin{figure}
    \centering
    \includegraphics[width=0.5\linewidth]{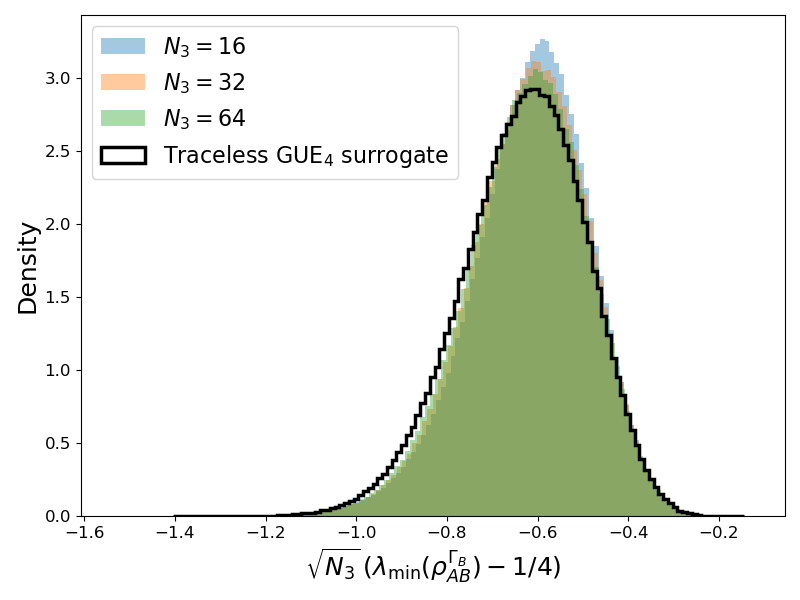}\includegraphics[width=.5\linewidth]{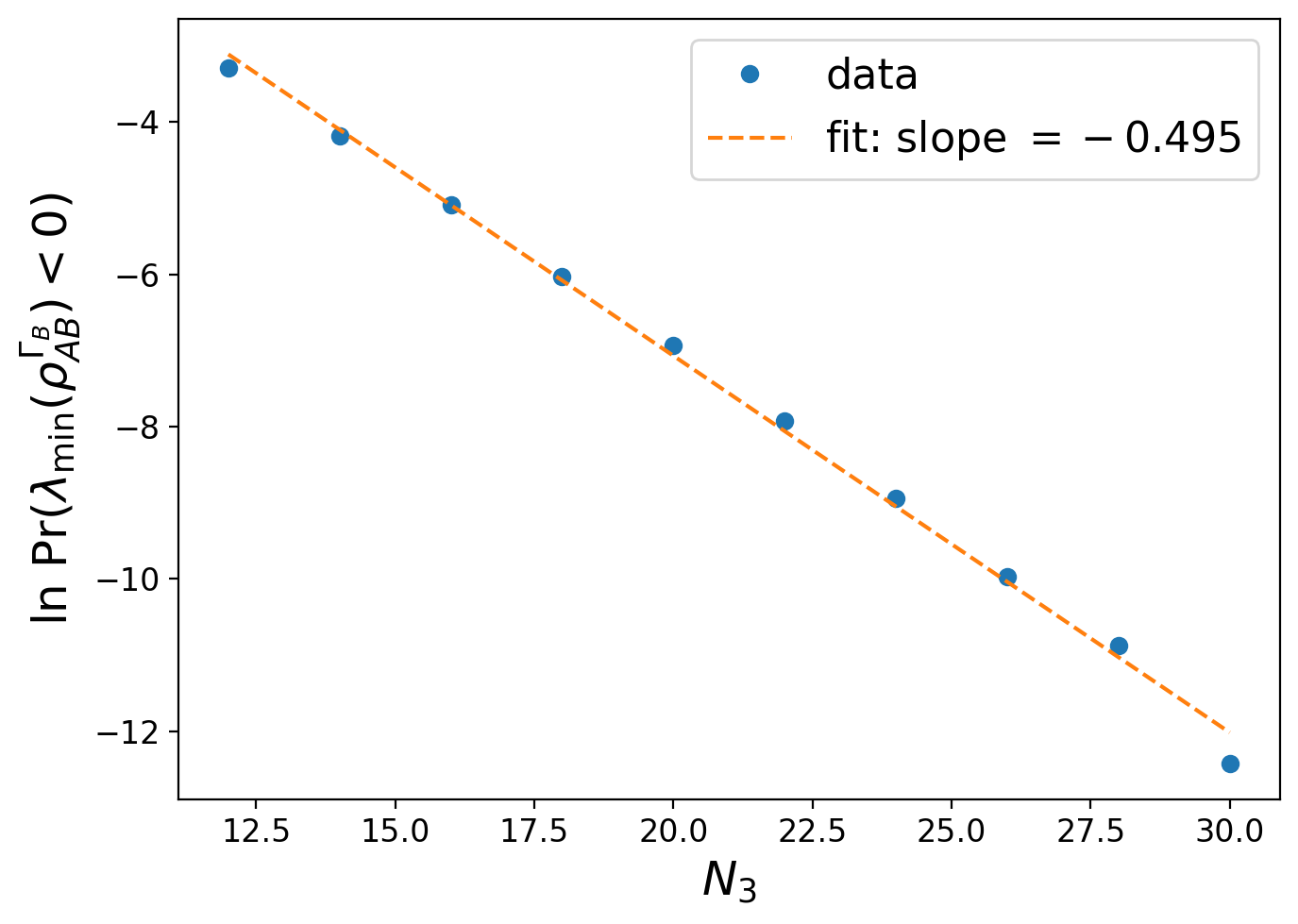}
    \caption{(Left) Data collapse of the density of a shifted and scaled smallest eigenvalue, $x_{\rm min}=\sqrt{N_3}(\lambda_{\rm min}(\rho^{\Gamma_B}_{AB})-1/4)$, of the reduced density matrix $\rho_{AB}$ of two qubits in Haar random  pure states of dimension $4N_3$. Shown also is the density of the smallest eigenvalue  of an ensemble of $4 \times 4$ GUE matrices with appropriate scale.  (Right) The probability that two qubits of such states are entangled, as a function of the "environment" dimension $N_3$. For both panels, a million pure states are sampled.}
    \label{fig:2qubitPT}
\end{figure}

A more detailed analysis of the distribution of the partial transpose spectrum of the $4\times 4$ matrix reduced  density matrix $\rho_{AB}$, indicates that the model $\rho_{AB}^{\Gamma_B}=I_4/4 + X/\sqrt{N_3}$, with the fluctuating matrix $X$ satisfying $\Tr X=0$, $\Tr X^2=15/16$, $\Tr X^3=0$ is a good surrogate for $\rho_{AB}^{\Gamma_B}$. This is justified using the first 3 known ensemble moments of the partial transpose, including Eq.~\eqref{eq:TrPT3}, which indicate that $\br X^3 \kt \sim 1/\sqrt{N_3}$. With increasing number of qubits in partition $C$, $N_3$, $\rho_{AB}$ becomes increasingly close to being maximally mixed, hence the shift. The scaling with $\sqrt{N_3}$, ensures that $ X \sim O(1)$. Fig.~\eqref{fig:2qubitPT} (left panel) shows the distribution of the shifted and scaled mininum eigenvalue of the partial transposed $\rho_{AB}$, $x_{\rm min}=\sqrt{N_3}(\lambda_{\rm min}(\rho^{\Gamma_B}_{AB})-1/4)$. The density of the smallest eigenvalue of traceless ${\rm GUE}_4$ matrices $M$ of size $4$, such that $\br \Tr M^2\kt=15/16$ is also shown. For various values of $N_3$ the scaled minimum eigenvalues $x_{\rm min}$  collapses and approximately converges to that of the ${\rm GUE}_4$. The probability that the density matrix $\rho_{AB}$ is NPT is the probability that $x_{\rm min}< -\sqrt{N_3}/4$. Thus, the problem of entanglement in two qubit subsystems becomes a large deviation problem of $4\times 4$ random matrices. As the tail of the distribution of $\lambda_{\rm min}$ is $ ~ \exp(-\alpha x^2)$, for some constant $\alpha$, this implies that the probability of NPT states and hence entangled qubit pairs decreases with $N_3$ as,
\beq
{\rm Pr}[\lambda_{\rm min}(\rho^{\Gamma_B}_{AB})<0]= {\rm Pr}[x_{\rm min}<-\sqrt{N_3}/4]\sim \exp(-\gamma N_3) \sim \exp(-\gamma 2^{L-2}).
\eeq
Here $\gamma>0$ and $L$ is the total number of qubits. This super-exponential decrease with the number of "environmental qubits" is responsible for the almost sudden quenching of entanglement as the $L$ crosses $6$, as in Fig.~\ref{fig:RandomQubitState_Concurrence}. Figure~\ref{fig:2qubitPT} (right panel) displays the numerical support of this exponential decay with $N_3$ with $\gamma \approx 0.5$.

These quantitatively illustrate the general principle of monogamy of entanglement \cite{NielsenChuang,Bengtsson2007}. Even as any two qubits lose entanglement with each other with increasing environment qubits in typical states, they get maximally entangled with the $L-2$ other qubits collectively, due to the large bipartite entanglement noted in Eq.~\eqref{eq:PageLubkin}. As random states are a proxy for quantum chaos, we often see in strongly non-integrable systems, entanglement being shared in a multipartite manner and local entanglement between small enough subsystems being lost. For an example of how this manifests in a nonintegrable many-body Ising chain, see \cite{Karthik2007}.

\section{Entanglement Generation by Chaotic Dynamics}

Bipartite systems are the natural and simplest starting point for studying the impact of quantum chaos on entanglement. Consider a product Hilbert spaces $\mcH_A \otimes \mcH_B$ with local dimensions $N_1$ and $N_2$. There are local operators such as $u_A \otimes \mbI_B$ and $\mbI_A \otimes u_B$ and nonlocal or global operators $U_{AB}$ that can have operator entanglement and when acting on product states $\ket{\phi_A}\otimes \ket{\phi_B}$ can result in entangled states. Only exceptional nonlocal operators do not entangle generic product states. 

Hamiltonians of the form $H=H_A+H_B+H_{AB}$ generate the time evolution of such bipartite systems. Consider kicked systems that could either be a Trotterization of time-independent Hamiltonians or arise exactly from Floquet dynamics. In particular, consider a general class of Hamiltonians
\beq
\label{eq:kickhamil}
H=H_A(t)\otimes \mbI_B + \mbI_A \otimes H_B(t) +  H_{AB}
\sum_{n=-\infty}^{\infty} \delta\left(\frac{t}{\tau} -n\right),
\eeq
where $\tau$ is the time between the kicks and we allow the local dynamics to be non-autononomous. If it is a result of Trotterization, then $\tau\rightarrow 0$ corresponds to the unkicked limit. The local unitary evolution between kicks $(j-1)$ and $j$ is given by the two unitary operators,
\beq
u_{A j}=\mathcal{T}\exp\left(-\frac{i}{\hbar} \int_{(j-1)\tau}^{j\tau} H_{A}(t) dt\right),\;\; u_{B j}=\mathcal{T}\exp\left(-\frac{i}{\hbar} \int_{(j-1)\tau}^{j\tau} H_{B}(t) dt\right),
\eeq  
where $\mathcal{T}$ denotes time ordering.
The complete time evolution operator between (just before) kicks $j-1$ and $j$, including the nonlocal interaction,  is 
\beq
\mathcal{U}_j= \left( u_{A\, j} \otimes u_{B\, j}\right)\, U, \;\; U=\exp \left(-\frac{i}{\hbar} \tau H_{AB}\right).
\label{eq:U_uAuB}
\eeq
This is the Floquet operator across the time period $\tau$ between kick numbers $(j-1)^{-}$ and $j^{-}$, monitored just before successive kicks.
The propagator across $n$ kicks is 
\beq
\mathcal{U}^{(n)}=\mathcal{T} \prod_{j=1}^n \mathcal{U}_j= \mathcal{T} \prod_{j=1}^n  (u_{A_{j}}\otimes u_{B_{j}})\, U.
\label{eq:Upower(n)}
\eeq
The brackets around the ``power" $n$ in $\mcU$ indicate that there are $n$ different terms in the product
in general and the time ordering will not be explicitly indicated in the following. In most studies $u_{Aj}$ and $u_{Bj}$ are independent 
of $j$, the local Hamiltonians are independent of time. This leads to a time-periodic
system with $\mathcal{U}$ as the Floquet operator or quantum map and one is then interested in their spectra and eigenfunctions or the powers $\mathcal{U}^n$ for time evolution. However, a product of unitary operators with different local operators can be useful, especially when there is local chaos. The entanglement content in the eigenstates of $\mcU$ as a function of the chaos present in the biparitite system has been studied, as well as the entanglement generated in the time evolved states $\mcU^n (\ket{\phi_A}\otimes \ket{\phi_B})$ has been, and continues to be, extensively studied, a sampling being  \cite{MilSar1999, Lakshminarayan2001,FujMiyAtu2003,Prosen_2003,Demkowicz_Kus2004,Trail_Madhok_2008,lombardi2011entanglement,Ghose, Tomsovic2018,
Jethin2023,LerPap2020}. A related, if less studied topic is the operator entanglement of $\mcU^{(n)}$ or $\mcU^n$, as well as their entangling powers that measure the average entanglement resulting from their application on random product states. They are a naturally interconnected set of questions that often provide complementary insights. It may also be mentioned that "bipartite" can also mean two parts of a multipartite collection of particles. In fact, even single particle dynamics is sometimes useful to view as collective many-particle dynamics, such as in the kicked top case \cite{Ghose}, or simply as abstract dynamics on a collection of qubits \cite{Scott2003}.

To start with, it is useful to recall basic notions of bipartite entangling power and operator entanglement and show how they thermalize under the action of 2-unitary designs

\subsection{Operator entanglement, entangling power, and gate-typlicality}

\paragraph{Operator Schmidt decomposition and Operator entanglement}

 Let $U$ be a bipartite unitary in $\mcH_{AB}$. Similarly to states, an operator Schmidt decomposition
 \cite{Zyczkowski2004} represents $U$ as an optimal sum of orthonormal product of $\{M^{A}_j\}_{j=1}^{N^2}$ and $\{M^{B}_j\}_{j=1}^{N^2}$ each forming an operator basis: $\Tr(M^{A \dagger}_j M^{A}_k)=  \Tr(M^{B \dagger}_j M^{B}_k)=\delta_{jk}$. Thus,
\beq
U= \sum_{j=1}^{N^2} \sqrt{\lambda_j} \, M^A_{j} \otimes M^{B}_j,\;\;\;
\frac{1}{N^2}\sum_{j=1}^{N^2} \lambda_j = 1, \qquad \ld_1\geq \cdots \ld_{N^2}\geq 0.
\label{eq:SchmidtU}
\eeq
The constraint on the Schmidt values $\ld_i$ follows from the unitarity of $U$ and implies that $\{\lambda_{i}/N^2\}_{i=1}^{N^2}$,
can be treated as a discrete probability measure that characterizes the nonlocality of the
operator $U$. For simplicity we consider the symmetric case, $N_1=N_2=N$, while the indicated references deal with the more general case. To elaborate, let $U \rightarrow U'= (u_{A} \otimes u_{B}) \, U  \,(v_{A} \otimes v_{B})$,
where $u_{A,B}$ and $v_{A,B}$ are local unitary operators. The nonlocality measures of $U$ and $U'$ must be identical, so they have to be invariant with respect to 
local unitary transformations. It is clear from the definition of the operator Schmidt
decomposition that the set $\{ \lambda_i\}_{i=1}^{N^2}$ are  such invariants.

Another set of $N^2$ invariants are constructed from the operator Schmidt decomposition of
the operator product $U \mbS$ where $\mbS$ is the
\textsc{swap} operator: 
\beq
\label{eq:SchmidtUS}
U\mbS= \sum_{j=1}^{N^2} \sqrt{\mu_j} \, \tilde{M}^{A}_j \otimes \tilde{M}^{B}_j,\;\;\; \frac{1}{N^2}\sum_{j=1}^{N^2} \mu_j=1.
\eeq
That the set $\{\mu_i\}_{i=1}^{N^2}$ constitutes $N^2$ invariants follows from the observation that the Schmidt eigenvalues of $U\mbS$, the $\mu_i$, are also those of $U'\mbS$. The product $\mbS U$ does not produce any new invariants. 
The moments and entropies of these invariants provide measures of how nonlocal the operator
$U$ is, leading to a class of \emph{operator entanglement entropies}. Convenient diagnostics  related to the second moment of the invariants are given by,
\beq
E(U)=1-\frac{1}{N^4}\sum_{j=1}^{N^2} \lambda_j^2, \;\;\text{and}\;
E(U\mbS)=1-\frac{1}{N^4}\sum_{j=1}^{N^2} \mu_j^2, \qquad 0 \leq E(U), \, E(U \mbS) \leq 1-\frac{1}{N^2}.
\eeq
Here $E(U)$ and $E(U\mbS)$ are the \emph{linear} operator entanglement entropies of the operators $U$ and $U\mbS$ respectively. The entropy $E(U)=0$ if and only if  the bi-partite gate is
of the product form, $U=u_A\otimes u_B$.

In order to find the Schmidt values of a bipartite operator it is useful to recall two definitions, that of matrix realignment, $\bra{ij}U^R\ket{kl}=\bra{ik}U\ket{jl}$, and the partial transpose, $\bra{ij}U^{\Gamma}\ket{kl}=\bra{il}U\ket{kj}$, which has already be mentioned above in the context of PPT states.
Taking the realignment
(also called reshuffling  \cite{Bengtsson2007})
of the Schmidt decompositions in Eq.~\eqref{eq:SchmidtU} and Eq.~\eqref{eq:SchmidtUS} give
\beq
U^R=\sum_{j=1}^{N^2}\sqrt{\ld_i} \,\ket{M_j^A}\bra{M_j^{B*}},\qquad
(U\mbS)^R=U^{\Gamma}\mbS=\sum_{j=1}^{N^2}\sqrt{\mu_i}\, \ket{\tilde{M}_j^A}\bra{\tilde{M}_j^{B*}}.
\eeq
Hence, $\lambda_i$ and $\mu_i$ are the eigenvalues of $U^R U^{R \dagger}$ and $U^{\Gamma} U^{\Gamma \dagger}$ respectively. In terms of these operators the linear operator entanglements are
\beq
E(U)=1-\frac{1}{N^4} \Tr[(U^R U^{R \dagger})^2], \qquad E(U\mbS)=1-\frac{1}{N^4} \Tr[(U^{\Gamma} U^{\Gamma \dagger})^2].
\eeq

Thus, $E(U)$ is maximized if and only if $U^R$ is unitary. These have recently drawn considerable attention as they are the same as being dual unitary, and circuits made of them are being extensively studied as models of maximal many-body quantum chaos that is also in some ways solvable \cite{BertiniKosProsen2019}. As $\mbS^R=\mbS$, it is dual unitary (indeed self-dual) and $E(\mbS)=1-1/N^2$ is the maximum possible. Finally we note an alternative form of $E(U)$  that is useful for Haar averaging over $U$
as well as its connection to the entangling power. This involves two copies of the bipartite unitary on spaces labeled $AB$ and $A'B'$:
\beq
E(U)=1-\frac{1}{N^4}\Tr\left[ U^{\otimes 2}\mbS_{AA'}U^{\dagger \otimes 2} \mbS_{AA'}\right],\qquad U^{\otimes 2} =U_{AB} \otimes U_{A'B'}. 
\label{eq:EU_swapform}
\eeq

\paragraph{Entangling power and gate-typicality}
The entangling power of a bipartite unitary $U$, $e_p(U)$
is the {\it average} entanglement created when $U$ acts on product states
$|\phi_A\kt|\phi_B\kt$ sampled according to the Haar measure on the subspaces:
% P. Zanardi, C. Zalka, and L. Faoro, Entangling power of quantum evolutions, Phys. Rev. A 62, 030301 (2000).
%
\beq
\label{eq:ep_symm}
e_p(U) = \left(\frac{N+1}{N-1} \right)\,
\overline{\mathcal{E}(U|\phi_A\kt|\phi_B\kt)}^{\phi_A,\phi_B}.
\eeq
If the entanglement measure is the linear entropy,
$\mathcal{E}(\phi_{AB}) = 1 -
\Tr(\rho^2_A)$, it
has been shown that the entangling power is related to the operator entanglements $E(U)$ and $E(U \mbS)$. Using the identity, $\Tr(\rho_A^2)=\Tr[(\rho_A \otimes \rho_{A'}) \mbS_{AA'}]$, where $\rho_{A'}$ is a copy of $\rho_A$, it follows that
\beq
\Tr(\rho_A^2)=\Tr\left[ U^{\otimes 2} \ketbra{\phi_A}^{\otimes 2} \ketbra{\phi_B}^{\otimes 2} U^{\dagger \otimes 2} \mbS_{AA'}\right].
\eeq
As $\phi_A$ and $\phi_B$ are independently sampled, it is possible to use the random state second moment averages from Eq.~\eqref{eq:Haar_second_moment}. Finally, recognizing the form of $E(U)$ in Eq.~\eqref{eq:EU_swapform}, results in the compact formula \cite{Zanardi2001} 

\beq
e_p(U)=\frac{1}{E(\mbS)}\left[E(U) + E(U\mbS) - E(\mbS)\right], \qquad 0\leq e_p(U) \leq 1.
\label{eq:ep}
\eeq
If $e_p(U)=0$ then $U$ is either a product of local operators or local unitary equivalent to the \textsc{swap} $\mbS$. The fact that \textsc{swap} does not create
any entanglement when acting on product states leads to $e_p(\mbS)=0$, although the operator entanglement $E(\mbS)$ is maximal. Indeed the \textsc{swap} gate interchanging two particles is a highly nonlocal operation. This forms one motivation for introducing the \textsl{gate typicality} as \cite{Bhargavi2017}
\beq\label{eq:gt}
g_t(U) := \frac{1}{2E(\mbS)}\left[E(U) - E(U\mbS) + E(\mbS) \right],\;\;
0 \le g_t(U) \le 1.
\eeq
and $g_t(U)=1$ if and only if $U$ is the \textsc{swap} gate $\mbS$ or is local unitarily equivalent to it. 
Thus, while entangling power does not distinguish the local operators from the \textsc{swap}, gate-typicality does. The second motivation is that it will arise naturally when thermalization is considered. It
turns out that rather than discussing the pair $\{E(U), E(US)\}$ it is more convenient  to work with the linearly related pair of  $\{e_p(U), g_t(U)\}$ \cite{Bhargavi2017,Bhargavi2020}. Recent work has shown how these determine the approach of structured random unitary circuits to unitary 2-designs \cite{suzuki2025globalrandomnessrandomlocal}.

Note that the maximal value $e_p^{\text{max}}(U)=1$ is reached  if and only if $E(U)=E(U\mbS)=E(\mbS)$. It turns out that there are no two-qubit gates that  maximize $E(U)$ and $E(U\mbS)$
simultaneously \cite{Higuchi2000},
 and as a result  for $N=2$, $e_p^{\text{max}}(U)=2/3$, a value reached, for example, by  the ${\rm CNOT}$ gate. On the other hand, it has been shown that for all $N>2$ there are $U$ such that $e_p(U)=1$. These have direct implication on the construction of the so-called absolutely maximally entangled states for four parties, see \cite{AME_review} for a recent review.

Just as typical bipartite states are highly entangled, typical bipartite unitaries from the Haar measure have nearly maximal operator entanglement and entangling power. The average operator entanglement $E(U)$ (which is the same as that of $E(U \mbS)$) can be found when $U$ is sampled from the Haar measure on $\mcU(N^2)$. Starting from the form in Eq.~\eqref{eq:EU_swapform}, using the twirling formula Eq.~\eqref{eq:twirl2}
with $X=\mbS_{AA'}\mbS_{BB'}$, and noting that the dimension of the unitaries is now $N^2$ gives the result. From this  we get the Haar average of the entangling power and gate-typicality using their definitions in Eq.~\eqref{eq:ep} and Eq.~\eqref{eq:gt}. These are
\beq
\overline{E}=\frac{N^2-1}{N^2+1},\;\; \overline{e_p}=\frac{N^2-1}{N^2+1}, \;\; \overline{g_t}=\frac{1}{2}.
\label{eq:EU_HaarAvg}
\eeq
More general formulas when $N_1\neq N_2$ are in \cite{Bhargavi2020}. Note that there is a lack of standardization of an overall factor in the definition of the entangling power. The scaling chosen here is such that the range of $e_p(U)$ is independent of the dimensionality $N$. Finally, we reiterate and emphasize the local unitary invariance of all these measures:
\beq
E(U')=E(U), \;\; e_p(U')=e_p(U), \;\; g_t(U')=g_t(U). \qquad U'= (u_{A} \otimes u_{B}) \, U  \,(v_{A} \otimes v_{B}).
\label{eq:LUI}
\eeq

\paragraph{Thermalization of entangling power}

As discussed above, for bipartite systems, one is generally interested in the iteration of $\mcU=(u_A\otimes u_B)U$, and the entanglement in $\mcU^n\ket{\phi_A \phi_B}$ as a function of time $n$. However this would in general be dependent on the initial product state. The entangling power, as it averages over the initial states, provides a more universal diagnostic, especially when it comes to chaotic dynamics. Thus, $e_p(\mcU^n)$ can be expected to be useful in understanding the production of entanglement \cite{Demkowicz_Kus2004}. Thermalization in this context implies that $e_p(\mcU^n)$ approaches the Haar average $\overline{e_p}$ as $n \rightarrow \infty$. 

Therefore, consider for $n=2$
\[ e_p(\mcU^2)=e_p[(u_A\otimes u_B)\, U\, (u_A\otimes u_B) \, U]= e_p[ U\, (u_A\otimes u_B) \, U].\]
While the second equality is due to local unitary invariance, the sandwiched local dynamics plays a crucial role as emphasized in \cite{Bhargavi2017}. However, at this stage, little can be said in generality. When there is quantum chaos in the dynamics of the subsystems, that is $u_{A,B}$ have random matrix properties, Haar averaging over $u_A$ and $u_B$ can provide meaningful insights. Thus using Eq.~\eqref{eq:Wein2_deltaform} and assuming that $u_A$ and $u_B$ are independently Haar distributed the following more general relations are derived \cite{Bhargavi2017,Bhargavi2020}:
\beq
\begin{split}
&\left \br e_p\left[ U (u_A\otimes u_B) V \right]\right \kt_{u_A, u_B} =\,
e_p(U)+e_p(V) -\, 
% e_p(U)e_p(V)/\overline{e_p},\\
\frac{e_p(U)e_p(V)}{\overline{e_p}},\\
& \left \br g_t\left[ U (u_A\otimes u_B) V \right]\right \kt_{u_A, u_B} =\,
g_t(U)+g_t(V) -\, \frac{g_t(U)g_t(V)}{\overline{g_t}}.
\end{split}
\label{eq:epUVavg}
\eeq
Few observations on these remarkably elegant forms are in order:
\begin{itemize}
    \item The role of the gate-typicality along with entangling power is highlighted. In particular, other invariants, such as the operator entanglement, satisfy more complicated relations that mix $E(U)$ and $E(U\mbS)$. See \cite{Faidon2024} for essentially a rederivation, but in terms of the operator entanglements to define an operator space entangling power.
    \item If $U$ is such that $e_p(U)=\overline{e_p}$, then for any other unitary operator $V$, $\left \br e_p\left[ U (u_A\otimes u_B) V \right] \right \kt_{u_A,u_B}=\overline{e_p}$. Similar
    property holds for
    $g_t(U)$.
  %  \item Although not highlighted previously, the reason for the existence of these elegant forms are related to the fact that the ensemble 
 %    \[ \mathcal{E}_U=\{(u_A \otimes u_B) U\; \; |\; u_A, u_B \sim {\rm Haar}, \;\; e_p(U)=\overline{e_p}, \, g_t(U)=\overline{g_t}\}\] forms an {\em exact} unitary 2-design in $\mcH\otimes \mcH$ using Haar unitaries on $\mcH$ alone.
 %    {\color{red} It is not clear how many unitaries are needed to assure an {\sl exact} unitary 2-design (and how to generate them).  K.}

%% Thanks Karol: well the ensemble had infinite members with local $u_A, u_B$

     \item Given any  bipartite unitary $U \in \mcU(N^2)$, consider the ensemble consisting of all its local unitary equivalent operators: \[ \mathcal{E}_U=\{(u_A \otimes u_B) U (v_A \otimes v_B)\; \; |\; u_A, u_B, v_A, v_B \in \mcU(N) \sim {\rm Haar} \}\] 
      Although not previously highlighted, the existence of the elegant forms in Eq. ~\ref{eq:epUVavg} is related to the nontrivial eigenvalues of the second moment operator $M_2^{\mathcal{E}}$, Eq.~\eqref{eq:unitary_t_design_def}, of the ensemble $\mathcal{E}_U$,   being $1-e_p(U)/\overline{e_p}$ and $1-g_t(U)/\overline{g_t}$. This emphasizes the unique role of $e_p(U)$ while revealing the deeper interpretation of gate-typicality.
\end{itemize}

Generalizations involving $\br e_p(\mcU^n)\kt_{u_A, u_B}$ are not known for $n>2$, and involve higher order Weingarten terms. However, if the local unitaries are independent and random in each time step, which could be relevant for systems as described in Eq.~\eqref{eq:Upower(n)}, a corollary of Eq.~\eqref{eq:epUVavg} immediately follows with $V=\mcU^{(n-1)}$
\begin{align}
\begin{split}
\br e_p{(\mcU^{(n)})}\kt_{\rm Loc} &= e_p(U) + \left[1- \frac{e_p(U)}{\overline{e_p}}\right]\br
e_p(\mcU^{(n-1)})\kt_{\rm Loc} %{W_{n-2}}
= \overline{e_p}\left[1 - \left(1 - \frac{e_p(U)}{\overline{e_p}}\right)^n\right].
\end{split}
\label{eq:epUavg_nloc_Theo}
\end{align}
An identical equation also follows for the gate-typicality. Thus, on averaging $e_p(\mcU^{(n)})$ tends exponentially to the Haar averaged entangling power. There is thermalization driven purely by local unitary operators, as long as $e_p(U)>0$, however small it may be. One may expect that even for Floquet systems with moderate interactions and strong local chaos, this exponential approach holds approximately without any averaging. This is demonstrated in some results below. It also informs us that the entangling power of even the single time unitary interaction  is a proxy for interaction strength in Floquet systems, controlling the growth of entanglement in the presence of strong chaos. 

\paragraph{Fault tolerant quantum computing and the need for magic}

In fault-tolerant quantum computation, a central structural distinction is between \emph{Clifford} and \emph{non-Clifford} operations. 
%{\color{blue} 
In the case of multi-qubit systems
%}
the finite set of Clifford gates map Pauli operators to Pauli operators under conjugation and preserve the set of stabilizer states. Circuits composed solely of Clifford gates acting on stabilizer inputs can be efficiently simulated classically via the Gottesman--Knill theorem \cite{NielsenChuang}. Consequently, Clifford operations are regarded as ``free'' resources, as they can typically be implemented with low overhead in quantum error-correcting codes due to the property of transversality. However, non-Clifford gates, whose implementation is substantially more expensive, are required for universal quantum computation. Quantifying the resourcefulness of non-Clifford operations—often referred to as "magic" or "non-stabilizerness"—is therefore central to understanding both computational universality and overhead \cite{Lorenzo2022}.

Characterizing a unitary by the average amount of magic it generates on stabilizer inputs—its \emph{non-stabilizing power} $m_p(U)$ parallels entangling power $e_p(U)$, where the "free" resource are product operators. Introducing this measure, a recent work \cite{Varikuti2025} shows that when non-Clifford gates are interspersed with random Clifford gates, the circuit's magic thermalizes toward its Clifford-averaged value $\overline{m_p}$, in exactly the same was a entangling power above. In particular, 
\beq
\br m_p(U C V)\kt_C=m_p(U)+m_p(V)-\frac{m_p(U)m_p(V)}{\overline{m_p}},
\eeq
where $C$ is the set of Clifford gates and $U$ and $V$ are arbitrary unitaries in $\mcU(N)$. Therefore, such "decoupling" under free resource averaging may be a more general feature of quantum information.

\subsection{Numerical illustration and universal transitions with coupled kicked rotors}

Several bipartite Floquet operators have been considered in the literature from the point of view of entanglement in eigenstates as well as time-evolving states. They all have generically the form in Eq.~(\ref{eq:U_uAuB}), with $u_j^A=u^A$, $u_j^B=u^B$, being independent of time $j$. Two popular models include coupled kicked tops
\cite{MilSar1999}
%\bibitem{PB26} A. V. Purohit and U. T. Bhosale. Quantum Kicked Top: A Paradigmatic Model, this volume
% \bibitem{BC26} G. Benenti, G. Casati, J. Gong, Z. Zou,
%The Quantum Kicked Rotor: A Paradigm of Quantum Chaos. Foundational aspects and new perspectives, 
%preprint arXiv:2604.abcde and in this volume}
%
%%%% Arul: @ Karol: I am restricting the references here only to **coupled** tops and rotors and so removed some of the suggested references. 
and coupled kicked rotors
\cite{Lakshminarayan2001,RicLanBac2014}, or standard maps. The Floquet operator for the latter reads
\beq
\mcU=\left[U_S(K_1) \otimes U_S(K_2) \right] U_b,
\eeq
where $U_S(K)$ are the quantum standard maps in Eq.~\eqref{eq:quantum_standard_map} and the parameter dependence is explicitly displayed. The interaction $U_b$ is often taken as diagonal in position representation and is determined by a potential $V(q_A, q_B)$.
The choice
\beq
\label{potential_V}
U_b=\exp(-i V/\hbar) \ \ {\rm with} \ \ 
V=(b/4 \pi^2)\cos[2\pi (q_A+q_B)]
\eeq
leads to a fairly well studied system. Its classical limit is a 4-dimensional (symplectic) map \cite{RicLanBac2014}. 
Unentangled initial states $\ket{\phi_{AB}(0)}$ of two types are considered, namely (i) a product of coherent states $\ket{q_0p_0} \otimes\ket{q_0'p_0'}$ and (ii) a product of Haar random states. The entanglement content in the time evolved states
\beq
\ket{\phi_{AB}(t)}=\mcU^t \ket{\phi_{AB}(0)}
\eeq
is of interest, as a function of the subsystem chaos parameter $K_1$, $K_2$, the interaction $b$ and the effective (inverse) Planck constant $N$. 

\begin{figure}
    \centering
    \includegraphics[width=.8\linewidth]{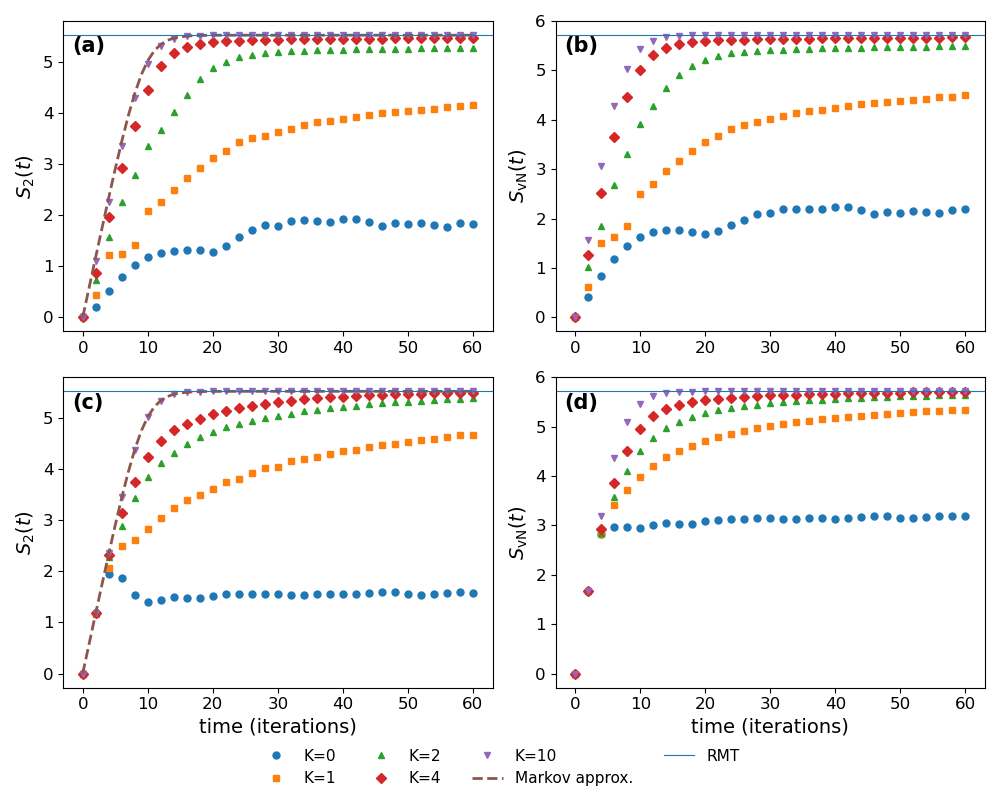}
    \caption{The Renyi-2 entropy and von Neumann entropy growth for two coupled quantum standard maps. In (a) and (b) the initial state is a product coherent state centered at the uncoupled fixed points ($(q_1,p_1, q_2,p_2)=(0.5,0,0.5,0)$. The initial state in (c) and (d) is a random product state. The individual subspaces have dimension $N=500$, the interaction is set as $b=0.01$, and the parity and TR symmetries are broken with phases. Shown are 5 different values of the single standard map parameters $K$. Also, indicated in (a) and (c) as the "Markov approx." is the expression for $S_2$ in Eq.~\eqref{eq:Renyi2-Markov}, based on the independently local averaged entangling power of Eq.~\eqref{eq:epUavg_nloc_Theo}.}
    \label{fig:Entanglement_vs_time}
\end{figure}

Fig.~\ref{fig:Entanglement_vs_time} shows the time evolution of both the von Neumann entropy, $S_{\rm vN}(t)$, and the R\'enyi-2 entropy, $S_2(t)$. The initial coherent state is taken to be centered at $q_0=0.5,\, p_0=0$ in both maps. Classically, in the uncoupled maps, this corresponds to an unstable fixed point for all $K>0$, and for simplicity we take $K_1=K_2=K$, and the interaction is $b=0.01$. The effective inverse Planck constant is determined by the dimension
of each subspace,  $N=500$. 

For these parameters, both types of initial states, coherent and random,  when evolved show  increasing entanglement with local instability and chaos. In particular for $K=0$, when the uncoupled maps are integrable, under this moderate interaction the entanglements saturate to a value that is significantly smaller than the RMT values (for large $N$, these are the Page value $\ln N-1/2$ for $S_{\rm vN}$ and $\ln (N/2)$ for $S_2$). On the other hand, when $K=10$, and the subsystem dynamics is fully chaotic with a Lyapunov exponent $\approx \ln 5$, after an initial linear growth, the entanglements saturate to the RMT values. Thus, for both initial states, there is thermalization of entanglement in the presence of strong chaos. For intermediate values of $K$, in the case of initial coherent states, we see  a dependence on the local instability at the fixed point $(0.5,0,0.5,0)$. This is in qualitative agreement with the expectation that until the Ehrenfest time entropy grows as $\sim  \Lambda_{\rm loc} t$, where $\Lambda_{\rm loc}$ is the local Kolmogorov-Sinai entropy \cite{Sin09}, which 
due to Pesin theorem is given by 
the sum of positive Lyapunov exponents  \cite{MilSar1999,LerPap2020}. However, this exponential increase is limited when $K$ is too large, around $K \sim \sqrt{N}$, when the Ehrenfest time becomes less than the kicking interval leading to instant scrambling. 

When this happens, the initial coherent state case is not very different from that of the random states, and the growth of $S_2(t)$ is approximated by the thermalization of the entangling power in Eq.~\eqref{eq:epUavg_nloc_Theo}. As $S_2(t)= -\ln \Tr(\rho_A^2(t))=-\ln (1-S_L(t))$, if there is sufficient decorrelation between consecutive time-steps that identical local dynamics is as good as independent random ones, we may replace $S_L(t)$ by the averaged entangling power and for large $N$,
\beq
S_2(t) \approx -\ln\left[ \frac{2}{N}+(1-e_p(U))^t\right].
\label{eq:Renyi2-Markov}
\eeq

This is indicated as the ``Markov approximation" in Fig.~\eqref{fig:Entanglement_vs_time}, and indeed serves even in this Floquet setting as a good guide for strongly chaotic subsystems. Even when the initial state is a coherent product state, the scrambling is strong enough to make this a good approximation. This gives a time scale for the thermalization of the entanglement as $t_*\sim -\ln N/\ln [1-e_p(U)]$ that is similar to the Ehrenfest time, but the role of the Lyapunov exponent is replaced by the entangling power of the interaction. The initial linear growth of entanglement is captured by  $S_2(t)\approx -t \ln[1-e_p(U)]$ to a good approximation. For the coupled standard maps shown in Fig.~\ref{fig:Entanglement_vs_time}, more explicit forms of this initial growth are possible, as the entangling power may be found as a function of the interaction parameter $b$
present in Eq.~(\ref{potential_V}), 
to a good large $N$ approximation as
\beq
e_p(\mcU)=e_p(U_b)\approx 1-\sum_{k=-\infty}^{\infty} J_k^4\left(\frac{bN}{2 \pi}\right)\approx 1- \int_0^1  J_0^2\left(\frac{bN}{\pi}\sin \pi x \right)\, dx \approx 1-\frac{C}{bN},
\eeq
where $J_k$ are Bessel functions and $C$ involve constants and $\ln(bN)$ terms.

%{\color{red}  what is meaning of $b$ in the formulae above ?}
%% AL: I have now emphasized here that the interaction strength is $b$.

Thus, the initial growth is $S_2(t) \approx t\, \ln (bN/C)$.
An ensemble of Haar-random states can behave quite differently from that of coherent states when the dynamics is not fully chaotic \cite{Demkowicz_Kus2004}, and the full complexity of the entanglement growth in the case of mixed dynamics (with both regular and chaotic regions) is yet to be fully understood.

\paragraph{Universal spectral and entanglement transitions}

The case of weakly interacting strongly chaotic systems has been studied with the help of perturbation theories \cite{FujMiyAtu2003,BanLak2004,Prosen_2003}. In this case, a dimensionless universal transition parameter, involving the interaction strength and effective Planck constant, has been identified. This determines spectral transitions and  entanglement growth with interaction as well as with time \cite{Tomsovic2018,Jethin2020,Jethin2023}.
For the standard map coupling $U_b$, the transition parameter is
\beq
\Lambda=\frac{N^2}{4 \pi^2}\left [1-J_0^2\left(\frac{Nb}{2\pi}\right) \right] \approx \frac{N^4 b^2}{32 \pi^4}. 
\eeq
 $\Lambda=0$ is an uncoupled limit with no entanglement, and $\Lambda \ll 1$  is a low entanglement perturbative regime, while $\Lambda \sim 1$ is a transition regime, and for $\Lambda \gg 1$ a fully entangled limit is reached with global chaotic properties. 
where the approximations are valid for $b\,N \ll 1, \; N \gg 1$. Summarizing key results:
\begin{itemize}
   % \item Spectral statistics transitions from the Poissonian for $\Lambda \ll 1$ to Wigner-Dyson when $\Lambda \gg 1$, the critical value being $\Lambda \sim 1$. As $N \rightarrow \infty$, increasingly small coupling ($b \sim 1/N^2$) is sufficient for globally chaotic spectra \cite{Srivastava2016}. 
    
    \item Eigenfunction entanglement: for $\Lambda \ll 1$, the average linear entropy and the von Neumann entropies are found using a regularized perturbation theory to be \cite{Tomsovic2018} \beq \br S_L\kt =\pi^{3/2}\sqrt{\Lambda}/2+O(\Lambda), \qquad \br S_{\rm vN}\kt =\pi^{3/2} \sqrt{\Lambda}+O(\Lambda).\eeq Thus, these indicate a lack of eigenstate thermalization and a violation of the Eigenstate Thermalization Hypothesis (ETH) \cite{Sr94}. ETH is seen to be  restored at strong coupling, $\Lambda \sim 1$, using so-called recursively embedded perturbation theory. This is found to agree well with the eigenstate entanglement for the coupled standard maps.
    
    \item For time evolution, a rescaled time $t=n D \sqrt{\Lambda}$ is to be used, where $D=2\pi/N^2$ is the mean-level spacing. If the initial states are products of uncoupled subsystem eigenstates, even strong subsystem chaos, does not lead to thermalization \cite{Jethin2020}. For example, in the case of linear entropy:
    \beq
    S_L(t; \Lambda)=C(2;t) \sqrt{\Lambda} +O(\Lambda) \approx 4 \pi t \sqrt{\Lambda}+O(t^3 \Lambda).\qquad S_L(\infty; \Lambda)=\sqrt{\Lambda}\begin{cases}
        5 \sqrt{\pi/8}  & {\rm for\;\; COE}\\5 \pi^{3/2}/8 & {\rm for \;\; CUE}.
    \end{cases}
    \label{eq:SL_quench}
    \eeq
    Here $C(2;t)$ contains the essential variation in time and is a known function that differs between the COE and CUE cases. While the short time linear slope is independent of the ensemble, it is interesting that the saturation value does depend with a slightly larger value for the CUE case. Again results for non-perturbative values of $\Lambda$ have been derived \cite{Jethin2020}.
    \item Time evolution of generic states leads to near maximal entanglement even for $\Lambda \ll 1$, unlike the negligible entropy in Eq.~\eqref{eq:SL_quench}. It has been pointed out that {\it quantum coherence in the uncoupled eigenbasis} organizes and determines these vastly different entanglement growths. Quantum coherence measures the strength of the  off-diagonal density matrix elements. This basis dependent quantity has been extensively studied as a resource in quantum information \cite{CoherenceRMP}. In the context of coupled strongly chaotic systems, coherence in the uncoupled eigenbasis is a resource for thermalization of the entanglement \cite{Jethin2023}. Random states have a large coherence in the uncoupled eigenbasis and lead to RMT entanglement values  even for $\Lambda \ll 1$, although it will take a long time to do so. The role of coherence has been studied in other contexts of quantum chaos and localization in \cite{Namit2021, Styliaris2019}. Using recursive perturbation theory, non-perturbative in $\Lambda$ expressions can be found for various cases of initial states, and is equivalent to the Markov approximation when using Eq.~\eqref{eq:epUavg_nloc_Theo} 
    for Floquet systems \cite{Jethin2023}.

\end{itemize}

%\begin{itemize}
%  \item Entanglement in random states
%  \cite{HHHH09}
%  and comparison with dynamical states/eigenstates of quantum chaos. (a) Page's formula, (b) concurrence and lower order entanglement (c) negativity in tripartite subsystems. 
%  \item Entangling power and operator entanglement. Definitions and uses in quantum chaos and thermalization in many-body systems. 
%  \item Implication for QI: chaos as an efficient generator of resource states for teleportation, dense coding, and quantum simulation. Monogamy of entanglement in typical states. Use for benchmarking quantum supremacy in unitary circuits.
 % \item Resource note: Wishart ensemble $\rightarrow$ eigenvalue distribution of reduced density matrices; Mar\v{c}enko--Pastur law; entanglement spectrum as a chaos diagnostic.predicted
%\end{itemize} 

%\subsubsection{Level 3 heading in sentence case}\label{chap1:subsubsec1}
%\paragraph{Level 4 heading in sentence case}

\section{Quantum channels and randomness}
%\label{chap3:sec1}
\label{sec:Channels}

\subsection{Quantum operations and chaotic dynamics}

Quantum dynamics of a perfectly isolated system is described by a unitary transformation of a pure state, $|\psi\rangle \to U|\psi\rangle$. In realistic settings, however, the intended unitary evolution $U$ is typically accompanied by additional perturbations, so that the actual transformation takes the form $U' = VU$, where $V = \exp(iHt)$ may be interpreted as a unitary noise contribution. Here $H = H^\dagger$ is a Hermitian operator, which can be viewed as a Hamiltonian corresponding to a quantized chaotic system or modeled as a random matrix drawn from the Gaussian orthogonal or unitary ensemble. The strength of the noise is controlled by the interaction time $t$, and the noiseless dynamics is recovered in the limit $t \to 0$.

To describe the evolution of a system interacting with an environment, one employs the notion of a quantum state $\rho$, also referred to as a density matrix. It is a positive operator, $\rho = \rho^\dagger \ge 0$, normalized by the trace condition, ${\rm Tr}\,\rho = 1$. Let $\Omega_N$ denote the set of all normalized states acting on an $N$-dimensional Hilbert space ${\cal H}_N$. Any physical process in discrete time is described by a linear map $\Psi:\Omega_N \to \Omega_N$, which is trace-preserving and {\sl completely positive}, meaning that the extended map $\Psi \otimes {\mathbbm I}_M$ remains positive for arbitrary dimension $M$ of the environment. Such transformations are called {\sl quantum operations}, or quantum stochastic maps, and can be represented by a matrix $\Psi$ of size $N^2$,
\begin{equation}
\label{super1}
\rho' = \Psi(\rho), \qquad \text{or equivalently} \qquad 
\rho_{m\mu} = \Psi_{m\mu,\, n\nu}\,\rho_{n\nu},
\end{equation}
with the summation over repeated indices understood.

It is often convenient to reorder the elements of the superoperator \cite{Bengtsson2007} by introducing the {\sl dynamical matrix} \(D\), also known as the {\sl Choi matrix},
\begin{equation}
\label{reshuff}
D_{\Psi} \equiv \Psi^R, \qquad \text{so that} \qquad 
D_{mn, \, \mu\nu} = \Psi_{m\mu,\, n\nu}
 \qquad \text{and} \qquad 
D = N(\Psi \otimes {\mathbbm I})\,|\phi^+\rangle \langle \phi^+|,
\end{equation}
where $|\phi^+\rangle = (\sum_{j=1}^N |j,j\rangle)/\sqrt{N}$. Here, the symbol \(X^R\) denotes the reshuffling (or realignment) operation applied to a four-index matrix \(X\), as introduced in Sec.~5.1; it corresponds to a representation-dependent permutation of matrix elements. The matrix \(D_{\Psi}\) is Hermitian whenever the map \(\Psi\) preserves Hermiticity.
According to the Choi theorem \cite{Bengtsson2007,Wa18}, a linear map \(\Psi\)
acting on a state of size $N$
is completely positive if and only if the corresponding dynamical matrix \(D\) 
of order $N^2$ is positive semidefinite, $D \geq 0$. Consequently, $D$ may be interpreted as a positive operator acting on the composite Hilbert space, 
$\mathcal{H} := \mathcal{H}_A \otimes \mathcal{H}_B$.
The final expression in  Eq. (\ref{reshuff}) is known as the 
{\sl Jamio{\l}kowski isomorphism}, which establishes a correspondence between a map $\Psi$ and the associated quantum state $D/N$ of an extended system.

Any completely positive map admits a {\sl Kraus representation},
\begin{equation}
\label{Kraus}
\rho \;\mapsto\; \rho' = \Psi(\rho) \; = \;  \sum_{i=1}^{M} \, K_i \, \rho \, K_i^\dagger.
\end{equation}
% I have changed A_i --> K_i to avoid conflict with partial trace Tr_A D 
In this representation, the set of Kraus operators $\{K_i\}$ allows one to express the superoperator in the form
\begin{equation}
\label{super2}
\Psi = \sum_{i=1}^{M} K_i \otimes \overline{K_i},
\end{equation}
where $\overline{K_i}$ denotes the complex conjugate of $K_i$, and the adjoint is given by $K_i^\dagger = \overline{K_i}^{\,T}$.
The theorem of Choi implies that taking 
the reshaped eigenvectors of $D$ as Kraus operators $K_i$
the number $M$ of terms in expression (\ref{Kraus})
is given by the rank of the Choi matrix, $r(D)\le N^2$.

The map \(\Psi\) is {\sl trace preserving} if $\operatorname{Tr}\,\Psi(\rho) = \operatorname{Tr}\,\rho = 1$, which holds if and only if $\sum_{i=1}^{M} K_i^\dagger K_i = \mathbb{I}_N$. This condition is equivalent to a partial trace constraint on the dynamical matrix, $\operatorname{Tr}_A D = \mathbb{I}_N$, which further implies $\operatorname{Tr}\,D = N$. If the dual condition $\operatorname{Tr}_B D = \mathbb{I}_N$ is also satisfied, the operation is {\sl unital}, since the maximally mixed state is preserved, $\Psi( \mathbb{I}_N/N) = \mathbb{I}_N/N$.

Any quantum channel $\Psi$, defined by its Kraus form (\ref{Kraus}), can also be represented as a partial trace over an auxiliary system describing an environment of sufficiently large dimension $M$,
\begin{equation}
\label{env2}
\rho'=\Psi(\rho)= {\rm Tr_B}  
\bigl(V(\rho \otimes |0\rangle \langle 0|)V^{\dagger}\bigr),
\end{equation}
where $|0\rangle \in {\cal H}_B$ belongs to an $M$-dimensional Hilbert space. The corresponding Kraus operators $K_j$ are determined by the blocks of size $N$ of a unitary matrix $V$ of size $NM$.

To generate a random operation $\Psi$ acting on a system of size $N$, it is sufficient to take $V$ to be a Haar-distributed random unitary matrix of order $NM$, or alternatively an evolution operator of a quantum chaotic system of the same dimension. The case $M = N^2$ is of particular importance, as it leads to a full-rank Choi matrix $D$ and induces the uniform (Hilbert--Schmidt) measure in the convex set of quantum operations \cite{BCSZ09}.

The same ensemble of random maps can be constructed in an alternative way using random Wishart matrices $W = GG^{\dagger}$, where $G$ denotes a Ginibre matrix with independent complex Gaussian entries. To generate a valid random Choi matrix, one must enforce the appropriate partial trace condition. Let $Y := {\rm Tr}_A GG^{\dagger} \ge 0$, so that the matrix $\sqrt{Y}$ is well defined. For a random Ginibre matrix $G$, the matrix $\sqrt{Y}$ is typically invertible, which allows one to introduce
\begin{equation}
\label{Gini_Choi}
D = \Bigl( {\mathbbm I}_N \otimes
\frac{1}{\sqrt{Y}} \Bigr)
GG^{\dagger}
\Bigl( {\mathbbm I}_N \otimes
\frac{1}{\sqrt{Y}} \Bigr),
\end{equation}
which serves as a random Choi matrix: it is positive definite, and the partial trace condition $\operatorname{Tr}_A D = \mathbb{I}_N$ is satisfied by construction. To recover the superoperator $\Psi$, it suffices to reshuffle the Choi matrix according to Eq.~(\ref{reshuff}).

To investigate spectral properties of random operations, it is convenient to work in the generalized Bloch representation. Any mixed state $\rho$ of size $N$ can be expanded in an operator basis $\{\Lambda_j\}$ consisting of $N^2 - 1$ generators of the group $SU(N)$, normalized as ${\rm Tr}\,\Lambda_i \Lambda_j = \delta_{ij}$, together with the rescaled identity $\Lambda_0 = \mathbb{I}_N/\sqrt{N}$. In the simplest cases, $N=2$ and $N=3$, one uses the rescaled Pauli and Gell-Mann matrices, respectively. The {\sl Bloch representation} then reads
\begin{equation}
\label{Bloch}
\rho = 
\frac{ \mathbb{I}_N}{N} +
\sum_{j=1}^{N^2-1}
  \tau_j \Lambda_j
  {\rm \ \ with \ \ }  
\tau_j = {\rm Tr} \; \rho \Lambda_j,
\end{equation}
so that any state $\rho$ is represented by a Bloch vector $\vec{\tau}$ with $N^2 - 1$ real components $\tau_j$.

In this representation, a quantum operation $\Psi$ acts as an affine transformation of the Bloch vector,
\begin{equation}
\label{Bloch_map}
{\vec \tau}\to {\vec \tau}'=\Psi(\vec \tau)
= C {\vec \tau}+ \vec {\kappa}
{\ \ \rm hence  \ \ }
\Psi= 
\left(
\begin{matrix}1 & 0 \cr
                         \kappa  & C \cr  
                 \end{matrix}
                  \right).
\end{equation}
The entries of the real transformation matrix $C$ of size $N^2 - 1$ can be expressed in terms of the Kraus operators as $C_{ij} = {\rm Tr} \sum_m \Lambda_i K_m \Lambda_j K_m^{\dagger}$, while the components of the translation vector $\vec{\kappa}$ are given by, $\kappa_i = {\rm Tr} \, \Lambda_i \Lambda_0 \sum_m K_m K_m^{\dagger}$, and vanish for unital maps.

The spectrum $\lambda_i$ of the superoperator $\Psi$ consists of the leading Frobenius--Perron eigenvalue $\lambda_1 = 1$, corresponding to the invariant state of the map, together with the eigenvalues of the transformation matrix $C$. For random operations defined by Eq. (\ref{env2}), with environment size $M$ determining the number of Kraus operators, the matrix $C$ can be approximated by a real Ginibre ensemble \cite{BCSZ09,KNPPZ21}. Its spectrum is asymptotically confined to the Girko disk, while the normalization condition implies \cite{BSCSZ10} that its radius scales as $R \sim 1/\sqrt{M}$. In particular, for random maps generated according to the uniform measure, corresponding to $M = N^2$, one obtains $R \sim 1/N$.

The spectral gap, $a=1-|\lambda_2|=1-R$,
determines the speed of convergence 
of any initial state $\rho$ to the invariant state,
$\omega=\Psi(\omega)$,
under $t$-fold action of the map.
Making use of the trace distance,
$d(\rho,\omega)={\rm Tr}|\rho-\omega|$
with $|X|=\sqrt{XX^{\dagger}}$,
one can show  \cite{BSCSZ10}
that for generic random maps $\Psi$
the distance
$d\bigl(\Psi^t(\rho),\omega\bigr)$,
averaged over the set of  initially pure random states,
$\rho=|\phi\rangle \langle \phi|$,
behaves as 
$d(t)\approx \exp(-\alpha t)$ with $\alpha=-\ln R$.
The same properties characterize spectrum
of a deterministic quantum system in the regime
of strong chaos subjected to a quantum measurement.

\begin{figure}[ht]
    \centering    
\includegraphics[width=.58\linewidth]{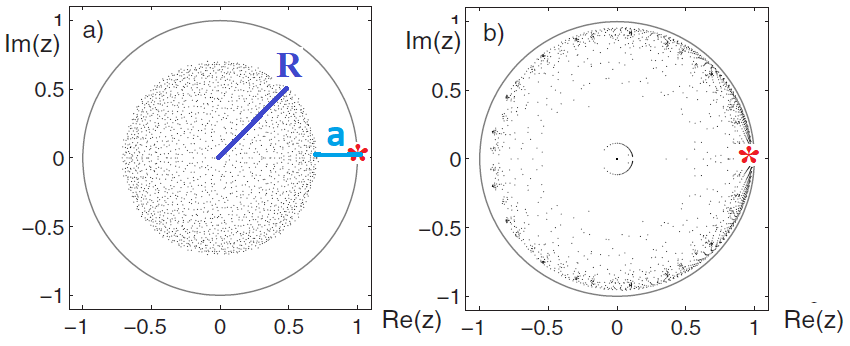}  
    \caption{Spectra of quantum operations 
      in unit disk for generalized quantum
      baker map with tunable degree of chaos 
        \cite{BSCSZ10}
        of dimension $N=64$
      subjected to $M=2$ orthogonal measurement operators.
        a) in strongly chaotic regime
      the spectra display properties
      typical of  random operations: Girko disk of  radius $R\sim 1/\sqrt{M}$ and spectral gap $a=1-R$;
      b) spectra in
      a weakly chaotic regime
        are non universal and system dependent - note lack of
        spectral gap.
      Red star ($\star$) denotes the leading Frobenius--Perron eigenvalue $\lambda_1=1$
      present for all quantum stochastic maps.
    }
    \label{fig:spectra1}
\end{figure}

Fig.\ref{fig:spectra1} shows the spectrum of
superoperator corresponding to quantum
baker map~(\ref{eq:quantum_baker})
modified to allow one to control the degree of chaos \cite{BSCSZ10} 
and subjected to a measurement with $M=2$ generic measurement operators.
As in the case of random maps,
the spectrum consists of the leading eigenvalue,
$\lambda_1=1$,
and the bulk of eigenvalues covering the disk of Girko
of radius $R\sim M^{-1/2}$ almost uniformly.
Note the spectral gap $a\approx 1-1/\sqrt{2}$
plotted in panel a) for strongly chaotic regime.
In the case b) of weak chaos
there is no spectral gap,  as $|\lambda_2|\approx 1$.
      
\medskip 

An important family of unital quantum channel
is formed by {\sl mixed unitary} channels
(\ref{R_rho}),
also called {\sl random external fields},
% cite {Alicki_Lendi} ??
\beq
\label{mixed-unitary}
\rho'= \Phi(\rho) = \sum_{j=1}^M p_j U_j \rho U_j^{\dagger},
{\ \rm with \  } \sum_{j=1}^M p_j =1
\eeq
This expression has a simple interpretation  
from the perspective of the theory of dynamical systems:
choosing at each step one of $M$ unitary dynamics $U_i$ with probability $p_i$
we arrive at a quantum analogue of {\sl iterated functions system} (IFS) 
\cite {LZS03}.
Although each unitary operation $U_i$ represents an isometry, 
in the generic case of $M\ge 2$, positive probabilities $p_i$
and generic, non-commuting  $U_i$, the system has an unique invariant state determined by the eigenvector corresponding 
to the unique leading eigenvalue $\lambda_1=1$ of  superoperator $\Phi$.

In the single qubit case, $N=2$, every unital channel
belongs to this class and is equivalent,
up to a unitary transformation \cite{Bengtsson2007},
to  a {\sl Pauli channel}: convex combination
of identity and three Pauli rotations. 
Such channels are often used to describe action of noise
on desired unitary dynamics $U_1$, 
so one sets $p_1=1-\epsilon$, while other three 
probabilities read  $p_j=\epsilon/3$.
For larger dimensions $N\ge 3$
there exist channel of Landau and Streater~\cite{LS93}
which is unital but not mixed unitary.
Taking a channel $\Psi$ 
and mixing it with complete depolarization,
$\Psi_*(\rho)={\mathbbm I }/N$,
we receive their convex combination,
$\Psi_{\alpha}= \alpha \Psi + (1-\alpha) \Psi_*$.
It was shown by Watrous \cite{Wa09} 
that for any unital  $\Psi$ of an arbitrary dimension $N$
there exist $\alpha >0$ such that 
$\Psi_{\alpha}$ is mixed unitary,
so that the neighborhood of maximal depolarizing channel $\Psi_*$
belongs to the set of mixed unitaries. 
This feature is similar to the separable ball
\cite{Karol1998}
around a maximally mixed state of a bipartite system:
A separable mixed state can be decomposed into a
mixture of product states, while the Choi matrix $D$
of a mixed unitary channel is formed by 
a mixture of maximally entangled states
(\ref{reshuff}).
Furthermore, 
for any unital $\Psi$ there exist natural $k$ such that
its $k$-fold concatenation
$\Psi^k$ becomes mixed unitary \cite{KLPR24}.

% watrous book 

\begin{figure}[ht]
    \centering    
 \includegraphics[width=.81\linewidth]{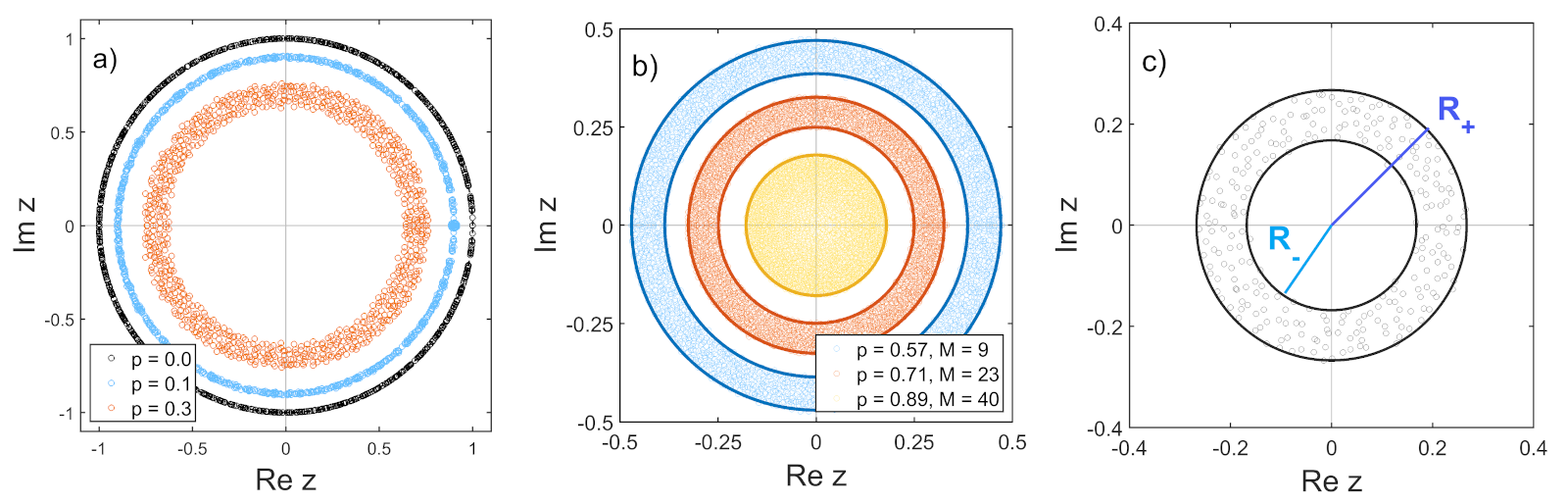} \ \
    \caption{Bulk of the spectra of superoperators for
    a) mixed unitary channel
     (\ref{mixed-unitary})
    acting on $N=30$ system with $M=2$ Haar random unitaries mixed with weights $p$ and $1-p$:  black $p=0$ -- unitary channel with
    spectrum at the unit circle;
     rings of eigenvalues for  $p=0.1$  (blue)    and $p=0.3$ (red);
    b) diluted unitary   channel 
    (\ref{diluted})
    with $p=0.57, M=9$ (blue ring), 
        $p=0.71, M=23$ (red ring),
         $p=0.89, M=40$ (orange disk)
      and radii of the rings given in 
      Eq.~(\ref{ring1}).
      In each case the data obtained for
    a single realization with $N=50$
    show $N^2-1$ eigenvalues,
    as the leading eigenvalue $\lambda_1=1$
    is not plotted here;
    c) random complementary channel
    $\tilde \Psi$
      which maps ${\cal H}_{N=14}$ into       ${\cal H}_{M=18}$
      with size of the ring determined by
      Eq.~(\ref{ring2}).
      }
    \label{fig:spectra2}
\end{figure}

The same form (\ref{mixed-unitary})
of channels is also used with $U_j$
representing random unitary matrices
or evolution operators of quantum chaotic 
systems. Superoperators
$\Psi$ representing such channels
(\ref{super2})
can be approximated as a weighted combination of 
$M$ random unitaries $V_i$ of order $N^2$,
\beq
\label{super3}
\Psi = \sum_{i=1}^{M} p_i U_i \otimes \overline{U_i}
\ \approx \ 
 \sum_{i=1}^{M} p_i V_i
\eeq
Exemplary  spectra of superoperators
obtained
as a mixture of $M=2$ random unitaries
of size $N=30$ with weights $p, 1-p$
are shown in Fig. \ref{fig:spectra2}c.
In the case $p=0$
superoperator (\ref{super2}) is unitary.
For a larger values of $p$ the spectrum 
consists of the leading eigenvalue
$\lambda_1=1$
and  a ring  of eigenvalues,
which  for $p\ge 1/2$ 
forms a disk centered at the origin.

In a particular case of symmetric mixture
and equal weights, $p_i=1/M$,
the superoperator (\ref{super3})
can be approximated by the 
sum of  independent random  unitary matrices,
$\Psi \approx (\sum_{j=1}^M V_j)/M$.
Hence the distribution of its squared singular values, i.e. eigenvalues $x$ of $\Psi\Psi^{\dagger}$,
is asymptotically described by the {\sl Kesten distribution} \cite{Ke59,ZPNC11} 
describing the sum of random unitaries,
\beq
\label{Kesten}
\  P_M(x) = 
\frac{1}{ 2 \pi}
\frac{\sqrt{4M(M-1)x - M^2x^2}}{Mx-x^2}
\  \ \ {\rm for } \ \ \  x\in \Bigl[0,4 \frac{M-1}{M}\Bigr].
\eeq
For a mixture of a large number $M$ of unitaries
this distribution converges to the 
Marchenko-Pastur law 
(\ref{eq:marcenko-pastur-law_Q1})
with support in $[0,4]$.

To describe  the influence of noise
for unitary quantum dynamics
one can use an ensemble
of {\sl diluted unitary} channels,
proposed in \cite{SRCP20}
and successfully applied to describe 
measured spectra of superoperators
corresponding to four-qubit 
circuits implemented on a quantum computer  \cite{WRS26}.
A desired unitary transformation $U$,
weighted by probability $p<1$,
is subjected to a noise modeled by $M$  independent random Kraus operators $K_j$, 
which satisfy the trace condition,
\beq
\label{diluted}
\ \Phi(\rho) = p U \rho U^{\dagger} +(1-p)
  \sum_{j=1}^M K_j \rho K_j^{\dagger},
{\  \ \rm with \  } \sum_{j=1}^M K_j^{\dagger} K_j
 = {\mathbbm I}_N
\eeq

Besides the Frobenius--Perron leading eigenvalue
$\lambda_1=1$,
the bulk of the spectrum forms a disk or an
annulus,
depending on the parameters of the model.
Approximating the superoperator by a 
weighted sum of random unitary and a 
Ginibre matrix, 
is it possible \cite{SRCP20}
to derive the inner and outer radii
of the ring,
\beq
\label{ring1}
R_{\pm} =  \sqrt{(1-p)^2 \pm p^2/M},
\eeq
so the ring reduces to a disk
as $R_-=0$, which happens for
$p\ge p_c=\sqrt{M}/(\sqrt{M}+1)$.
Such exemplary spectra are shown in Fig. 
\ref{fig:spectra2}b
for $N=50$ and three sets of parameters
$p$ and $M$. 

Any trace preserving map $\Psi$
can be written in the environmental form
(\ref{env2}), so that 
the principal state $\rho$ of size $N$
is coupled to the environment $B$ 
of size $M$. If in this expression,
the partial trace 
is performed over the system $A$,
one obtains a {\sl complementary map},
${\tilde\Psi}(\rho)= {\rm Tr_A}  
\bigl(V(\rho \otimes |0\rangle \langle 0|)V^{\dagger}\bigr)$,
which modifies the size of the system,
$\tilde \Psi: {\cal H}_N \to {\cal H}_M$.
Spectral properties
of such maps can be modeled
with {\sl induced Ginibre ensemble} \cite{FBKS12},
which leads to a similar annulus--disk transition of the spectrum. The inner and outer radii of the ring read
\beq
\label{ring2}
R_- =  \sqrt{ 1/N -N/M^2}, \
R_+ =  1/\sqrt{N} \ \
{\rm for } \ \ M\ge N \ \
{\rm and } \ \ 
R_- =  \sqrt{ N/M^2 -1/N}, \
R_+ =  \sqrt{N}/M \ \
{\rm for } \ \ M\le N,
\eeq
so the ring reduces to the disk if
$M=N$ so that $R_-=0$.
Note that in this case
the radius $R_+=1/\sqrt{N}$
differs from the one $R \sim 1/N$
for a generic random operation,
which is obtained 
by setting  in
Eq. (\ref{env2})
$M=N^2$
to assure the full rank of the Choi matrix.

In the ensembles  of mixed unitary operations, diluted unitary channels and complementary random maps,
the spectra form a disk or a ring of eigenvalues, related to
the {\sl single ring theorem} 
 \cite{FSZ01,GKZ11}.
 The outer radius $R_+$
determines the spectral gap, $a=1-R_+$,
and the convergence rate
to the invariant state.
Note that in 
all the models
discussed above 
random unitary matrices can be 
replaced by evolution
operators of quantized
chaotic systems.

% 
% Chaotic dynamics as highly mixing channels

\subsection{Concentration of measure and
non-additivity of channel capacity}

The effect of {\sl concentration of measure},
used in Sec.  \ref{chap2:sec1}
to explain small fluctuations of entropy of a random quantum state 
around the average value, 
is implied by a 1919  
lemma of L{\'e}vy  \cite{Ta96,Le01},
which deals with high-dimensional geometry.
A random point drawn from a ball ${\bf B}^n$ in $n$ dimensions is likely to lie close to the boundary ${\bf S}^{n-1}$, so most of the mass of a high-dimensional ball is concentrated near its surface. If one likes geometric metaphors, one may say that the skin of a high-dimensional fruit contains most of its substance. Fortunately, this observation does not require us to peel an orange in seven dimensions and dispose of the leftovers.

\begin{figure}[ht]
    \centering
\includegraphics[width=.59\linewidth]{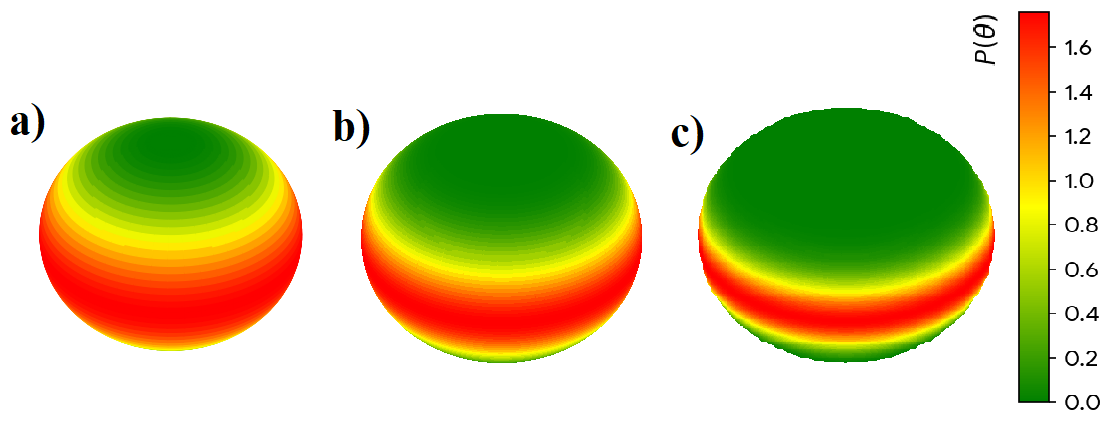}
    \caption{Concentration of measure and {\it fat equator} effect: density on the sphere with distribution of the polar angle, 
    $P(\theta)\sim \sin^{n-1} \theta$,
    corresponding to $n$-sphere 
    is peaked along the equator 
    --
    a) $n=4$, b) $n=8$, c) $n=20$.
    }
    \label{fig:conennr1}
\end{figure}

A closely related phenomenon occurs on the sphere itself. Choose an arbitrary point on the high-dimensional sphere ${\bf S}^{n}$ and declare it to be the north pole. If another point is chosen uniformly at random on the sphere, it is overwhelmingly likely to lie close to the equator. 
In other words, absolute value of the
scalar product of two random unit vectors
in high dimension is typically small, so these vectors are likely to be close to
orthogonal.
This follows directly from the integral representation for the volume of the unit $n$–sphere. In geographic language, the latitude of a random point typically takes values close to zero, so the equator may appropriately be described as {\sl fat} - see Fig.~\ref{fig:conennr1}.

Before continuing with geometric intuition, it is useful to examine a simple probabilistic example that illustrates a similar phenomenon. Consider $n$ independent random variables $\xi_1,\dots,\xi_n$, each taking the values $\pm1$ with equal probability. The expectation value of their sum $\zeta_n=\sum_{i=1}^n\xi_i$ is zero. Indeed, in a long sequence of fair coin tosses the number of heads is typically close to the number of tails.
Moreover, the central limit theorem implies that the fluctuations of $\zeta_n$ are of order $\sqrt{n}$. Although $\zeta_n$ may in principle attain values as large as $n$, it usually remains close to its mean value $E(\zeta_n)=0$. This statement can be quantified by the {\it Hoeffding exponential inequality} -- see \cite{Ta96}.
\begin{equation}
P\Bigl(\;\frac{|\zeta_n|}{n}\ge\epsilon\;\Bigr)
\ \le \
2\exp\; (-n\epsilon^2/2).
\label{exponbound}
\end{equation}
Thus, the probability of observing deviations of size $\epsilon$ decreases exponentially with the number of variables. For large $n$, significant deviations from the mean become extremely unlikely.

At first glance, the geometric observations about spheres and the probabilistic behavior of sums of random variables appear unrelated. In fact, they are manifestations of the same fundamental principle, known as the {\it concentration of measure phenomenon}
\cite{Ta96,Le01,AS17}.
 In high-dimensional spaces, random effects tend to average out, and many quantities become sharply concentrated around typical values.
 Roughly speaking, if a function depends on many independent random variables in a sufficiently balanced and smooth manner, the contributions of randomness compensate each other. As a result, the function fluctuates only weakly around its mean value.

To make this idea precise, one introduces the concept of Lipschitz continuity:
{\sl A function $f$ defined on a set $Y\subset\mathbbm{R}^n$ is called {\it Lipschitz} with constant $\eta$ if for all $x,y\in Y$}
\begin{equation}
|f(x)-f(y)| \ \le\ \eta \; |x-y|.
\label{Lip}
\end{equation}
Intuitively, a Lipschitz function cannot vary too rapidly: changes in its value are bounded by proportional changes in its argument. For example, the linear function $y=ax$ is globally Lipschitz with constant $a$. The quadratic function $f(x)=x^2$ is locally Lipschitz on the interval $[0,b]$ with constant $\eta=2b$, whereas the function $\sqrt{x}$ is not Lipschitz on $[0,1]$ because its derivative diverges at the origin.
The importance of the Lipschitz property becomes clear in the following fundamental result.

{\bf L{\'e}vy’s Lemma.}
{\sl Let $f:{\bf S}^{n-1}\rightarrow\mathbbm{R}$ be a Lipschitz function with constant $\eta$. Let $\langle f\rangle$ denote is mean value. If a point $x$ is chosen uniformly at random from the sphere ${\bf S}^{n-1}$ with $n>2$, then}
\begin{equation}
P\bigl(|f(x)-\langle f\rangle|>\epsilon\bigr)
\ \le \
2\exp \left(-\frac{(n-1)\epsilon^2}{2\eta^2}\right).
\label{Levy00}
\end{equation}
To see L{\'e}vy’s lemma in action, consider the function given by the first Cartesian coordinate of a point on ${\bf S}^{n-1}$,
In this case the lemma quantifies the “fat equator’’ phenomenon visualized in
Fig. \ref{fig:conennr1}. The probability that a random point lies outside an equatorial band of width $2\epsilon$ is bounded by $2\exp(-n\epsilon^2/2)$ and therefore decreases exponentially with the dimension.

The relevance of these ideas becomes particularly striking in quantum theory. A pure quantum state in a Hilbert space of dimension $N$ can be represented as a Hopf circle on the sphere ${\bf S}^{2N-1}$. This observation allows one to apply 
L{\'e}vy’s lemma to study typical properties of quantum states.
Consider, for example, a random pure state $|\psi\rangle$ of a bipartite system of dimensions $N\times N$. 
The partial trace of the projector,
$\rho={\rm Tr}_N
|\psi\rangle \langle \psi|$
forms a state of order $N$
distributed according to the Hilbert–Schmidt measure. 
Its average von Neumann entropy 
$\langle S\rangle $
behaves asymptotically as
follows from Eq.~(\ref{eq:PageLubkin}).
 
L{\'e}vy’s lemma implies that the probability of large deviations from this average is exponentially small,
\begin{equation}
P\Bigl(|S(\rho)-\langle S\rangle|\ge\epsilon\Bigr)
\le
2\exp\left(
-\frac{(N^2-1)\epsilon^2}{8(\pi\ln N)^2}
\right).
\label{Leventropy}
\end{equation}

Hayden, Leung, and Winter  \cite{HLW06} 
derived this bound by considering the function $f(\psi)=S(\rho(\psi))$, estimating its Lipschitz constant $\eta\sim\ln N$, and applying L{\'e}vy’s lemma. Thus for large $N$ the entropy of a random state is overwhelmingly likely to be close to its mean value. Since this entropy measures the entanglement of the corresponding bipartite pure state, the result implies that a generic bipartite pure state is highly entangled.
Similar arguments can be applied to other quantities, such as the purity $\langle\mathrm{Tr}\rho^2\rangle$ of random mixed states \cite{DOP07}. In this formulation the problem naturally connects with random matrix theory.

\begin{figure}[ht]
    \centering
\includegraphics[width=.55\linewidth]{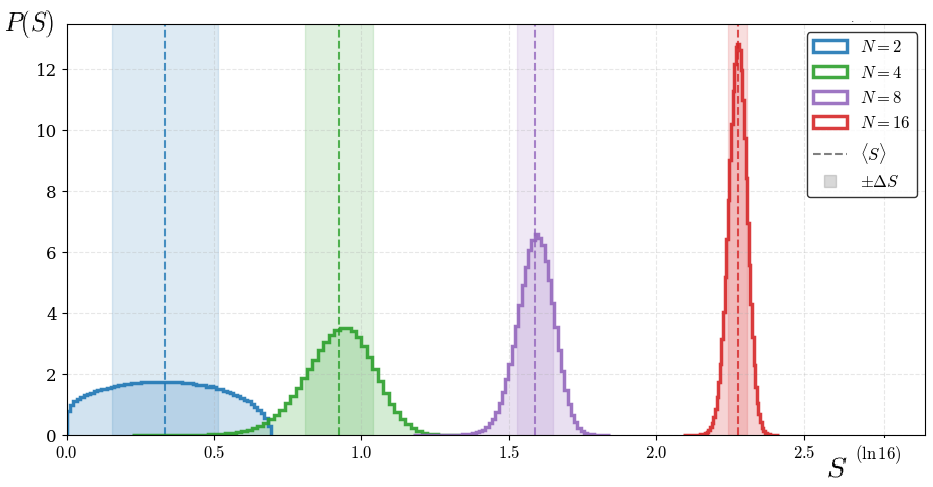}
    \caption{Concentration of measure for entanglement entropy $S$ for random pure states of $N \times N$ system
    for $N=2,4,8$ and $N=16$.
    The distributions are peaked 
     around the mean value 
     (\ref{eq:PageLubkin}),
     % ${\rangle S} \approx ln N -1/2$,
     while their
     concentration grows with $N$ 
     according to Eq. (\ref{Leventropy}).}
    \label{fig:conennr2}
\end{figure}

Consider the fixed-trace Wishart matrix (\ref{Wishart}), distributed according to the Hilbert–Schmidt measure. The eigenvalue density of such states converges asymptotically to the Marchenko–Pastur distribution (\ref{eq:marcenko-pastur-law}). Concentration of measure allows us to strengthen this statement: a single large random matrix $\rho$ almost surely exhibits 
similar spectral distribution.
Analogous reasoning applies to quantum channels. A random completely positive trace-preserving map $\Phi$, distributed according to the Hilbert–Schmidt measure, can be represented by a Choi matrix $D$ obtained from a Wishart matrix of size $N^2$ and normalized so that 
$\mathrm{Tr}_A D=\mathbb{I}_N$ -- see Sec. 6.1.
For a large $N$ the complementary partial trace $D_A=\mathrm{Tr}_B D$ approaches the identity. Consequently, a typical random channel becomes nearly bistochastic and maps the maximally mixed state $\rho*=\mathbbm{1}/N$ close to itself. The output entropy is therefore almost surely close to $\ln N$.

%{ \color{blue}  
%a relation to quantum chaos will be added here:

Quantum chaotic systems give rise to operators whose properties are well described by random matrix theory. In other words, they represent the {\sl mean behavior}, averaged over the entire space of operators. 
Due to L{\'e}vy’s lemma, large deviations from this typical behavior become increasingly unlikely as the dimension grows. 
% fine K  {\color{blue}
This is consistent with the Bohigas--Giannoni--Schmit conjecture \cite{BGS84}
and  the starting phrase of the introduction:
Within a given symmetry class, quantum chaotic systems are all alike:
they exhibit universal features and share the same spectral properties.
% }

Beyond these deep geometric insights,
which explain observed universality, concentration of measure has had profound consequences for quantum information theory. A notable example concerns the long-standing additivity conjecture for the {\sl Holevo  quantity} $\chi$ of a quantum channel,
%$\Phi$  defined as
\begin{equation}
\label{chi}
\chi(\Phi) = \max_{\{p_i,\rho_i\}} \left[ 
S\!\left( \sum_i p_i \,\Phi(\rho_i) \right)
- \sum_i p_i \, S\!\left(\Phi(\rho_i)\right)
\right],
\end{equation}
where the maximum is taken over all ensembles $\{p_i,\rho_i\}$ of input states. 
This quantity, based on
concavity of the von Neumann entropy $S$,  measures
 the maximal amount of classical information that can be transmitted through a quantum channel \cite{Wa18}.
 It had been conjectured that $\chi(\Phi^{\otimes n})=n\chi(\Phi)$, which would imply that the classical capacity of a quantum channel equals its single-use Holevo information.
Shor showed \cite{Sh04} that this conjecture is equivalent to additivity of the minimal output entropy,
$S_{\rm min}(\Phi_1 \otimes \Phi_2) \ = \ S_{\rm min}(\Phi_1)+S_{\rm min}(\Phi_2)$,
  where $S_{\rm min}(\Phi)=\min_{\rho} S(\Phi(\rho))$.

The latter quantity is easier to handle and 
it allowed Hastings to obtain the following  breakthrough result \cite{Ha09}.

{\bf Hastings’ nonadditivity theorem.}
{\sl There exist quantum channels $\Phi_1$ and $\Phi_2$ 
for which the minimal output entropy is strictly subadditive}
\begin{equation}
S_{\rm min}(\Phi_1 \otimes \Phi_2) \ < \ S_{\rm min}(\Phi_1)+S_{\rm min}(\Phi_2),
 \label{Hastings2}
\end{equation} 

The proof is nonconstructive and relies on generic mixed unitary channels
(\ref{mixed-unitary})
 combined with concentration-of-measure arguments. It shows that correlated input states may produce lower output entropy—and thus greater resistance to noise—than any product input state. 
The original  probabilistic counterexample 
can be related \cite{AS17}
to a version of the
known theorem by Dvoretzky, which states 
that a generic cross-section of a high-dimensional  convex body is looks almost like an Euclidean ball. 

%\textcolor{blue}{
We review the key elements of a simplified formulation due to  Brand{\~a}o 
and M.~Horodecki \cite{BH10}.
Consider a random operation $\Phi$, complementary to (\ref{env2}), acting on a density matrix $\rho_A$ of size $N$ as
$\Phi(\rho_A)={\rm Tr}_A \bigl[U(\rho \otimes |0\rangle\langle 0|)U^{\dagger}\bigr]$,
where $U\in U(NM)$. Since the original subspace ${\cal H}_A$ is traced out, the output state $\sigma_B$ has dimension $M$. We focus on the regime $N\gg M$. The conjugated  map $\bar \Phi$ is defined by replacing $U$ with its complex conjugate $\bar U$,
so that superoperator (\ref{super2})
is obtained by exchanging both subsystems.
%%%
%%% I replaced Tr_N by Tr_A, rho by rho_A and \sigma by \sigma_B. I also removed "environment" as referring to B subsystem, as it is interpreted sometimes (as in BH10) as the subsystem that is traced out: here A.
%also replaced the B states as |0X0|.
%% OK - fine KZ !
%rewrote a little
%%
%At first, one establishes the lower bound,
The proof starts by observing that 
%% I removed the lower bound typo, it is anyway upperbound.
\begin{equation}
S_{\rm min}(\Phi \otimes {\bar \Phi})\  
    \le \  2 \ln M - \frac {\ln M}{M}.
 \label{BH1}
\end{equation}
This is achieved by bounding the operator norm of the image of the maximally entangled state
$\ketbra{\phi^+}$ 
with respect to the tensor map,
$||(\Phi \otimes { \bar \Phi})\ketbra{\phi^+}||_{\infty} \ge 1/M$.
%where 
%$|\phi^+\rangle= \frac{1}{\sqrt{N}} \sum_{i=1}^N |i,i\rangle$.
% We have used this notation before, so I removed this. 
In the second (and crucial) step, 
L{\'e}vy's lemma is applied, together with a careful estimate of the Lipschitz constant, to show that the probability that the map $\Phi$ sends a random pure state to a low-entropy state is negligible.
One selects a positive constant $c_0>0$ and demonstrates that for any $c\ge c_0$ the following lower bound for the minimal output entropy of a single map holds almost surely,
\begin{equation}
  \ln M - \frac {c}{M} \ \le \  S_{\rm min}(\Phi) .
 \label{BH2}
\end{equation}
%% AL: I rearranged the para and put the Levy part first and stated the lower bound in the end.
%% fine K
To complete the proof, one combines inequality (\ref{BH1}) with twice inequality (\ref{BH2}). Using the identity for dual maps,
$S_{\rm min}(\Phi)= S_{\rm min}
( {\bar \Phi})$,
the resulting expression can be rewritten as
\begin{equation}
S_{\rm min}(\Phi \otimes {\bar \Phi}) \ \le \
S_{\rm min}(\Phi) + S_{\rm min}({\bar \Phi})
- \frac {(\ln M -2c)}{M}.
 \label{BH3}
\end{equation}
By choosing the dimension $M$ sufficiently large, the term in parentheses becomes strictly positive, which implies that additivity is generically violated in this setting.

Note that the above reasoning is non-constructive, in line with Hastings’
%% AL Replaced "analogy" wit "line"
%% OK K
original proof: we conclude that there exist quantum channels $\Phi_1$ and $\Phi_2$ for which inequality (\ref{Hastings2}) holds,  
without explicitly constructing them. 
%\textcolor{blue}{
This provides another instance of a hay-in-a-haystack problem, where proving that a generic rule is true is easier than a specific instance. Thus, while explicit constructions have been found when the R\'enyi-$q$ entropies, Eq.~\eqref{eq:Renyi-Tsallis_defn}, for $q \neq 1$ is used  \cite{Derksen_constructive_counter} it still remains open for the crucial case of the 
von~Neumann entropy, $q=1$.

Since the minimal output entropy is strictly subadditive (\ref{Hastings2}), the Holevo information $\chi$,
which involves maximizing an entropy difference, is strictly superadditive \cite{Ha09}. That is, there exist quantum channels for which
\beq
\label{chi2}
\chi(\Phi_1 \otimes \Phi_2) > \chi(\Phi_1)+\chi(\Phi_2).
\eeq
This crucial  inequality implies that
while sending information through two 
quantum channels it  may be 
superior  to use both channels 
in parallel and act on a correlated input state of the composed system.
Determining the ultimate classical capacity of a quantum channel therefore involves optimization over spaces of arbitrarily high dimension—one of the key conceptual differences between quantum and classical information theory.

It is 
tempting to look at the additivity problem from the point of view of quantum chaos. As it is known that unitary operators
describing single step quantized chaotic dynamics displays features of Haar random matrices and $t-$designs, one could try to find dynamical systems
%in this manner constructive counterexample for the
that violate the additivity conjecture. Consider, for instance
two strongly coupled kicked tops, with spins $j_1$ and $j_2$, respectively, in a chaotic regime.
As partial traces over the first and the 
second subsystem serve as examples of pseudo-random
operations 
\cite{MilSar1999,BanLak2004,PPZ16},
one might fix the classical parameters of the system in the chaotic regimes
and look for dimensions $N=2j_1+1$ and $M=2j_2+1$,
for which the non-additivity could be confirmed.
This task looks difficult without new ideas 
to bound the entropy, as 
rough estimations show \cite{FKM10}
that to achieve a visible gain of entropy $\Delta S$ 
% of order $10^{-5}$ 
one has to work  with huge dimensions still.
%{\color{blue}
However, the above result demonstrates that the interplay between quantum information theory and quantum chaos is deeper and more intricate than one might have initially anticipated.%}
%% Karol, do  you want ... at the end, I simply made it a fullstop!
%OK  K.

\medskip

%\begin{itemize}
%  \item 
%  \item Higher-order statistics: Weingarten calculus, unitary $k$-designs.
%  \item Implication for QI: certifying device randomness via chaotic evolution; chaotic systems as natural resources for approximate $t$-designs.
% 
% \item  
%\end{itemize}
%
%a link to resource theory \cite{CG19}
%% perhaps redundant right now
% as we removed resources from the title

\section{Outlook and Open Questions}

The interplay between quantum chaos and quantum information is the subject of
vigorous and increasingly multidisciplinary investigation. In this contribution
we have focused on a few, hopefully representative, aspects of this rich
tapestry, whose central unifying theme is entropy in its many avatars—from
disorder to information, from classical chaos to quantum entanglement.

Classical dynamical indicators such as the Kolmogorov--Sinai (KS) entropy~\cite{ozorio1989,ott-2002}
and Lyapunov exponents have found partial quantum counterparts. In particular,
the short-time growth of entanglement entropy in chaotic systems has been linked
to the KS entropy, while the largest Lyapunov exponent manifests in the growth
of phase-space measures such as the Wehrl entropy and in out-of-time-order
correlators (OTOCs)~\cite{Roberts2017Chaos}. Nevertheless, it remains fair to say that a
fully satisfactory and unambiguous quantum analogue of the KS entropy is still
lacking, despite longstanding efforts~\cite{CNT87,SZ94,
alicki1996quantum,RobertAlicki2004}.

Quantum chaos, through its intrinsic tendency to scramble information across
degrees of freedom, has long been viewed as an obstacle to coherent quantum
information processing~\cite{NielsenChuang}. More recently, its competition with
localization phenomena, even in the absence of explicit computation, has been highlighted~\cite{Berke_2022}. At the same time, as emphasized in this
review, scrambling and randomness are valuable resources in quantum information
protocols. Chaotic dynamics can generate highly entangled states and approximate
Haar randomness, suggesting a dynamical route to state and unitary $t$-designs.

This perspective is particularly compelling in systems with well-defined
classical limits, such as quantum maps, where the onset of chaos can be tuned
and analyzed semiclassically. This stands in contrast to generic many-body
systems, where randomness is typically inferred indirectly through spectral
statistics consistent with random matrix theory~\cite{EWSL03}. Moreover, random
subspaces are known to achieve the capacity of noisy quantum channels~\cite{Hayden_Decoupling_2008},
and chaotic dynamics may provide physically realizable approximations to such
encodings. The extreme scrambling exhibited by chaotic systems also underlies
information-theoretic phenomena such as the Hayden--Preskill recovery protocol
for black holes~\cite{Hayden_2007}.

Beyond these foundational connections, recent work has identified unexpected
applications of chaos. For instance, the quantum Fisher information (QFI) has
revealed that chaotic dynamics can enhance parameter estimation precision in
quantum sensing tasks, outperforming integrable systems~\cite{Fiderer2018}. Apart from entanglement, other measures of nonlocality, as measured by violation of Bell inequalities, are of interest. In this context, intriguing connections have been observed between the fluctuation properties of the spectra of Bell operators and maximal violations \cite{Aloy2026}.
Open quantum systems provide another frontier, where the characterization of
quantum chaos and the formulation of analogues of the
Bohigas--Giannoni--Schmit conjecture 
\cite{BGS84}
remain active areas of research.

\medskip

Despite these advances, several questions, such as the following, remain open.

\paragraph{Quantum analogues of classical  
dynamical
entropy.}
While signatures of classical chaos appear in short-time dynamics, a universally
accepted quantum analogue of the KS entropy that governs long-time behavior is
still missing. Clarifying this connection, and understanding the crossover from
semiclassical to random matrix regimes, remains a central challenge.

\paragraph{Dynamical generation of designs.}
Chaotic systems appear to generate approximate state and unitary designs, but
the scaling with system size, evolution time, and perturbation strength is not
well understood. In particular, determining the minimal resources required for
design formation, and the role of integrability or mixed phase space in
hindering it, are important open problems.

\paragraph{Structure beyond randomness.}
While random matrix theory captures universal spectral fluctuations, deviations
arising from symmetries, arithmetic structure, or nonuniform hyperbolicity
suggest a richer landscape. Understanding how such structures influence
entanglement, scrambling, and pseudorandomness is an ongoing challenge.

\paragraph{Open and noisy systems.}
Realistic quantum systems are inherently open. How decoherence, measurements,
and environmental coupling modify signatures of chaos, scrambling, and
information flow—particularly in connection with measurement-induced phase
transitions—remains to be fully understood.

\paragraph{Experimental realization.}
Although simple quantum chaotic systems such as the baker's map have been
implemented in small-scale platforms~\cite{Weinstein2002}, scalable realizations on
modern quantum architectures are still limited. Identifying experimentally
accessible diagnostics of chaos and randomness that are robust to noise is an
important direction for future work.

\medskip

{\sl Quantum chaos} occupies a dual role: it acts both as a mechanism for the rapid
decoherence of quantum information and as a generator of entanglement and
pseudorandomness—key resources for quantum technologies. Developing a unified
framework that reconciles these opposing aspects, while incorporating both
universal features and system-specific structures, remains a central goal.
The resolution of these questions promises not only deeper insight into the
foundations of quantum mechanics, but also practical advances in quantum
information science.

\vskip 0.5cm

%{\color{red} 
%  do we wish to discuss the topics below?
%  }
%\begin{itemize}
%{\color{blue} 
%\item observation entropy and dynamical KS entropy  (K)

%\item OTOC + information scrambling (A)

%\item  hyper sensitivity  - Shack Caves  (K)
%}

% possible citation to 
% {Pl25} M. P{\l}odzie{'n},  Lecture Notes on Information Scrambling, Quantum Chaos, and Haar-Random States,
%preprint  arXiv:2511.14397

 % \item Black holes as ultimate ``chaotic quantum computers.''
%  \item Resource note: open problems (universality classes of entanglement growth, chaos--error correction dualities, chaos in hybrid quantum--classical learning).
%\end{itemize}

% {\color{blue} We conclude this contribution with some general remarks on the impact of quantum chaos on quantum information. On one hand, coupling a principal system to an environment modeled by a quantum chaotic system inevitably leads to the decoherence of quantum information. On the other hand, quantum chaotic systems can generate entanglement—the fundamental resource of quantum information science. It is the quantum entanglement which allows one for  encoding of a single qubit into larger quantum systems, making the information more robust against environmental interactions that affect only a small subset of qubits. As several key questions at the intersection of quantum chaos and quantum information remain unresolved, we have the pleasure of inviting the reader to contribute to this challenging and evolving field.}

\begin{ack}[Acknowledgments]

  It is a pleasure to thank
  Sergey Denisov and
  Mykhailo Hontarenko 
  for preparing Figs.
  \ref{fig:spectra2} and
  \ref{fig:conennr2}, 
  respectively.
  We are grateful to 
 Silvia Pappalardi 
 %{\color{red} ADD} 
 for valuable discussions
and remarks on the preliminary version of this work.
\end{ack}

%\seealso{contribution {\sl Chaos and random matrices} - in this volume}
% I do not think we need to place this remark in the first arxive version.
% (It is NOT a 'volume' so far :)
% We shall see all the contributions 
% submitted and then decide what to cite
% in the final version

\bibliographystyle{JHEP}%
\bibliography{reference}

\end{document}